\documentclass[11pt,paper]{article}
\usepackage{jheppub}

\usepackage[usenames,dvipsnames,table]{xcolor}
\usepackage{graphicx,amsmath,amssymb,multirow,array,bm,mathrsfs}
\usepackage{epsf,amsfonts}
\usepackage[numbers,sort&compress]{natbib}

\usepackage{slashed}

\newcommand{\fixform}[1]{\texorpdfstring{#1}{~}}

\newcommand{\beq}{\begin{equation}}
\newcommand{\eeq}{\end{equation}}
\newcommand{\bea}{\begin{eqnarray}}
\newcommand{\eea}{\end{eqnarray}}

\newcommand{\half}{\frac{1}{2}}

\newcommand{\tr}{\mbox{tr}}

\newcommand{\form}[1]{\bm{#1}}

\newcommand{\hodge}{{}^\star}

\newcommand{\fomega}{\form{\omega}}

\newcommand{\fA}{\form{A}}

\newcommand{\fF}{\form{F}}

\newcommand{\fGamma}{\form{\Gamma}}

\newcommand{\fR}{\form{R}}

\newcommand{\fP}{{\form{\mathcal{P}}}}

\newcommand{\ICS}{{\form{I}}_{\rm CS}}

\newcommand{\fL}{\form{L}}

\newcommand{\fK}{\form{\K}}

\newcommand{\GN}{G_{_N}}

\newcommand{\zbar}{\bar{z}}

\newcommand{\Panom}{\fP_{\rm anom}}
\newcommand{\fPanom}{\Panom}

\newcommand{\ve}{\epsilon}
\newcommand{\fr}{\form{r}}
\newcommand{\fsigma}{\form{\sigma}}

\newcommand{\epsUUab}{\ve^{ab}}
\newcommand{\epsddab}{\ve_{ab}}
\newcommand{\epsddcd}{\ve_{cd}}

\newcommand{\iepsUUab}{i\ve^{ab}}
\newcommand{\iepsUdab}{i\ve^{a}{}_{b}}
\newcommand{\iepsddab}{i\ve_{ab}}

\newcommand{\miepsab}{(-i\ve_{ab})}

\newcommand{\fw}{\form{w}}
\newcommand{\pa}{\partial}
\newcommand{\m}{\mu}
\renewcommand{\r}{\rho}
\newcommand{\n}{\nu}
\newcommand{\s}{\sigma}
\newcommand{\lt}{\left}
\newcommand{\rt}{\right}
\newcommand{\SeeCS}{S_{\text{ent,CS}}}
\newcommand{\See}{S_{\text{ent}}}

\newcommand{\Sdenanom}{\form{{\cal S}}_{\rm ent}}

\newcommand{\be}{\begin{equation}}
\newcommand{\ee}{\end{equation}}

\def\KK{\bar{\fK}\cdot\fK}

\def\GN{(\Gamma_N)}
\def\RN{(R_N)}
\def\fGN{\form{\Gamma}_N}
\def\fRN{\form{R}_N}

\def\SEEWald{S_{\rm ent}^{\rm Wald}}
\def\SEEKK{S_{\rm ent}^{KK}}

\def\fK{\form{K}}
\def\fKbar{\bar{\form{K}}}
\def\Dhat{\hat{D}}
\newcommand{\Seeetr}{S_{\rm ent}^{KK,\, {\rm extra}}}
\newcommand{\SeeWaldetr}{S_{\rm ent}^{{\rm Wald},\, {\rm extra}}}
\def\Sigmatilde{{\Sigma\rtimes I
}}

\def\onCone{\text{Cone}_{\Sigmatilde, n}}
\def\onConeCS{\text{Cone}_{\Sigma, n}}

\title{Holographic Entanglement for Chern-Simons Terms}

\author[a]{Tatsuo Azeyanagi,}
\author[b]{R. Loganayagam}
\author[c]{and Gim Seng Ng}

\affiliation[a]{
D\'{e}partement de Physique, Ecole Normale Sup\'{e}rieure, CNRS, 24 rue Lhomond,
75005 Paris, France.}
\affiliation[b]{School of Natural Sciences, Institute for Advanced Study, Princeton, NJ 08540, USA.}
\affiliation[c]{Department of Physics, McGill University, Montr\'{e}al, QC H3A 2T8, Canada.}

\emailAdd{tatsuo.azeyanagi@phys.ens.fr}
\emailAdd{nayagam@ias.edu}
\emailAdd{gim.ng@mcgill.ca}
\abstract{
We derive the holographic entanglement entropy contribution from pure and mixed gravitational Chern-Simons(CS) terms in AdS$_{2k+1}$. 
This is done through two different methods : first, by a direct evaluation of CS action in a holographic replica geometry and second by a
descent of Dong's derivation applied to the corresponding anomaly polynomial.  In lower dimensions $(k=1,2)$, the formula
coincides with the Tachikawa formula for black hole entropy from gravitational CS terms. New extrinsic curvature corrections appear for $k\geq 3$ : 
we give explicit  and concise expressions for the two pure gravitational CS terms in AdS$_7$ and present various consistency checks, including agreements with the black hole entropy formula when evaluated at the bifurcation surface.
}

\begin{document}
\maketitle

\section{Introduction} 
\label{sec:intro}

\subsection{Overview}
AdS/CFT takes a state of CFT and recasts into geometry on the AdS side. While we understand a lot about how this dictionary works,
a clear cut algorithm on the field theory side to construct the dual geometry is missing. Whatever the final algorithm be, it is becoming 
increasingly clear that entanglement measures in the CFT state should play a crucial role in the final 
answer \cite{VanRaamsdonk:2010pw,Bianchi:2012ev}. 

Recent years have seen an expanding interest in entanglement measures in holography. Consequently, we will refrain from 
giving an exhaustive list and confine ourselves to the works crucial to understanding our paper.  
We begin with the work by Lewkowycz-Maldacena \cite{Lewkowycz:2013nqa} which clarified the subtleties in holographic replica trick and proved Ryu-Takayanagi
formula \cite{Ryu:2006bv,Ryu:2006ef} within their framework.\footnote{See also an earlier work \cite{Fursaev:2006ih} toward the proof of Ryu-Takayanagi formula 
and a criticism of this earlier proof can be found in \cite{Headrick:2010zt}.}
This was followed by Dong \cite{Dong:2013qoa} (and  partly by Camps in \cite{Camps:2013zua}) 
who used the holographic replica trick 
to derive an interesting formula for the entanglement entropy of higher derivative theories with a Lagrangian density depending only on Riemann tensor.\footnote{Works preceding Dong dealing with higher derivative actions include \cite{Ogawa:2011fw,deBoer:2011wk, Casini:2011kv, Hung:2011xb, Bhattacharyya:2013jma,Bhattacharyya:2013gra, Fursaev:2013fta, Miao:2013nfa}.
For some recent works relevant to the higher-derivative correction to holographic entanglement entropy formula, 
see \cite{Bhattacharyya:2014yga,Banerjee:2014oaa, Bueno:2015xda,Camps:2014voa,Erdmenger:2014tba,Wall:2015raa,Bhattacharjee:2015yaa,Astaneh:2014sma,Astaneh:2014wxg,Astaneh:2015tea,Miao:2014nxa,Miao:2015iba,Huang:2015zua}. See  \cite{Miao:2014nxa}
for a generalization to the case involving derivatives of the Riemann tensor.
} 
The holographic entanglement entropy formula due to Dong takes the schematic form
\bea
 \See&=& \int_{\Sigma}\left[ \left(\frac{\partial \fL}{\partial \fR}\right)
 +\fK \left(\frac{\partial^2 \fL}{\partial \fR \partial \fR}\right)\fK
 \right]\,,
\eea  where the $\fL$ is the gravity Lagrangian density, $\fR$ is the Riemann curvature and $\fK$'s denote the extrinsic curvatures.

The above formula and its descendants end up producing a bewildering series of extrinsic curvature terms which fall 
out at the end of a somewhat technical computation. Unlike the leading Wald term, which Wald \cite{Wald:1993nt} and Iyer-Wald \cite{Iyer:1994ys} 
famously interpreted as a Noether charge and can be written in a very simple form, it is unclear how  
should one physically interpret these extrinsic corrections. It is also pertinent to ask what would be a good way of organizing and
rewriting these extrinsic terms which makes their physical content transparent. Ideally, we would like to have a 
structural understanding of these terms  using  general properties of (holographic) entanglement entropy and give alternate derivations
which clarify why they should be there.\footnote{Recently it has been suggested in \cite{Wall:2015raa} 
that linearized second law can give an alternative derivation of  some of the extrinsic corrections.} It would also be useful to construct independent checks which does not 
rely on holographic replica trick.

In this article, we will focus on a special class of higher derivative terms : the pure and mixed gravitational Chern-Simons (CS) terms.
The universal nature of Chern-Simons terms along with their relation to anomalies of the dual CFT make them a perfect laboratory
to study the 
extrinsic corrections mentioned above. Given that the Wald-Tachikawa entropy \cite{Tachikawa:2006sz,Bonora:2011gz}\footnote{We note that the original derivation by Wald was applicable only to covariant Lagrangian densities. This derivation was later generalized to 3d and higher dimensional CS terms on the work by Tachikawa\cite{Tachikawa:2006sz}. The special case of 3d gravitational CS terms was however already known from other points of view\cite{Solodukhin:2005ah,Kraus:2006wn,Sahoo:2006vz} .  The covariance issues in  Tachikawa's original derivation were pointed out in \cite{Bonora:2011gz} and eventually resolved in \cite{Azeyanagi:2014sna} where the  general BH entropy formula (see Eq.~\eqref{eq:tachikawa-entropy}) for arbitrary CS terms was written down. We will refer to this formula as Tachikawa formula in the following.
} of large charged rotating AdS black holes has been 
reproduced by analyzing CFT anomalies \cite{Jensen:2013kka,Azeyanagi:2015gqa}, 
it is plausible that the extrinsic corrections \`a la  Dong \cite{Dong:2013qoa}  can also be explained by a careful
study of entanglement entropy in the presence of anomalies. 
This work was initiated in the hope that if we examine the extrinsic curvature corrections in the case of Chern-Simons terms,  it would help us clarify the structure of these corrections and possibly even derive them from CFT considerations.

One way to think about these extrinsic curvature correction terms  
in the context of Noether method is to think of them as a particular choice of pre-symplectic
current that is natural from the entanglement entropy point of view. From this point of view, the question can be rephrased as to why this
particular choice is privileged over others. In a previous paper \cite{Azeyanagi:2014sna}, we have pointed out that in the case of Chern-Simons terms Wald's prescription 
for pre-symplectic current needs to be modified to preserve covariance under diffeomorphisms and gauge transformations. This, apart from 
resolving conceptual issues about covariance in previous works  \cite{Tachikawa:2006sz,Bonora:2011gz}, also reproduces the correct 
odd parity  Cardy type formula for higher dimensional black holes \cite{Azeyanagi:2013xea,Azeyanagi:2015gqa}. Thus, another
motivation for studying  extrinsic corrections for Chern-Simons terms is to compare these two deformations of the pre-symplectic current.
Given the explicit charged rotating black holes with Chern-Simons terms constructed in \cite{Azeyanagi:2013xea} and the interesting way in which 
anomalies are encoded in their geometry, it is plausible to hope that detailed studies of entanglement probes and the extrinsic corrections
in these geometries would give us more  physical insight into both anomalies and entanglement entropy.

In this paper, we will take the first step towards these goals by repeating the holographic replica calculation for a gravitational theory with 
CS terms and leave a detailed analysis of its structure to future work. In particular, we will postpone the important 
problem of how to determine the entangling surface from the bulk equations of motion. Our focus here would be to get the 
entanglement functional which when integrated over an appropriate  surface would give us the entanglement entropy.
This computation, while a slight generalization of \cite{Dong:2013qoa}, 
presents its own peculiar subtleties due to the non-covariance of CS terms. The Dong formula thus does not apply directly to our case. This state
 of affairs can be remedied in two ways - the first is to just repeat the holographic replica method, face the subtleties carefully and get the corresponding formula with Chern-Simons terms. The case of gravitational CS term on  AdS$_3$ has already been studied in \cite{Castro:2014tta}\footnote{Previous discussions of this problem include \cite{Sun:2008uf, Alishahiha:2013zta}.} 
 and it was argued there was no extrinsic curvature dependence over and above
the Tachikawa formula for BH entropy \cite{Tachikawa:2006sz,Bonora:2011gz,Azeyanagi:2014sna} .  We will extend their computation to higher dimensions : mixed gravitational  CS 
term in AdS$_5$ along with pure/mixed gravitational  CS terms in AdS$_7$. While the mixed CS terms are a straightforward generalization
of AdS$_3$ case and show no corrections to the  Tachikawa formula, the pure gravitational CS terms in AdS$_7$ show an interesting dependence on extrinsic curvatures.

While this first method is a direct generalization of Dong's original derivation and a simple abstract formula can be written down for any CS term (See Eq.~\eqref{eq:CSDong}),
the complicated Christoffel connection dependence in higher dimensional CS  terms makes it more and more tedious to evaluate our formula
explicitly as we move higher in dimensions.\footnote{Perhaps a clever reader can exploit special properties of CS terms  to simplify our formula and its evaluation.}
Keeping this in mind, we also present an indirect but an easier method to arrive at the same answer for any CS term by 
using the corresponding anomaly polynomial (or the characteristic class). The idea here is to imagine the CS term $\ICS$ as descending from a covariant theory 
living in one higher dimension with an action $d\ICS= \fPanom$. In the first step, we calculate the holographic replica action in one higher dimension.
As we verify in our Appendix \ref{sec:anomalymethod2} , because of covariance, the answer can also be directly obtained from Dong's formula. 
In the second step, we write the answer as a pure boundary contribution and identify that boundary contribution
with the holographic replica action of CS terms. In all cases computed, our answers by this method matches with a direct evaluation of CS action.

\subsection{Summary of Main Results}
For reader's convenience, 
 we summarize here the  main results obtained in this paper ---
the derivation of holographic formulae for 
Chern-Simons contribution to entanglement entropy.
Our results for gravitational and mixed Chern-Simons terms 
 in AdS$_{d+1}$ with $d=2k$ ($k=1, 2, 3$) are as follows :\footnote{Due to the simplicity of the expression, we write down 
 the anomaly polynomials $d\ICS= \fPanom$ instead of Chern-Simons terms themselves.}
  \begin{eqnarray}
&& {\rm AdS}_3 \,\, {\rm with}\,\, \fPanom= c\, \tr(\fR^2) \qquad\,\,\,\,\,\quad: \qquad  
  \SeeCS= 2(2\pi) (2c)\int_{\Sigma }  \fGN\,, \label{eq:trR2} \\
&&  {\rm AdS}_5 \,\, {\rm with}\,\, \fPanom= c\, {\fF}\wedge \tr(\fR^2) \,\,\,\,\,\quad: \qquad 
  \SeeCS =2(2\pi) (2 c)\int_{\Sigma }\fF\wedge 
\fGN\,,  \\
&&   {\rm AdS}_7 \,\, {\rm with}\,\, \fPanom= c\, {\fF}^2\wedge \tr(\fR^2) \,\,\,\quad: \qquad 
   \SeeCS =2(2\pi) (2 c) \int_{\Sigma }\fF^2\wedge 
\fGN\,,  \\
&&    {\rm AdS}_7 \,\, {\rm with}\,\, \fPanom= c\, \tr(\fR^4) \qquad\,\,\,\,\,\,\quad: \qquad  \nonumber \\
 &&\quad\SeeCS \nonumber\\
 &&\quad =2 (2\pi)(4c)  \int_{\Sigma}
\Biggl[
\fGN\wedge \fRN^2
+[-2\fRN +3 (\fKbar \cdot \fK)]\wedge [(\fK\cdot \Dhat \fKbar)+ (\fKbar\cdot \Dhat \fK)]
\nonumber \\
&&\qquad\qquad\qquad\qquad\qquad\qquad\qquad\qquad
+
(\Dhat \fK \cdot \fr \cdot  \fKbar)-
(\Dhat \fKbar \cdot\fr \cdot  \fK)
\Biggr]\,, \label{eq:trR4} \\
&&    {\rm AdS}_7 \,\, {\rm with}\,\,\fPanom= c\, \tr(\fR^2)\wedge \tr(\fR^2)\,\,\,\,: \qquad  \nonumber \\
&& \quad \SeeCS \nonumber \\
& &\quad= 2^2(2\pi)(4c) \int_{\Sigma} 
\Biggl[
\fGN\wedge \fRN^2
+\frac{1}{2}\fGN\wedge{{\rm tr}\left({\fr}^2\right)} \nonumber\\
&&\qquad\qquad\qquad\qquad\qquad\qquad
+2[-\fRN+(\fKbar\cdot \fK)]\wedge [(
\fK\cdot \Dhat \fKbar)
+(\fKbar \cdot \Dhat \fK)]
\Biggr]\,.\label{eq:trR2trR2}
 \end{eqnarray}
In the above expression, $\Sigma$ denotes the  bulk entangling surface with  co-dimension two, $\fF$ is $U(1)$ field strength, $\fR$ is the curvature two-form, 
${\bf K}_i$ and  $\bar{\bf K}_i$ are extrinsic curvatures of $\Sigma$ and ${\fr}$ is the intrinsic curvature two-form of $\Sigma$. The $\fGN$ and $\fRN=d\fGN$ are the normal bundle connection and field strength while $\Dhat$ is the covariant derivative associated with $\fGN$ 
defined such that\footnote{We  provide explicit and detailed expressions for these notations and various other  definitions in our appendices.}
\begin{eqnarray}
&& \Dhat{\bf K}_i \equiv  D{\bf K}_i - \fGamma_N\wedge {\bf K}_i \, ,
\qquad
\Dhat\bar{\bf K}_i \equiv  D\bar{\bf K}_i + \fGamma_N\wedge \bar{\bf K}_i \, , 
\end{eqnarray}
where  $D$ is the covariant derivative on $\Sigma$. The `$\cdot$' in the above equation 
denotes the contraction of the indices using the intrinsic metric on $\Sigma$. 
The anomaly coefficient $c$ is a numerical constant. For simplicity, 
 in the course of the computation starting from the next section, 
 we will set the anomaly coefficient $c$ to be unity. It is straightforward to restore the anomaly coefficients 
by  multiplying our results  with $c$. 

The first three formulae are exactly the corresponding Wald-Tachikawa formula for black hole entropy in the presence of 
CS terms. As explained before, the interesting extrinsic curvature dependence appears only for the pure gravitational
CS terms in AdS$_7$.

\subsection{Outline} 
The outline of this paper is as follows. We start in \S\ref{sec:buildingblocks} by 
introducing the setup we will investigate throughout this paper: Chern-Simons terms, anomaly polynomials 
and regularized cone geometry. We also provide 
a brief review of \cite{Dong:2013qoa} and then explain how the holographic 
entanglement entropy formulae can be obtained from the anomaly polynomials and Chern-Simons terms. 
In \S\ref{sec:anomalymethod1}, we summarize our derivation of the formulae from 
the anomaly polynomials corresponding to  
3d and 7d gravitational Chern-Simons terms. 
In \S\ref{sec:directchernsimons}, we present an alternative derivation of the same results directly from the Chern-Simons terms.
Finally in \S\ref{sec:consistency}, we explain a straightforward generalization of the holographic 
entanglement entropy formula to certain class of mixed Chern-Simons terms, 
consistency check with the black hole entropy formulae  and
the frame-dependence of the entanglement entropy in the presence of 
quantum anomalies. The final section \S\ref{sec:conclusion} of the main text is 
devoted for the conclusions and discussion. 

The appendices of this paper are organized as follows. Appendix \ref{sec:useful} 
collects some explicit forms of geometric quantities evaluated on the regularized cone geometry,  
such as connection one-form, curvature two-form as well as their wedge products.   
Details of the computation in \S\ref{sec:anomalymethod1} are given in Appendix \ref{sec:anompolydetail} while
Appendix \ref{sec:anomalymethod2} 
summarizes another way to derive the holographic entanglement entropy formulae 
by directly applying Dong's formula to anomaly polynomials, 
without doing the expansion near the regularized conical defect explicitly.  
Finally, Appendix \ref{app:7dCSdetails} summarizes the direct derivation of the holographic entanglement entropy formula from the 7d single-trace gravitational Chern- Simons terms.

\vskip 0.1in

\noindent
{\bf Note Added}: While we were finishing this paper, \cite{Guo:2015uqa} appeared and have substantial overlap in the derivation of holographic entanglement entropy formula 
for 7d single-trace gravitational Chern-Simons term from 
the anomaly polynomial. Their published results are consistent with ours.
We also note that  \cite{Nishioka:2015uka, Iqbal:2015toa}  have some overlap with our paper.\footnote{We would like to thank N.~Iqbal and A.~C.~Wall for correspondence.}
 
\section{Setup and Outline of Computation}\label{sec:buildingblocks}
In this section, we begin by summarizing the setups we use throughout this paper. We start with 
a brief review of Chern-Simons terms  and their corresponding anomaly polynomials. This is followed by a
discussion of the regularized cone geometry on which we will evaluate 
the anomaly polynomials and Chern-Simons terms to obtain 
the holographic entanglement entropy formula.\footnote{See
Appendix~\ref{sec:useful} for a list of  geometric quantities evaluated on this background.}
Following \cite{Dong:2013qoa}, we then will focus on two types of contributions (the Wald term and the extrinsic correction term)
to the holographic entanglement entropy formula with a particular focus on our case of interest.
We will conclude by summarizing our computational strategy.

\subsection{Chern-Simons Term and Anomaly Polynomial}\label{subsub:csterm}
A Chern-Simons term $\ICS$ is an odd dimensional form made by wedging together Christoffel connection 1-forms $\fGamma^\mu{}_\nu$,
vector potential 1-forms $\fA$ and their exterior derivatives $d\fGamma^\mu{}_\nu$ and $d\fA$.
The combination is chosen such that  under  gauge transformation and diffeomorphism, it transforms by  an exact term. 
The most straightforward way to characterize the Chern-Simons term is by the corresponding anomaly polynomial defined by $\fPanom=d\ICS$. This anomaly polynomial is 
by far simpler and is written only in terms of covariant quantities (i.e. field  strength two-form and 
Riemann curvature two-form).  

The Chern-Simons term corresponding to any given anomaly polynomial  can be written down  in a compact way by using the
transgression formula (see for example \cite{Bonora:2011gz}). We will begin our discussion with pure gravitational Chern-Simons terms
whose anomaly polynomial  depends only on the Riemann curvature two-form.  We will later generalize to mixed Chern-Simons terms
in \S\ref{sec:consistency}.

Here are the explicit forms of the gravitational Chern-Simons terms and the corresponding anomaly polynomials 
we will need in \S\ref{sec:anomalymethod1} and \S\ref{sec:directchernsimons}. 
In three-dimensional case, the gravitational Chern-Simons term and 
the corresponding anomaly polynomial are written as 
\begin{eqnarray}
\ICS^{3d} = \tr\left({\bf \Gamma}\wedge {\fR}- \frac{1}{3}{\bf \Gamma}^3 \right)\, ,
\qquad \fPanom^{3d} = \tr( {\fR}^2)\, , 
\end{eqnarray}
while for $7d$ case there are two types of gravitational Chern-Simons terms: 
the first one is in the form of the single-trace 
\begin{eqnarray}\label{eq:Ics7dsingletrace}
&&\ICS^{7d,\,single} = 4\int_0^1 dt \, \tr \left[
{\bf \Gamma}\wedge (t{\fR}+t(t-1){\bf \Gamma}^2)^3 \right]
\,,  \\
&& \fPanom^{7d,\, single} =  \tr( {\fR}^4)\, , 
\end{eqnarray}
while the second one is double-trace 
\begin{eqnarray}
\ICS^{7d,\,double} = \tr\left({\bf \Gamma}\wedge {\fR}- \frac{1}{3}{\bf \Gamma}^3 \right)\wedge 
\tr( {\fR}^2)\,, \qquad \fPanom^{7d,\,double} =  \tr( {\fR}^2)\wedge \tr( {\fR}^2)\, . 
\end{eqnarray}
Here ${\bf \Gamma}^{\mu}{}_\nu$ is the connection one-form and 
${\fR}^\mu{}_\nu=d{\bf \Gamma}^\mu{}_\nu+ {\bf \Gamma}^\mu{}_\rho\wedge{\bf \Gamma}^\rho{}_\nu$ is the
curvature two-form.  The trace in the above expression is taken for the matrix indices 
carried by the wedge products of these quantities. For example, 
$\tr({\fR}^2)= {\fR}^\mu{}_\nu\wedge {\fR}^\nu{}_\mu$.

\subsection{Deriving Holographic Entanglement Entropy Formulae: A Review}
Our goal in this paper is to derive holographic entanglement entropy formulae for 
Chern-Simons terms by using the argument of Lewkowycz-Maldacena \cite{Lewkowycz:2013nqa}. 
The main idea of \cite{Lewkowycz:2013nqa} is the following :  start with the geometry near the regularized conical geometry
with conical deficit $2\pi(n-1)$, 
expand the action on this background around $n=1$ (equivalently 
around $\epsilon =0$ where $\epsilon\equiv 1-1/n$) 
and then pick up the linear terms in $n-1$ (equivalently, linear term in $\epsilon$), 
which corresponds to the holographic entanglement entropy functional. 
In \cite{Dong:2013qoa} (see also \cite{Camps:2013zua}), this argument is generalized to 
any general higher-derivative theory whose Lagrangian is a polynomial of Riemann tensor only. 

Since the argument has been very well discussed in \cite{Lewkowycz:2013nqa} and reviewed in \cite{Dong:2013qoa,Castro:2014tta}, we will just be brief here and refer the interested readers to these papers for more details. The essential idea is that in the dual CFT, one starts by computing the R\'enyi entropy for any positive integer $n$ and then take the $n\rightarrow 1$ limit to obtain the entanglement entropy:
\be
\See= \lim_{n\rightarrow 1} \frac{1}{1-n}\left[
\log Z_n - n \log Z_1
\right]\,,
\ee where the partition function on a replica manifold $M_{n}$ 
 is given by $Z_n$. In the AdS/CFT context, $Z_n$ is interpreted as the bulk gravitational partition function for a geometry 
 $B_n$ that asymptotes to $M_n$, that is, $Z_n=\exp(I_{grav}[B_n])$
where $I_{grav}[B_n]$ is the on-shell gravitational action on $B_n$ in the semiclassical limit. 
One convenient method introduced in  \cite{Lewkowycz:2013nqa} is to extract $\See$ using a regularized cone  with a conical deficit 
 $2\pi(n-1)$ along a co-dimension two surface in the bulk:\footnote{We note that our notation for the Euclidean action deviates from 
 \cite{Dong:2013qoa} by a minus sign.}
\be
\See= \left.\partial_\epsilon \left(
I_{cone}^{inside}
\right)\right|_{\epsilon=0}\, , 
\ee where $I_{cone}^{inside}$ is the action evaluated near the tip of the regularized cone.

The problem now is  to evaluate the right-hand-side of this equation.
Below, after introducing the explicit metric of the regularized 
cone geometry following the notation of \cite{Dong:2013qoa}, 
we will briefly review the argument in \cite{Dong:2013qoa} 
and then explain the origin of non-trivial contributions to holographic entanglement entropy.
We also argue that, for the setups in this paper, 
we can simplify the evaluation thanks to the simple wedge-product structure of Chern-Simons terms and anomaly polynomials.  


\subsection{Metric of Regularized Cone Geometry}
Following the notation of  \cite{Dong:2013qoa}, we write down the metric near the tip of the regularized cone
geometry 
in $(D+2)$-dimensional Euclidean spacetime up to the second order of $(z, \bar{z})$-expansion
as 
\begin{eqnarray} \label{eq:metric_conical}
&&ds^2 = e^{2A}\left(dzd\bar{z} + e^{2A}T (\bar{z}dz-zd{\bar{z}} )^2 \right)
+\left(g_{ij} + 2K_{aij}x^a + Q_{abij}x^ax^b \right)dy^i dy^j \nonumber \\
&&\qquad\qquad+ 2ie^{2A}\left(U_i + V_{ai}\,x^a\right)
\left(\bar{z}dz- z d\bar{z}\right)dy^i +\ldots
 \, , 
\end{eqnarray}
or, in terms of components,\footnote{In a general higher derivative gravity theory, there exists an ambiguity in the definitions of $T,Q$ and $V$. 
This has been termed the ``splitting problem'' by \cite{Miao:2015iba, Camps:2016gfs}. It is expected that equations of motion will resolve such an ambiguity. 
In our case, as we will show in the bulk of the paper, $T, Q$ and $V$ will not contribute to the entanglement entropy and hence the splitting problem does not arise in our setup.
} 
\begin{eqnarray}
&&G_{zz} = e^{4A}\,T\bar{z}^2\, , \qquad 
G_{\bar{z}\bar{z}} = e^{4A}\,T z^2\, , \qquad 
G_{z\bar{z}} = \frac{1}{2}e^{2A} - e^{4A}\,T\,|z|^2\, , \nonumber \\
&&\quad
G_{iz} = ie^{2A}(U_i+V_{ai}\,x^a)\,\bar{z}\, , \qquad 
G_{i\bar{z}} = -ie^{2A}(U_i+V_{ai}\,x^a)\,z\, ,  \\
&&\qquad\qquad\qquad
G_{ij} = g_{ij} + 2K_{aij}\,x^a + Q_{abij}\,x^ax^b\, . \nonumber 
\end{eqnarray} 
 Here $(x^1, x^2)=(z, \bar{z})$ are the coordinates transverse to the co-dimension two conical defect, 
 while $y^i$ ($i=1, 2, \ldots, D$) are the ones along it.
 When we evaluate some quantities like the curvature two-form on this metric, 
 we will simply say that we evaluate them on the regularized cone geometry. 
In the evaluation of the anomaly polynomials on this geometry, 
 as we will explain shortly and work out in \S\ref{sec:anomalymethod1}, 
$D$ is taken to be $d$ (where $d$ is a positive even integer) and $y^d$ is set to the coordinate of a half line
(i.e. $y^d\in I=[0,\infty)$). 
When we evaluate
the Chern-Simons terms directly as in \S\ref{sec:directchernsimons}, 
on the other hand,  
$D$ is set to $(d-1)$.\footnote{Throughout this paper, spacetime indices are denoted with Greek 
letters $\mu, \nu, \rho,\ldots$. For simplicity, we use this notation for 
both before and after we uplift the spacetime by one-dimension to deal with the anomaly polynomials.}
 In the above metric, the functions
 $T, g_{ij}, K_{aij}, Q_{abij}, U_i, V_{ai}$ depend only on $y^i$,  not on $z$ and $\bar{z}$.  
 The regularization function  $A$ is given by
 \begin{eqnarray}
 A = -\frac{\epsilon}{2}\log(|z|^2+a_{reg}^2)\, , 
 \end{eqnarray}
where $\epsilon=1-1/n$ and $a_{reg}$ is a regularization parameter taken to $a_{reg}\to0$ at the end of the computation. 

The inverse metric up to the first order in the $(z, \bar{z})$-expansion 
is given by 
\begin{eqnarray} \label{eq:metinv}
&&G^{zz}= G^{\bar{z}\bar{z}}=0, \qquad G^{z\bar{z}} = 2e^{-2A}\, , \nonumber \\
&&G^{iz}= 2\,i\,g^{ik}U_k\, z\, , \qquad G^{i\bar{z}}= -2\,i\,g^{ik}U_k\, \bar{z}\, ,  \\
&&\qquad G^{ij} = g^{ij} - 2\,g^{ik}g^{jl}K_{akl}\,x^a\, . \nonumber 
\end{eqnarray}
Here $g^{ij}$ is the inverse of $g_{ij}$ (i.e. $g_{ij}\,g^{jk}=\delta_i^{k}$).  
As we will see, the inverse metric up to this order is enough for our purpose. 

It is also useful to define
the normal-bundle connection $\GN_i$ and its field strength as
\be
\GN_j \equiv(-2i U_j) \, , \quad
\RN_{jk}\equiv 2 \partial_{[j} \GN_{k]}\,. 
\ee We also define the differential forms for them as 
$\fGamma_N \equiv \GN_i dx^i$ and $\fR_N \equiv d\fGamma_N$.
See Appendix \ref{sec:usefulconversion} for more details on their standard definitions.

\subsubsection{Review of \cite{Dong:2013qoa}}
\label{subsubsec:reviewDong}
To derive the holographic entanglement entropy formula, the central question is  
how one can practically collect the terms linear in $\epsilon$ 
when the action on the regularized cone background is evaluated. 
When the Lagrangian is a polynomial of Riemann tensor only,  
generalization of the argument by \cite{Lewkowycz:2013nqa}
is carried out in \cite{Dong:2013qoa}. 
The point of \cite{Dong:2013qoa} is that there are two types of sources for the order-$\epsilon$ contribution:
\begin{enumerate}
\item \underline{$\partial\bar{\partial}A$ Term}\\
On the regularized cone background, the Riemann tensor contains terms proportional to $\partial \bar{\partial}A$ 
(here $\partial$ and $\bar{\partial}$ respectively are derivatives with respect to $z$ and $\bar{z}$). 
Since $\partial \bar{\partial}A = - \pi\epsilon \delta^2(z, \bar{z})$, we have 
the order $\epsilon$ terms of the form 
\begin{eqnarray}
&&\int d^2 z\,d^{D}y 
\left[\sqrt{+G}\left(\frac{\partial L}{\partial R_{\mu\nu\rho\sigma}}\right)\right ]_{\epsilon=0} 
(\partial \bar{\partial}A\,\,{\rm terms \,\, in\,\,}  R_{\mu\nu\rho\sigma})\, .   \nonumber 
\end{eqnarray}

\item \underline{$\partial A \bar{\partial}A$ Term}\\
Since $\partial A = - \epsilon/(2{z})$ and $\bar{\partial} A = - \epsilon/(2\bar{z})$, 
the product $\partial  A \bar{\partial}A$ is of order $\epsilon^2$. However, 
the integration with respect to $(z,\bar{z})$ can compensate to generate order-$\epsilon$ terms. 
To see this, let us define the polar coordinate $(\rho, \tau)$ for the 2d space 
transverse to the conical defect such that $z=\rho e^{i\tau}$. 
Then, due to the integration (here $C$ is a positive constant)
\begin{eqnarray} \label{eq:integraldada}
\int d^2z \partial A \bar{\partial}A e^{-CA} 
= \pi\epsilon^2 \int d\rho \rho  \frac{\rho^2}{(\rho^2+a_{reg}^2)^{2-(\epsilon/2)C}}
=-\frac{\pi \epsilon}{C}+\ldots\, , 
\end{eqnarray}
the following type of order $\epsilon$ terms can show up: 
\begin{eqnarray}
2\int d^2 zd^{D}y
\left[ \sqrt{+G} \left(\frac{\partial^2 L}{\partial R_{\mu\nu\rho\sigma} \partial R_{\alpha\beta\gamma\delta}}\right)  \right]_{\epsilon=0}
(\partial A\,\,{\rm terms \,\, in\,\,}  R_{\mu\nu\rho\sigma})\times
(\bar{\partial}A\,\,{\rm terms \,\, in\,\,}  R_{\alpha\beta\gamma\delta})\, . \nonumber 
\end{eqnarray}
\end{enumerate}  
To compute these two contributions, what is needed to keep in the Riemann tensor is: 
\begin{itemize}
\item Order-$\epsilon$ terms of the form $\partial A$, $\bar{\partial}A$ or 
$\partial \bar{\partial} A$ with zeroth order terms in $(z, \bar{z})$-expansion multiplied,  
\item Order-$\epsilon^0$ terms which are 
at the same time the zeroth order in $(z,\bar{z})$-expansion. 
\end{itemize}

\subsection{Holographic Entanglement Entropy from Anomaly Polynomial: Strategy}
\label{sec:anompolystrategy}
Due to its simplicity,  we first describe  the indirect derivation using  anomaly polynomials $\fPanom=d\ICS$, instead of the
direct evaluation of Chern-Simons terms $\ICS$ (which will be described in the next subsection). More concretely, we first uplift by one-dimension and consider 
the regularized cone geometry in $(d+2)$-dimensional (Euclidean) spacetime 
(i.e. $y^d$ in Eq.~$\eqref{eq:metric_conical}$ is the coordinate of a half line $I=[0,\infty)$,  
corresponding to the uplifted direction). 
We then take the  higher dimensional  Lagrangian  to be the anomaly polynomial $\fPanom$  and 
evaluate the action  on the $(d+2)$-dimensional regularized cone geometry. After carrying out $\epsilon$-expansion and
collecting the linear terms, we will explicitly see the action to be an exact term. We can then write
\begin{eqnarray}
\int_{\onCone} \fPanom
 = \epsilon \int_{\Sigmatilde} d (\Sdenanom^{CS}) +\ldots
 =  \epsilon \int_{\Sigma}\Sdenanom^{CS}|_{y^d=0} +\ldots\, , 
\end{eqnarray}
where we will find compute $\Sdenanom^{CS}$ explicitly later in the text.
The holographic entanglement entropy functional for Cher-Simons terms
can then be obtained through $\SeeCS=\int_{\Sigma}\Sdenanom^{CS}|_{y^d=0}$.\footnote{We note that, when we uplifted to $(d+2)$-dimensions, we extended $\fR^\mu{}_\nu$ etc. in the following way: 
the components with at least one of the subscript or superscript set to $y^d$ vanish at $y^d=0$.
We also assumed that $\fR^\mu{}_\nu$ vanishes at $y^d\to\infty$ sufficiently fast.}
Here, $\Sigma$ represents the co-dimension two
 holographic entangling surface in the $(d+1)$-dimensional bulk where the Chern-Simons term is defined, while
$\onCone$ denotes the region near the tip of the regularized cone geometry \eqref{eq:metric_conical} corresponding to  
the uplifted conical defect $\Sigmatilde$.

As we have explained the prescription to obtain the holographic entanglement entropy formula 
from the anomaly polynomials, we next argue how to collect the order-$\epsilon$ terms 
in the (integral of) anomaly polynomials. Since the anomaly polynomials are 
written with the curvature two-form only (made of Riemann tensor), as in the case of \cite{Dong:2013qoa}, 
there are two types of contributions at order $\epsilon$. That is,
when we evaluate the anomaly polynomials on the regularized cone background, 
what we need to keep are the terms of the form 
\begin{itemize}
\item $\partial \bar{\partial}A$ times  zeroth order terms in $(z,\bar{z})$-expansion, 
\item $\partial A \bar{\partial} A$ times zeroth order terms in $(z,\bar{z})$-expansion. 
\end{itemize}

To evaluate these two types of terms, one needs to explicitly compute the curvature two-form $\fR$ on the regularized 
cone background.  
Related to this, here we provide one remark which reduces some intermediate computations. 
First of all, from Eqs.~\eqref{eq:metric_conical} and \eqref{eq:metinv}, we can compute 
the Christoffel connection up to order-$\epsilon$ in the $\epsilon$-expansion 
and up to first order in the $(z,\bar{z})$-expansion. The explicit result is 
summarized in Eq.~\eqref{eq:christoffel}. 
From this expression, we can 
easily confirm the following statement for the connection one-form $\fGamma$: 
Let us consider the terms of the form $\partial A$ or $\bar{\partial}A$ times zeroth order terms in $(z,\bar{z})$-expansion. 
Such terms appear only in $\fGamma^z{}_z$ and $\fGamma^{\bar{z}}{}_{\bar{z}}$ as 
$2e^{-0\times A}(\partial A dz) $  and $2e^{-0\times A}(\bar{\partial} A d\bar{z})$, respectively.  Therefore, in the curvature two-form, the following is true of  leading order terms in $(z,\bar{z})$-expansion: (1) there is no term of the form $\partial A\bar{\partial}A$ times zeroth order terms in $(z,\bar{z})$-expansion, (2) as for the terms of the form $\partial A$, $\bar{\partial}A$ or $\partial\bar{\partial} A$, they always occur as $\partial Adz$, $\bar{\partial} Ad\bar{z}$ or $ \partial\bar{\partial}Adz \wedge d\bar{z}$, times zeroth order terms in $(z,\bar{z})$-expansion. To avoid confusion, reader should note that here and elsewhere we count $dz, d\bar{z}$ as zeroth order objects in $(z,\bar{z})$-expansion.
 Because of this simple profile, we can drop the other terms proportional to $dz$ and/or $d\bar{z}$.  
 In particular, we note that $Q$ and $V$ terms of \cite{Dong:2013qoa} do not contribute 
 to our final answer. 
 
To summarize, starting from Eq.~\eqref{eq:christoffel}   
what we need to keep in the evaluation of the connection one-form and curvature two-form is:
\begin{enumerate}
\item 
For $\fGamma$, keep terms of the form
\begin{itemize}
\item Order-$\epsilon^0$ and zeroth order terms in $(z,\bar{z})$-expansion, 
\item $\partial Adz$ or $\bar{\partial} Ad\bar{z}$ times zeroth order terms in $(z,\bar{z})$-expansion.  
\end{itemize} 
\item 
For $d\fGamma$, keep terms of the form
\begin{itemize}
\item Order-$\epsilon^0$ and zeroth order terms in $(z,\bar{z})$-expansion, 
\item $\partial Adz$, $\bar{\partial} Ad\bar{z}$ or $\partial\bar{\partial}Adz\wedge d\bar{z}$ 
times zeroth order terms in $(z,\bar{z})$-expansion.  
\end{itemize} 
\item
For $\fR$, keep terms of the form
\begin{itemize}
\item Order-$\epsilon^0$ and zeroth order terms in $(z,\bar{z})$-expansion, 
\item $\partial Adz$, $\bar{\partial} Ad\bar{z}$ or $\partial\bar{\partial}Adz\wedge d\bar{z}$ 
times zeroth order terms in $(z,\bar{z})$-expansion.  
\end{itemize} 
\end{enumerate} 
Following this, we have computed $\fR$ etc.  
The result is summarized in Appendix \ref{sec:useful}. 

Here we have explained the strategy to derive the holographic formulae for the entanglement entropy 
by evaluating the anomaly polynomial on the regularized cone background. 
This is explicitly worked out in \S\ref{sec:anomalymethod1} and Appendix \ref{sec:anompolydetail}. 
In Appendix \ref{sec:anomalymethod2}, on the other hand, we have also explained an 
alternative way to derive the holographic formulae for entanglement entropy 
from anomaly polynomials. There, instead of doing the explicit expansion of 
the anomaly polynomial on the regularized cone geometry, 
we apply the Dong's entanglement entropy formula in \cite{Dong:2013qoa} 
to the anomaly polynomial (since the anomaly polynomial is covariant, Dong's formula is applicable) and reproduce all the results in \S\ref{sec:anomalymethod1}. 

\subsection{Holographic Entanglement Entropy from Chern-Simons Term: Strategy} 
\label{sec:csstrategy}
Although tedious computation is in general involved at the practical level, 
 in principle, we can also derive the holographic entanglement entropy formulae 
by evaluating the Chern-Simons terms on the regularized cone background. 
This computation is worked out in \S\ref{sec:directchernsimons} for some examples. 

Since the Chern-Simons terms depend on the curvature two-form $\fR$ 
as well as the connection one-form $\fGamma$, the Dong's formula is not applicable, 
and one needs to take into account some singularities coming from 
the connection one-form. Here, we will briefly explain and list the nontrivial 
contributions to the holographic entanglement entropy when the 
Chern-Simons terms are directly evaluated on the regularized conical cone
geometry. This can be seen as a generalization of \cite{Castro:2014tta} to higher dimensional Chern-Simons terms.

As in the case of the anomaly-polynomial-based computation, 
we have first of all the following two types of sources for order-$\epsilon$ term:
\begin{itemize}
\item $\partial \bar{\partial}A$ times  zeroth order terms in $(z,\bar{z})$-expansion, 
\item $\partial A \bar{\partial} A$ times zeroth order terms in $(z,\bar{z})$-expansion. 
\end{itemize} 
Since the connection one-form $\fGamma$ contains one derivative of $A$ only in the form of 
$\partial A dz$ and $\bar{\partial} Ad\bar{z}$, the singularities in $\fGamma$ do not 
contribute to the former, but will contribute to the latter. 

On top of these, as carefully treated in \cite{Castro:2014tta} (see also \cite{Solodukhin:2005ah}) 
for 3d gravitational Chern-Simons term, 
another type of source for the order-$\epsilon$ term exists 
for the Chern-Simons-term-based computation. 
Detailed explanation will be given in \S\ref{sec:directchernsimons}
through concrete examples.  Below, we will provide a brief explanation of this type of subtle terms.

To see this new type of contribution, one needs to treat the $(z, \bar{z})$-expansion and regularization carefully. 
Let us consider the following terms that in general 
appear in the Chern-Simons term evaluated on the regularized cone background: 
\begin{eqnarray}\label{eq:integrationbypart}
\int  \partial A dz \wedge d\fGamma \wedge(\ldots)
&=& \int  \partial A dz \wedge d\bar{z}\, \bar{\partial} \fGamma \wedge(\ldots)+(\ldots) \nonumber \\
&=& -\int  \partial \bar{\partial} A dz \wedge d\bar{z} \,  \fGamma \wedge(\ldots)+(\ldots)\nonumber \\ 
&=& \pi\epsilon [\fGamma \wedge(\ldots)]|_{0th\,\,in\,\,(z,\bar{z})-expansion} \, . 
\end{eqnarray} 
(and similarly for $ \bar{\partial} A d\bar{z}$). 
Here we note the $d\fGamma$ in the above expression
is the one before $(z, \bar{z})$-expansion is carried out. In the final line we have used the fact that 
$\partial \bar{\partial}A = - \pi\epsilon \delta^2(z, \bar{z})$. 
Therefore, as a result of the integration-by-part, 
this type of terms can show up nontrivially and we need to take them into account as well. 

Here is one remark. Even when this new type of contribution is taken into consideration, we can still neglect a lot of terms proportional to $dz$ and/or $d\bar{z}$ 
in $\fR$ and $\fGamma$ in the same way as the evaluation based on anomaly polynomials. 
This can be easily seen from the fact that 
$\partial A$ (resp. $\bar{\partial} A$) always show up in $\fR$ and $\fGamma$ 
in the form of $\partial Adz$ (resp. $\bar{\partial} A d\bar{z}$), and  $d\bar{z}$ 
(resp. $dz$) comes from the derivative $d=d\bar{z}\bar{\partial}+ dz\partial+...$ that we move by integration-by-part (see above). 
This means that we can still neglect a lot of terms proportional to $dz$ and/or $d\bar{z}$ 
in the same way as the evaluation based on anomaly polynomials. 
Therefore, even when we work out with the Chern-Simons terms, 
the terms in $\fR$ etc. summarized in Appendix  \ref{sec:useful}  are sufficient.

\section{Holographic Entanglement Entropy from Anomaly Polynomial}\label{sec:anomalymethod1}
Equipped with the expression of the building blocks $\fGamma, \fR$  and their wedge products summarized 
in Appendix \ref{sec:useful}, 
following the strategy given in \S\S\ref{sec:anompolystrategy}
now we explicitly evaluate the anomaly polynomials 
on the regularized cone geometry
and extract the terms linear in $\epsilon$. 
We first start with the anomaly polynomial $\fPanom^{3d}$ for 3d gravitational Chern-Simons term 
and reproduce the result obtained in \cite{Castro:2014tta}. 
We then proceed to 7d case in which we can consider two types of the gravitational 
Chern-Simons terms and thus two types of anomaly polynomials. 
We shall be terse and give only the essential intermediate results in this section.
The details of the computations are provided in Appendix~\ref{sec:anompolydetail}.
\subsection{\fixform{$\tr({\fR}^2)$}}
Let us first start with the anomaly polynomial $\fPanom^{3d}=\tr({\fR}^2)$ corresponding to 
3d gravitational Chern-Simons term and evaluate this polynomial on the regularized cone 
geometry. As explained in \S\S\ref{sec:anompolystrategy}, there are two types of potential contributions 
to the holographic entanglement entropy: the terms proportional to $\partial A \bar{\partial}A$ and 
those proportional to $\partial \bar{\partial}A $. We will first evaluate them separately and 
combine to get the holographic entanglement entropy formula from the anomaly polynomial. Below are the results:
\vspace{0.1in}\\
\noindent
\underline{(1) {$\partial\bar{\partial } A $ Term}}
\begin{eqnarray}
\int_{\onCone} \fPanom^{3d}|_{\partial\bar{\partial}A}
&=& 8\pi\,\epsilon \int_{\Sigmatilde} \left[\fRN - 2 (\bar{\bf K}\cdot {\bf K})\right]\, . 
\label{eq:3danompolyppbarA}
\end{eqnarray}

\noindent 
\underline{(2) $\partial A \bar{\partial} A$ Term }
\begin{eqnarray}
\int_{\onCone} \fPanom^{3d}|_{\partial A\bar{\partial}A} 
&=& 16\pi\epsilon \int (\bar{\bf K}\cdot {\bf K})\, .\label{eq:3danompolypApbarA}
\end{eqnarray}
By summing the above two contributions, we obtain the order-$\epsilon$ terms 
of $\fPanom^{3d}$ on the regularized cone geometry as  
\begin{eqnarray}\label{eq:finaltrR2}
8\pi\,\epsilon \int_{\Sigmatilde} \fRN= 8\pi\,\epsilon \int_{\Sigma} \fGN\, . 
\end{eqnarray}
Therefore, the holographic entanglement entropy formula for 3d gravitational Chern-Simons term is computed from the anomaly polynomial as 
\begin{eqnarray}
  \SeeCS^{3d}
 = 8\pi\, \int_{\Sigma} \fGN\, , 
 \end{eqnarray}
 which reproduces the result obtained in \cite{Castro:2014tta}.

\subsection{\fixform{$\tr({\fR}^4)$}}
As a next example, we proceed to 7d and consider the single-trace type of 
gravitational Chern-Simons term $\ICS^{7d,\, single}$. We will derive the 
holographic entanglement entropy formula from the corresponding 
anomaly polynomial $\fPanom^{7d, \, single}= \tr({\fR}^4)$. 
In the same ways as in 3d case, we first evaluate the terms proportional to 
$\partial\bar{\partial} A$ and $\partial A\bar{\partial}A$ separately and then 
combine to get the holographic entanglement entropy formula:

\vspace{0.1in}

\noindent
\underline{(1) $\partial \bar{\partial} A$ Term}\\
By collecting the singular terms proportional to $\partial A\bar{\partial}A dz\wedge d\bar{z}$ 
from $\fR^z{}_z$ and $\fR^{\bar{z}}{}_{\bar{z}}$, we obtain 
\begin{eqnarray}
&&\int_{\onCone} \fPanom^{7d, single}|_{\partial\bar{\partial}A} \nonumber \\
&&= 16\pi\epsilon\int_{\Sigmatilde} \Biggl[
\fRN^3 -6\fRN^2\wedge(\bar{\bf K}\cdot {\bf K}) 
- 4\fRN\wedge (\Dhat \fKbar\cdot \Dhat \fK)
+12 \fRN\wedge(\bar{\bf K}\cdot {\bf K})^2
\nonumber \\
&&\qquad\qquad\qquad\qquad 
+4\,(\Dhat\bar{\bf K}\cdot {\bf K})\wedge(\Dhat{\bf K}\cdot \bar{\bf K}) 
+4(\Dhat\bar{\bf K}\cdot \bar{\bf K})\wedge(\Dhat{\bf K}\cdot {\bf K})
-2\,(\Dhat\bar{\bf K}\cdot {\fr}\cdot \Dhat{\bf K}) \nonumber \\
&&\qquad\qquad \qquad\qquad
+8\,(\fKbar\cdot \fK)\wedge(\Dhat \fKbar\cdot \Dhat \fK)
-8\,(\bar{\bf K}\cdot {\bf K})^3
\Biggr]\, .  \label{eq:7danompolyppbarA}
\end{eqnarray}
\underline{(2) {$\partial A \bar{\partial}A$ Term}}\\
By combining some terms proportional to $\partial A \bar{\partial}Adz\wedge d\bar{z}$
as worked out in detail in Appendix~\ref{sec:anompolydetail}, we have
\begin{eqnarray}
&&\int_{\onCone}\fPanom^{7d, single}|_{\partial A\bar{\partial }A} \nonumber \\
&&=16\pi\epsilon\int \Biggl[
2\fRN^2\wedge(\fKbar\cdot \fK)
+\fRN\wedge[-6\,(\bar{\bf K}\cdot {\bf K})^2
+2(\bar{\bf K}\cdot{\fr}\cdot {\bf K})] \nonumber \\
&&\qquad\qquad\qquad
+2(\Dhat\bar{\bf K}\cdot {\bf K})\wedge(\Dhat{\bf K}\cdot \bar{\bf K})
-4(\Dhat\bar{\bf K}\cdot \bar{\bf K})\wedge(\Dhat{\bf K}\cdot {\bf K})
+2(\bar{\bf K}\cdot{\fr}\cdot{\fr}\cdot {\bf K}) \nonumber \\
&&\qquad\qquad\qquad
-2(\bar{\bf K}\cdot {\bf K})\wedge(\Dhat\bar{\bf K}\cdot \Dhat{\bf K})
-6(\bar{\bf K}\cdot {\bf K})\wedge(\bar{\bf K}\cdot{\fr}\cdot {\bf K})
+8(\bar{\bf K}\cdot {\bf K})^3
\Biggr]\, . \label{eq:7danompolypApbarA}
\end{eqnarray}
Summing the contributions above, we obtain 
\begin{eqnarray}\label{eq:trR4final}
&&16\pi\epsilon \int_{\Sigmatilde} d
\Biggl[
\fGN\wedge \fRN^2
+[-2\fRN +3 (\fKbar \cdot \fK)]\wedge \left[ (\fK\cdot \Dhat \fKbar) + (\fKbar\cdot \Dhat \fK)\right]
\nonumber \\
&&\qquad\qquad\qquad\qquad\qquad 
+
(\Dhat \fK \cdot \fr \cdot  \fKbar)-
(\Dhat \fKbar \cdot\fr \cdot  \fK)
\Biggr]\, . \nonumber
\end{eqnarray}
Therefore, up to total-derivative terms, the holographic entanglement entropy formula for the 7d single-trace 
gravitational Chern-Simons term is obtained from the anomaly polynomial as 
\bea
&&  \SeeCS^{7d,\, single} \nonumber  \\
 &&\quad =2 (2\pi)(4)  \int_{\Sigma}
\Biggl[
\fGN\wedge \fRN^2
+[-2\fRN +3 (\fKbar \cdot \fK)]\wedge \left[(\fK\cdot \Dhat \fKbar) + (\fKbar\cdot \Dhat \fK)\right]
\nonumber \\
&&\qquad\qquad\qquad\qquad\qquad \qquad\qquad\qquad\qquad\qquad\qquad
+
(\Dhat \fK \cdot \fr \cdot  \fKbar)-
(\Dhat \fKbar \cdot\fr \cdot  \fK)
\Biggr]\,.  \nonumber\\
\eea

\subsection{\fixform{$\tr({\fR}^2)\wedge \tr({\fR}^2)$}}
As a final example, here we will derive the holographic entanglement entropy 
formula for 7d double-trace gravitational Chern-Simons terms by evaluating the anomaly polynomial $\fPanom^{7d, double} = \tr({\fR}^2)\wedge \tr({\fR}^2)$ on the regularized cone background.  

\vspace{0.1in} 

\noindent
\underline{(1) {$\partial \bar{\partial} A$ Term}}\\
The terms of the form $\partial \bar{\partial} Adz\wedge d\bar{z}$  in  $\tr(\fR^2)\wedge\tr(\fR^2)$ 
can be collected easily since this contribution is essentially the 3d result wedged by 
the 3d anomaly polynomial with ${}^{(0)}$ terms in Eq.~\eqref{eq:Rdecomposed} substituted. 
Because of this ``factorization", 
 using the expression of $\fPanom^{3d}|_{0th}$ given in 
Eq.~\eqref{eq:panom3d0th},  
the result can be obtained easily:
\begin{eqnarray}
&&\int_{\onCone} \fPanom^{7d, double}|_{\partial\bar{\partial} A} \nonumber \\
&&=16\pi\, \epsilon\,  \left[\fRN-2(\bar{\bf K}\cdot {\bf K})\right]\wedge
\left[
2\fRN^2-8\fRN\wedge (\bar{\bf K}\cdot {\bf K})
-8(\Dhat\bar{\bf K}\cdot \Dhat{\bf K})
+ \tr({\fr}^2)+8(\bar{\bf K}\cdot {\fr}\cdot {\bf K})
\right]\, . \label{eq:7danomdoubleppbarA} 
\nonumber\\
\end{eqnarray}

\noindent
\underline{(2) {$\partial A \bar{\partial} A$ Term}} \\
For the terms proportional to $\partial A \bar{\partial}A dz \wedge d\bar{z}$, 
since $e^{-A}$ shows up both $\tr(\fR^2)$'s in $\tr(\fR^2)\wedge \tr(\fR^2)$ 
and one needs to integrate as in Eq.~\eqref{eq:integraldada}, the ``factorization" 
do not occur as on the contrary to  the case of $\partial \bar{\partial} A$ term. After some 
classification summarized in Appendix \ref{sec:anompolydetail}, we obtain
\begin{eqnarray}
&& \int_{\onCone} \fPanom^{7d, double}|_{\partial A\bar{\partial}A} \nonumber \\
&&=32\pi\epsilon \int_{\Sigmatilde}
\Biggl[
2\fRN^2\wedge(\bar{\bf K}\cdot {\bf K})-4
\fRN\wedge(\bar{\bf K}\cdot {\bf K})^2
-4(\bar{\bf K}\cdot {\bf K})\wedge(\Dhat\bar{\bf K}\cdot \Dhat{\bf K}) \nonumber \\
&&\qquad\qquad\qquad
+4(\Dhat\bar{\bf K}\cdot {\bf K})\wedge(\Dhat{\bf K}\cdot \bar{\bf K})
+4(\bar{\bf K}\cdot {\bf K})\wedge(\bar{\bf K}\cdot {\fr}\cdot {\bf K})
+(\bar{\bf K}\cdot {\bf K}) \wedge \tr({\fr}^2)
\Biggr]\, . \nonumber \\
&&\label{eq:7danomdoublepApbarA}
\end{eqnarray}
By summing the above two contributions, we obtain 
\begin{eqnarray}
&&32\pi\epsilon\int_{\Sigmatilde} d 
\Biggl[
\fGN\wedge \fRN^2
+\frac{1}{2}\fGN\wedge{{\rm tr}\left({\fr}^2\right)}
+2\left[-\fRN+(\fKbar\cdot \fK)\right]\wedge [
(\fK\cdot \Dhat \fKbar)
+(\fKbar \cdot \Dhat \fK)]
\Biggr]\, .\nonumber\\
\end{eqnarray}
Therefore, the holographic entanglement entropy formula for 7d double-trace gravitational Chern-Simons 
term is obtained from the anomaly polynomial as
\bea
&& \quad \SeeCS^{7d,\, double} \nonumber \\
&&\quad = 2^2(2\pi)(4) \int_{\Sigma} 
\Biggl[
\fGN\wedge \fRN^2
+\frac{1}{2}\fGN\wedge{{\rm tr}\left({\fr}^2\right)}
\nonumber\\
&&\qquad\qquad\qquad\qquad\qquad\qquad\qquad
+2\left[-\fRN+(\fKbar\cdot \fK)\right]\wedge [
(\fK\cdot \Dhat \fKbar)+(\fKbar \cdot \Dhat \fK)]
\Biggr]\,.\label{eq:trR2trR2v2}
\nonumber \\
\eea 


\section{Direct Derivation from Chern-Simons Term}\label{sec:directchernsimons}
The purpose of this section is to reproduce (some of) 
the holographic entanglement entropy formulae obtained in \S\ref{sec:anomalymethod1} 
by directly applying the argument of 
 \cite{Lewkowycz:2013nqa} and \cite{Dong:2013qoa} to the gravitational Chern-Simons terms, 
instead of the anomaly polynomials. 
Since the gravitational Chern-Simons terms depend on the connection one-form explicitly, 
we will first discuss the extension and modification of 
Dong's formula in \cite{Dong:2013qoa} to incorporate the Chern-Simons terms.
As mentioned briefly in \S\S\ref{sec:csstrategy}, in the Chern-Simons-term-based 
derivation of the holographic entanglement entropy formulae, one needs to 
take into account a new type of order-$\epsilon$ contribution,  
which, as we will explain, never shows up in the anomaly-polynomial-based computation. 
To explain this type of terms through a concrete example, 
we will provide a brief review of the 3d case worked out by \cite{Castro:2014tta}. 
After this,  we will present the computations for the 7d gravitational Chern-Simons terms and reproduce the result we obtained in \S\ref{sec:anomalymethod1}. 
We will also explain why this new type of contribution never shows up nontrivially 
in the  anomaly polynomials based computation in \S\ref{sec:anomalymethod1}. 

\subsection{Holographic Entanglement Entropy for Chern-Simons Term}
Let us first write down the generalization of Dong's holographic entanglement entropy formula 
to a general purely gravitational Chern-Simons term 
and then give explanation of each term. For a general 
$(d+1)$-dimensional purely gravitational Chern-Simons term $\ICS[\fGamma, \fR]$ ($d$: an positive even integer), 
the formal holographic entanglement entropy formula is given by\footnote{Here are two remarks on the derivatives with respect to $\fR^\mu{}_\n$ and $\fGamma^\mu{}_\n$ 
in this formula. 
For the derivative $\partial/\partial \fR^\mu{}_\nu$,  all components of $\fR^\mu{}_\nu$ are treated as independent. For 
example,  $\partial \fR^z{}_i / \partial \fR^i{}_z=0$. This is purely conventional.  
We also note that the derivative
$\partial/\partial\fGamma^\mu{}_\nu$ acts like an exterior derivative operator 
(this is because $\fGamma^\mu{}_\nu$ is a one-form). 
That is, for a wedge product of a $p$-form $\form{A}$ with a $q$-form $\form{B}$, we have
\be
\frac{\partial}{\partial\fGamma^\mu{}_\nu} (\form{A}\wedge \form{B})
=\left(\frac{\partial \form{A}}{\partial\fGamma^\mu{}_\nu}  \right)\wedge \form{B}
+(-1)^p \form{A}\wedge
\left(\frac{\partial \form{B}}{\partial\fGamma^\mu{}_\nu}  \right)\,.\nonumber
\ee
}    
\bea \label{eq:CSDong}
  \SeeCS&=& 
\SEEWald+\SeeWaldetr+\SEEKK+\Seeetr\, \quad {\rm where} \nonumber\\
\SEEWald&\equiv &
(2\pi) \int_{\Sigma} \left( \frac{\partial \ICS}{\partial \fR^z{}_z}\right)-(z\leftrightarrow\zbar)\, , \nonumber \\
 \SeeWaldetr &\equiv& 
\left. (2\pi) \int_{\Sigma} \left(\frac{\partial \ICS}{\partial \fGamma^z{}_z}\right)
\right |_{K=\bar{K}=0,\,{\rm integrate\,\, by \,\,part}}-(z\leftrightarrow\zbar) \, , 
  \nonumber\\
\SEEKK&\equiv &
+(8\pi)  \int_{\Sigma} \fK^i\left[\int_0^1 dt~t
\left(
 \frac{\partial^2 \ICS}{\partial \fR^i{}_z \partial \fR^z{}_j}
\right)_t\right]
\bar{\fK}_j-(z\leftrightarrow\zbar)\nonumber\\
&&-(4\pi)  \int_{\Sigma} \fK^i\left[\int_0^1 \frac{dt}{t}~ 
\left(
 \frac{\partial^2 \ICS}{\partial \fR^i{}_z \partial \fR^j{}_{\bar{z}}}
\right)_t\right]
\bar{\fK}^j-(z\leftrightarrow\zbar)
\,, \nonumber\\
\Seeetr&\equiv &
{-}(8\pi)  \int_{\Sigma}\left[\int_0^1 {dt}~t
\left(
 \frac{\partial^2 \ICS}{\partial \fGamma^z{}_z \partial \fR^z{}_{j}}
\right)_{t}\right]
\bar{\fK}_j-(z\leftrightarrow\zbar)
\nonumber\\
&&
{+}(4\pi)  \int_{\Sigma}\left[\int_0^1 \frac{dt}{t}~
\left(
 \frac{\partial^2 \ICS}{\partial \fGamma^z{}_z \partial \fR^j{}_{\bar{z}}}
\right)_{t}\right]
\bar{\fK}^j
-(z\leftrightarrow\zbar)
\nonumber\\
& &
+(4\pi)  \int_{\Sigma}\left[\int_0^1 \frac{dt}{t}~
\left(
 \frac{\partial^2 \ICS}{\partial \fGamma^z{}_z \partial \fGamma^{\bar{z}}{}_{\bar{z}} }
\right)_{t}\right]\,.
\eea
 
 In the above formula, essentially, the classification of the terms is done based on 
 the origin of the singularities that show up in the Chern-Simons term on 
 the regularized cone background.  Comments and explanations of each term are in order: 
\begin{itemize}
\item The terms $\SEEWald$ and $\SEEKK$ are those appearing in Dong's formula of \cite{Dong:2013qoa}: 
 $\SEEWald$ originates from the $\partial\bar{\partial}A$ terms in the curvature two-form $\fR$, 
 while $\SEEKK$ is from the $\partial A \bar{\partial}A$ terms whose $\partial A$ and $\bar{\partial} A$ are 
 both coming from $\fR$\,.  Therefore these terms only involve the derivatives of the 
 Chern-Simons term with respect to $\fR$\, . 

\item The term 
$\Seeetr$ originates in the $\partial A \bar{\partial}A$ terms
where at least one of  the $\partial A$ and $\bar{\partial} A$ is from the connection one-form $\fGamma$\,.  

\item In the evaluation of $\SEEWald$, we first take a derivative with respect to $\fR$ and
then substitute ${}^{(0)}$ terms in Eqs.~\eqref{eq:fgammaexpanded}  and \eqref{eq:Rdecomposed} 
with $A=0$. 

\item  In $\SEEKK$ and $\Seeetr$, we have introduced the integral with respect to a parameter $t$. 
The evaluation of these terms is carried out as follows. First, calculate the derivatives of the Chern-Simons terms, 
substitute all ${}^{(0)}$ terms in Eqs.~\eqref{eq:fgammaexpanded} and \eqref{eq:Rdecomposed}
setting $e^{-2A}=1$. Then replace all $\fK_i$ and $\fKbar_j$ by $t^{1/2} \fK_i$ and $t^{1/2}\fKbar_j$ respectively. Finally, do the integral with respect to $t$. 
This is a convenient way to capture the parameter `$q_{\alpha}$' of \cite{Dong:2013qoa}, 
which is introduced to  count half the number of extrinsic curvatures in the integrands. 
The origin of this parameter is the integral \eqref{eq:integraldada} 
which needs to be carried out to compute the contributions 
from $\partial A \bar{\partial}A$ term. 

\item The term $\SeeWaldetr$ is the new type of contribution special to the Chern-Simons-term-based 
computation that we mentioned briefly in \S\S\ref{sec:csstrategy}. As we will soon explain in \S\S\ref{sec:integrationbypart}, 
this type of terms do not generate any non-trivial extrinsic curvature dependent contribution. 
In addition, as explained in Eq.~\eqref{eq:integrationbypart}, either $\partial A$ or $\bar{\partial}A$ 
needs to be supplied from $\fGamma$ or $\fR$. 
Moreover, since $\partial A$ and $\bar{\partial}A$ are accompanied by the extrinsic curvatures
in $\fR$, this one derivative of $A$ should originate in $\fGamma$. 

In the following, we provide a way to compute this type of terms through some simple examples. 
For 3d gravitational Chern-Simons term $\ICS^{3d} = \tr(\fGamma \fR -(1/3)\fGamma^3) 
= \tr(\fGamma d\fGamma+(2/3)\fGamma^3)$,  we first evaluate the derivative of the Chern-Simons 
term and pick up the terms which contain $d\fGamma$:
\be
\left.(2\pi) \int_{\Sigma} \left(\frac{\partial \ICS}{\partial \fGamma^z{}_z}\right) \right|_{K=\bar{K}=0}-(z\leftrightarrow\zbar)
=(2\pi)\int_{\Sigma}  (d \fGamma^z{}_z)-(z\leftrightarrow\zbar) \, . 
\ee 
The reason to keep the terms with $d\fGamma$ is that one needs 
to integrate it by part as in Eq.~\eqref{eq:integrationbypart}.
After this,  we get rid of $d$ of $\fGamma^z{}_z$ (and its complex conjugalte) 
and substitute ${}^{(0)}$ terms in Eqs.~\eqref{eq:Rdecomposed} and \eqref{eq:fgammaexpanded}  
with $A=0$ and ${\bf K}=\bar{\bf K}=0$. We then finally obtain
\be
\SeeWaldetr = \left.(2\pi) \int_{\Sigma} \left(\frac{\partial \ICS}{\partial \fGamma^z{}_z}\right) \right|_{K=\bar{K}=0, \, 
{\rm integration\,\, by\,\, part}}-(z\leftrightarrow\zbar)
=(4\pi)\int_{\Sigma}\fGamma_N \, .   
\ee 
This removal of $d$ corresponds to the integration by part of $\partial$ or $\bar{\partial}$ 
explained in Eq.~\eqref{eq:integrationbypart}. More generally, the same procedure gives 
\begin{eqnarray}
&& {\rm For}\,\, \fPanom = \tr(\fR^{2k})\, , \qquad  \qquad\qquad 
\SeeWaldetr=(4\pi)\int_{\Sigma}
\fGN
\wedge \fRN^{2k-2} \, , \nonumber \\
&&{\rm For}\,\, \fPanom = \tr(\fR^{2k})\wedge\tr(\fR^{2l})\, , \nonumber\\
&&{\qquad\SeeWaldetr=(4\pi)\int_{\Sigma} \left(
\fGN\wedge \fRN^{2k-2}
\right) \wedge \tr (\fR^{2l}) |_{0th, K=\bar{K}=0, A=0}\, }\nonumber \\ 
&&{\qquad\qquad \qquad\qquad 
+(4\pi)\int_{\Sigma} \left(
\fGN\wedge \fRN^{2l-2}
\right) \wedge \tr (\fR^{2k}) |_{0th, K=\bar{K}=0, A=0}\, }\,.
\end{eqnarray}
In the second line, $\tr (\fR^{2l}) |_{0th, K=\bar{K}=0, A=0}$ means that  
$\tr (\fR^{2l})$ is evaluated by substituting 
${}^{(0)}$ terms in Eqs.~\eqref{eq:Rdecomposed} with the extrinsic curvatures and $A$ set to zero.

\end{itemize}

In the next subsections, in the course of the evaluation of 3d and 7d multi-trace gravitational Chern-Simons 
terms on the regularized cone geometry, 
we shall keep $\partial A, {\bar \partial A}$ and $\partial {\bar \partial A}$ all explicit  such that the origin of each term in the formula \label{eq:CSDong} will become clearer. 

We note that, as we will explain in \S\S\ref{sec:mixanom}, generalization of Eq.~\eqref{eq:CSDong} to 
a general mixed $U(1)$-gravitational Chern-Simons term is straightforward 
since $U(1)$ field strength evaluated on the regularized cone does not 
generate extra singularities to be taken into account. 
That is, this formula \eqref{eq:CSDong} is still valid for the mixed Chern-Simons term.

\subsection{3d Gravitational Chern-Simons Term}
\label{sec:3DCScompute}
Let us first review the derivation of holographic entanglement entropy 
for the 3d gravitational Chern-Simons term which is worked out by \cite{Castro:2014tta}. 
From the expressions computed in Appendix \ref{sec:ics3dcomp}, 
naively, one might think that the nontrivial contribution at order $\epsilon$ comes only from $\ICS^{3d}|_{\partial\bar{\partial}A}$, 
which after integration becomes 
\begin{eqnarray}\label{eq:ICSint_1st}
\int_{\onConeCS} \ICS^{3d}|_{\partial\bar{\partial}A} = \int_{\Sigma}
4\pi \epsilon~ \fGN\, . 
\end{eqnarray} 
However, as pointed out in \cite{Castro:2014tta} (see also \cite{Solodukhin:2005ah}), a little more careful treatment 
is needed for  
$\ICS^{3d}|_{\partial A, \bar{\partial}A}$. That is,  the integration-by-part of the form
 \eqref{eq:integrationbypart}  generates new type of contribution. 
Since we have already reviewed the rationale around Eq.~(\ref{eq:integrationbypart}), here we shall illustrate explicitly in this example what one gets from this term. In the current case, this term comes from
\begin{eqnarray}\label{eq:integrationbypartwo}
&&\int_{\onConeCS}
({}^{(d)} \fGamma^z{}_z )\wedge d\fGamma^z{}_z
=\int_{\onConeCS}
(-2\partial {\bar \partial} A dz \wedge d\zbar )\wedge \fGamma^z{}_z\, +\ldots\,, 
\end{eqnarray} where we have applied Eq.~(\ref{eq:integrationbypart}).
There is also a similar term coming from the integration by part of 
$({}^{(d)} \fGamma^{\bar{z}}{}_{\bar{z}} )\wedge d\fGamma^{\bar{z}}{}_{\bar{z}}$. 

Altogether, after integration, what we get is 
\begin{eqnarray} \label{eq:ICSint_2nd}
&&\int_{\onConeCS}\ICS^{3d}|_{\partial A, \bar{\partial}A} = \int_{\onConeCS} (-2\partial \bar{\partial} A dz\wedge d\bar{z})\wedge
\fGN\times 2 = \int_{\Sigma} 4\pi\epsilon ~\fGN \, . 
\end{eqnarray}
Therefore, we finally obtain the holographic entanglement entropy formula 
for 3d Chern-Simons term as 
\begin{eqnarray}
  \SeeCS^{3d} =  8\pi \int_{\Sigma} \fGN \, . 
\end{eqnarray}
This result by \cite{Castro:2014tta} is consistent with what we obtained by starting with the anomaly polynomial 
$\fPanom^{3d}$ in \S\ref{sec:anomalymethod1}. 

\subsection{Remark on Integration-by-Part Terms}
\label{sec:integrationbypart}
As we have seen in Eq.~\eqref{eq:integrationbypartwo} for 3d gravitational Chern-Simons term, the integration-by-part under an appropriate regularization 
can give non-trivial contribution to the holographic entanglement entropy formula. 
Before discussing 7d case, here we explain some property of this ``integration-by-part" type of terms. 

The first remark is that this type of contribution does not depend on 
extrinsic curvatures. The reason is as follows: in the Chern-Simons term evaluated 
at the regularized cone geometry, the extrinsic-curvature-dependent terms are always accompanied 
by the factor $e^{-CA}$ with nonzero positive constant $C$. Then when we consider ``integration-by-part" 
type of terms, the integration by part gives 
\begin{eqnarray}
&&\int dzd\bar{z} e^{-CA} (\partial A) d\fGamma(\ldots)
= 
-\int dzd\bar{z} e^{-CA} (\partial \bar{\partial} A-C\partial A\bar{\partial}A) \fGamma (\ldots) +\ldots =0 \, , 
\end{eqnarray} and similarly for $\bar{\partial} A$.
Here we have used $\partial \bar{\partial}A = - \pi\epsilon \delta^2(z, \bar{z})$ and 
the integration of the form \eqref{eq:integraldada}. 

Another remark is that, when one evaluates the anomaly polynomial, we did not take into account
this type of contribution. This can be justified as follows. When the anomaly polynomial 
is evaluated on the regularized cone background, since the terms proportional to $\partial Adz$ and $\bar{\partial} {A}d\bar{z}$ in $\fR$ are 
always accompanied by ${\bf K}_i$ and/or $\bar{\bf K}_i$, these ``integration-by-part"
terms appearing in the anomaly polynomial are always proportional to $e^{-CA}$ with nonzero $C$. 
Thus, from the above argument, this type of terms never contributes to holographic entanglement 
entropy formula so far as one start with the anomaly polynomials. On the other hand, when 
starting with the Chern-Simons terms, this type of term gives nontrivial contribution and thus we need to keep track of them carefully. 

 \subsection{7d Double-Trace Gravitational Chern-Simons Term}
 Now we return to the direct derivation of holographic entanglement entropy formula from 
 Chern-Simons terms. Here we consider the 7d double-trace type of gravitational Chern-Simons term $\ICS^{7d, double}$. We divide into the following three possible contribution and evaluate one by one : 
 \begin{eqnarray}
&&({\rm Case}\,\, 1)\qquad  \ICS^{3d}|_{singular} \wedge  \fPanom^{3d}|_{0th} \, , \nonumber \\
&&({\rm Case}\,\, 2)\qquad  \ICS^{3d}|_{0th} \wedge \fPanom^{3d}|_{singular} \, , \nonumber \\
&&({\rm Case}\,\, 3)\qquad  \ICS^{3d}|_{singular} \wedge \fPanom^{3d}|_{singular} \, . \nonumber  
 \end{eqnarray}
Within each of Case 1 and Case 2, there are three possibilities depending on whether the singular term is 
 proportional to $\partial A$ (or $\bar{\partial} A$), 
 $\partial\bar{\partial}A$,  or $\partial A\bar{\partial} A$. 
 Here the contribution proportional to $\partial A$ (or $\bar{\partial} A$) means 
 the one giving nontrivial contribution due to the integration-by-part \eqref{eq:integrationbypart}.
 On the other hand, in Case 3, we only need to consider $\partial Adz$ and $\bar{\partial} Ad\bar{z}$ terms 
 in $\ICS^{3d}|_{singular}$  and $\tr({\fR}^2)|_{singular}$, which after multiplication 
 gives terms proportional to $\partial A \bar{\partial} Adz\wedge d\bar{z}$. We note that all the ingredients 
 needed for this evaluation is summarized in Appendix \ref{sec:useful}.

 \subsubsection{Case 1} 
 We first consider Case 1. 
 As mentioned above, Case 1 contains the following three possibilities:\\
 
 \noindent
 \underline{(1) $\partial\bar{\partial} A$ Term} \\
 Since the this type of terms essentially ``factorizes" into the computation of $\partial \bar{\partial} A$ term 
 in 3d gravitational Chern-Simons term and the non-singular term of 
the anomaly polynomial for 3d Chern-Simons term, we can easily obtain  
 \begin{eqnarray}
&& \int_{\onConeCS} \ICS^{3d}|_{\partial\bar{\partial}A}\wedge \fPanom^{3d}|_{0th}  \\
 &&\qquad\qquad = \int_{\Sigma} 4\pi \epsilon \left[
 2\fRN^2 +\tr({\fr}^2) -8\fRN\wedge(\bar{\bf K}\cdot {\bf K})
 -8(\Dhat\bar{\bf K}\cdot \Dhat{\bf K})+8(\bar{\bf K}\cdot {\fr}\cdot{\bf K})  
 \right]\wedge \fGN\, . \nonumber 
 \end{eqnarray}
  \noindent
 \underline{(2) $\partial A\bar{\partial} A$ Term} \\
 This type of contribution vanishes because $\ICS^{3d}|_{\partial A \bar{\partial}A}=0$ :
 \begin{eqnarray}
 && \int_{\onConeCS} \ICS^{3d}|_{\partial A\bar{\partial}A}\wedge \fPanom^{3d}|_{0th}=0  \, .
 \end{eqnarray}
 \noindent
 \underline{(3) $\partial A$ or $\bar{\partial} A$ Term} \\
 By integrating by part as in Eq.~\eqref{eq:integrationbypart}, we can see that this type of contribution gives 
 nontrivial result:
  \begin{eqnarray}
&& \int_{\onConeCS} \ICS^{3d}|_{\partial A, \bar{\partial}A}\wedge \fPanom^{3d}|_{0th}
=
 \int_\Sigma 
 4\pi\epsilon
 \left[
 2\fGamma_N \wedge \fR_N^2 + \fGamma_N\wedge \tr(
 {\fr}^2)
 \right] \, . 
 \end{eqnarray}
 
 \subsubsection{Case 2} 
 As in Case 1, we can evaluate the three possibilities in Case 2 one by one as follows :\\
 
 \noindent
 \underline{(1) $\partial\bar{\partial} A$ Term} \\
 By using some results in Appendix \ref{sec:useful}, we can straightforwardly obtain the following result:
 \begin{eqnarray}
&&\int_{\onConeCS} \ICS^{3d}|_{0th}\wedge \fPanom^{3d}|_{\partial\bar{\partial}A}  \nonumber \\
&&= \int_{\Sigma}(8\pi \epsilon)\Biggl[
2\fGN \wedge\fRN^2 -4\fGN  \wedge\fRN\wedge(\bar{\bf K}\cdot {\bf K})
+\fRN\wedge \tr\left({\bf \gamma }{\fr}-\frac{1}{3}{\bf \gamma}^3\right) \nonumber \\
&&\qquad\qquad \quad
-4\fRN\wedge[({\bf K}\cdot \Dhat\bar{\bf K}) +(\bar{\bf K}\cdot \Dhat{\bf K})] 
\nonumber \\
&&\qquad\qquad\quad 
-2(\bar{\bf K}\cdot{\bf K})\wedge \tr\left({\bf \gamma }\fr-\frac{1}{3}{\bf \gamma}^3\right) 
+ 8(\bar{\bf K}\cdot{\bf K})\wedge[({\bf K}\cdot \Dhat\bar{\bf K})+(\bar{\bf K}\cdot \Dhat{\bf K})]
\Biggr]\, . 
 \end{eqnarray}
  \noindent
 \underline{(2) $\partial A \bar{\partial} A$ Term} \\
 In the same way as $\partial \bar{\partial} A$ term above, 
 by using the results in Appendix \ref{sec:useful}, 
 the direct evaluation gives 
 \begin{eqnarray}
 &&\int_{\onConeCS} \ICS^{3d}|_{0th}\wedge \fPanom^{3d}|_{\partial A \bar{\partial}A}  \nonumber \\
 &&= \int_{\Sigma}
 16\pi\epsilon\Biggl[
 2\fGN\wedge \fRN \wedge(\bar{\bf K}\cdot {\bf K}) +(\bar{\bf K}\cdot {\bf K}) \wedge 
 \tr\left({\bf \gamma }\fr-\frac{1}{3}{\bf \gamma}^3\right)  \nonumber \\
 &&\qquad\qquad\qquad 
 -2(\bar{\bf K}\cdot {\bf K})\wedge[({\bf K}\cdot \Dhat\bar{\bf K}) + 
 (\bar{\bf K}\cdot \Dhat{\bf K})]
 \Biggr] \, . 
 \end{eqnarray}

  \noindent
 \underline{(3) $\partial A$ or $\bar{\partial} A$ Term} \\
 After integration by part as in Eq.~\eqref{eq:integrationbypart}, 
 we can confirm that this type of term does not generate any nontrivial contribution : 
  \begin{eqnarray}
 &&\int_{\onConeCS} \ICS^{3d}|_{0th}\wedge \fPanom^{3d}|_{\partial A,  \bar{\partial}A} =0 \, . 
 \end{eqnarray}
 
  \subsubsection{Case 3} 
 By using some results in Appendix \ref{sec:useful}, we can easily obtain
 \begin{eqnarray}
 &&\int_{\onConeCS} \ICS^{3d}|_{\partial A, \bar{\partial}A}\wedge \fPanom^{3d}|_{\partial A,  \bar{\partial}A}  \nonumber \\
 && = \int_{\Sigma}
 (-16\pi\epsilon)
 \Biggl[
\fRN\wedge [(\bar{\bf K}\cdot \Dhat{\bf K}) +({\bf K}\cdot \Dhat\bar{\bf K})]
 -2(\bar{\bf K}\cdot{\bf K})\wedge[(\bar{\bf K}\cdot \Dhat{\bf K})+({\bf K}\cdot \Dhat\bar{\bf K})]
 \Biggr]\, . \nonumber \\
 \end{eqnarray}
 \subsubsection{Final Result}
 Summing results from all these cases,  we obtain the total contribution as
\begin{eqnarray}
&&32\pi\epsilon\int_{\Sigma} 
\Biggl\{
\fGN\wedge \fRN^2
+\frac{1}{2}\fGN\wedge \tr({\fr}^2)
+
2\left[-\fRN+(\fKbar\cdot \fK)\right]\wedge [(\Dhat\bar{\bf K}\cdot {\bf K})+(\Dhat{\bf K}\cdot \bar{\bf K})]
\nonumber \\
&&\qquad\qquad\qquad
+d\left[
\frac{1}{4}\fGN\wedge \tr\left(\gamma {\fr}-\frac{1}{3}\gamma^3 \right)
+\frac{1}{2}\fGN\wedge [({\bf K}\cdot \Dhat\bar{\bf K})+(\bar{\bf K}\cdot \Dhat{\bf K})]
\right]
\Biggr\}
\, . \nonumber
\end{eqnarray}

Therefore, by neglecting the total derivative term (we assume that 
there is no homological obstruction), we finally obtain the holographic entanglement entropy formula 
for 7d double-trace gravitational Chern-Simons term as follows:
\begin{eqnarray}
&&  \SeeCS^{7d,\, double} \nonumber \\
&&=32\pi \int_{\Sigma} 
\Biggl\{
\fGN\wedge \fRN^2
+\frac{1}{2}\fGN\wedge \tr({\fr}^2)
+
2\left[-\fRN+(\fKbar\cdot \fK)\right]\wedge[(\Dhat\bar{\bf K}\cdot {\bf K})+(\Dhat{\bf K}\cdot \bar{\bf K})]
\Biggr\}
\, ,  \nonumber\\
\end{eqnarray}
which agrees with the result from the anomaly polynomial in \S\ref{sec:anomalymethod1}.

A similar computation can be repeated for the single trace gravitational CS term in AdS$_7$ (i.e. Eq.~(\ref{eq:Ics7dsingletrace})).
The details are provided in Appendix \ref{app:7dCSdetails} and the result is
\begin{eqnarray}
&&\left[\int_{{\rm Cone}_{\Sigma, n}} \ICS^{7d\, single}\right]_\epsilon \nonumber \\
&&\qquad = (16\pi\epsilon) \int_\Sigma
\Biggl[
\fGamma_N\wedge \fR_N^2 + 
[-2\fR_N + 3(\bar{\bf K}\cdot{\bf K})]\wedge [({\bf K}\cdot \Dhat\bar{\bf K})+(\bar{\bf K}\cdot\Dhat{\bf K})]
\nonumber \\
&&\qquad\qquad\qquad\qquad
+ (\Dhat{\bf K}\cdot \fr \cdot \bar{\bf K}) - (\Dhat\bar{\bf K}\cdot \fr \cdot {\bf K})   
\nonumber \\
&&\qquad\qquad \qquad \qquad
+ d\Bigl[
\frac{1}{2}\fGamma_N\wedge[(\bar{\bf K}\cdot \Dhat{\bf K})+({\bf K}\cdot\Dhat\bar{\bf K})]
-\frac{3}{20}\fGamma_N\wedge[(\bar{\bf K}\cdot\gamma\cdot{\bf K})+({\bf K}\cdot \gamma\cdot\bar{\bf K})]
\nonumber \\ 
&&\qquad\qquad \qquad\qquad \qquad  
+ \frac{1}{4}[(\Dhat\bar{\bf K}\cdot\gamma\cdot {\bf K})-(\Dhat{\bf K}\cdot\gamma\cdot\bar{\bf K})]
-\frac{3}{20}[(\bar{\bf K}\cdot\gamma^2\cdot{\bf K})+({\bf K}\cdot\gamma^2\cdot\bar{\bf K})]
\Bigr]
\Biggr] \, . \nonumber \\
&&
 \end{eqnarray}
 Neglecting the total derivative term (we assume that 
there is no homological obstruction), we finally obtain the holographic entanglement entropy formula 
for 7d single-trace gravitational Chern-Simons term as follows:
\begin{eqnarray}
&&\SeeCS^{7d,\, single} = (16\pi) \int_\Sigma
\Biggl[
\fGamma_N\wedge \fR_N^2 + 
[-2\fR_N + 3(\bar{\bf K}\cdot{\bf K})]\wedge [({\bf K}\cdot \Dhat\bar{\bf K})+(\bar{\bf K}\cdot\Dhat{\bf K})]
\nonumber \\
&&\qquad\qquad\qquad\qquad\qquad\qquad
+ (\Dhat{\bf K}\cdot \fr \cdot \bar{\bf K}) - (\Dhat\bar{\bf K}\cdot \fr \cdot {\bf K})   \Biggr]\, ,
 \end{eqnarray}
which agrees with the result from the anomaly polynomial in \S\ref{sec:anomalymethod1}.


\section{Consistency Checks and Applications}\label{sec:consistency}
Now that we have obtained the holographic entanglement entropy 
formulae of purely gravitational Chern-Simons terms, 
we next generalize
to some mixed Chern-Simons terms. After this, we 
will explain the consistency of our formulae with the black hole entropy formula 
for Chern-Simons terms. We will also comment on the Lorentz-frame dependence 
of Chern-Simons contribution to (holographic) entanglement entropy. 
At the end of this section,  we will briefly comment on the application 
of our holographic entanglement entropy formula to 6d $(2,0)$
theories.


\subsection{Generalization to Mixed Chern-Simons Term}\label{sec:mixanom}
Let us consider the cases with $U(1)$-gravitational mixed Chern-Simons terms. In these cases, 
the corresponding anomaly polynomials contain the $U(1)$ field strength. 
Here we follow the notation in the Appendix of \cite{Dong:2013qoa} 
and explain a straightforward generalization of holographic entanglement 
entropy formulae to the mixed Chern-Simons terms.
The most general $U(1)$ gauge potential one-form $\fA_{U(1)}$ on $B_n$ 
(Here $B_n$ is the bulk extension of $n$-replicated boundary geometry $M_n$. 
We use $(\tilde{\rho}, \tilde{\tau}, y^i)$ for the coordinates of $B_n$) can be written as 
\begin{eqnarray}
{\fA}_{U(1)} &=&  d\tilde{\rho}\left[
a_{\rho}(y) +{\cal O}(\tilde{\rho}^2, \tilde{\rho}^n e^{\pm i n \tilde{\tau}})  
\right]+
 d\tilde{\tau}\tilde{\rho} \left[
a_{\tau}(y) +{\cal O}(\tilde{\rho}^2, \tilde{\rho}^n e^{\pm i n \tilde{\tau}})  
\right] \nonumber\\
&&\qquad\qquad\qquad\qquad\qquad\qquad\qquad\qquad\qquad
 +dy^i \left[
a_{i}(y) +{\cal O}(\tilde{\rho}^2, \tilde{\rho}^n e^{\pm i n \tilde{\tau}})  
\right]\, . 
\end{eqnarray}
By introducing the new coordinate $(\rho, \tau, y^i)$ (which is a natural coordinate after the orbifolding by ${\mathbb Z}_n$)  
 defined by $\tilde{\rho} = n \rho^{1/n}$ and $\tilde{\tau}=\tau/n$, 
we can rewrite this as 
\begin{eqnarray}
{\fA}_{U(1)} = \rho^{-\epsilon} d\rho \left[
a_{\rho}(y) +{\cal O}(z, \bar{z})  
\right] +  \rho^{-\epsilon} \rho d\tau \left[
a_{\tau}(y) +{\cal O}(z, \bar{z})  
\right] + dy^i \left[
a_{i}(y) +{\cal O}(z, \bar{z})  
\right]\, .  
\end{eqnarray}
Here $\epsilon = 1-(1/n)$ and $(z, \bar{z}) =(\rho e^{i\tau}, \rho e^{-i\tau})$.
Now we carry out the regularization of this cone geometry by introducing $A=-(\epsilon/2)\log(\rho^2+a_{reg}^2)$ as 
\begin{eqnarray}
{\fA}_{U(1)} = e^{A} d\rho \left[
a_{\rho}(y) +{\cal O}(z, \bar{z})  
\right] +  e^A \rho d\tau \left[
a_{\tau}(y) +{\cal O}(z, \bar{z})  
\right] + dy^i \left[
a_{i}(y) +{\cal O}(z, \bar{z})  
\right]\, .  
\end{eqnarray}
From this gauge potential one-form on the regularized cone, we can compute the $U(1)$ field strength as 
\begin{eqnarray}
{\fF}_{U(1)} &=& d {\fA}_{U(1)}  \nonumber \\
&=& 
\left[
\partial A dz\wedge d\bar{z} \times (\ldots) + \bar{\partial} A dz\wedge d\bar{z} \times (\ldots)   
\right] \nonumber \\
&& \qquad + 
({\rm terms}\propto dz\,\, {\rm or}\,\, d\bar{z}) + {\fF} + {\cal O}(z, \bar{z})\,.  
\end{eqnarray}
Here the second line is the terms of order $\epsilon$ and the third line is of order one. 
In the third line we have also defined $\fF = d (a_i(y)dy^i) = \partial_i a_j dy^i\wedge dy^j$.
Combined with the observation in \S\ref{sec:buildingblocks} (recall that only nontrivial contribution to 
the holographic entanglement entropy formulas comes from the terms proportional to $\partial\bar{\partial}A dz\wedge d\bar{z}$ 
or $\partial A\bar{\partial}A dz\wedge d\bar{z}$ in the anomaly polynomials), 
the order $\epsilon$ terms in 
${\fF}_{U(1)}$ do not generate any non-trivial contributions after the integration over $(z, \bar{z})$, 
and only non-trivial contribution to the holographic entanglement entropy formula comes from  
the singular terms of $\fR$-dependent part wedged with $\fF$'s. 
In other words, whenever we start with the anomaly polynomials with ${\fF}$, what we need to do is: 
(1) first replace ${\fF}_{U(1)}$'s by ${\fF}$, (2) for the gravitational part of the anomaly polynomial (i.e. ${\fR}^\mu{}_\nu$-dependent part), 
carry out the $\epsilon$-expansion and extract the linear terms in $\epsilon$-expansion. Therefore, once the purely gravitational cases are done, 
the mixed cases follow almost automatically. 
For example, for a general positive integer $k$, we have
 \begin{eqnarray}
&&  {\rm AdS}_{2k+3} \,\, {\rm with}\,\,  \fPanom= c\, \fF^k \wedge \tr(\fR^2) \qquad\ : \qquad  
  \SeeCS= 2(2\pi) (2c)\int_{\Sigma }  
\fF^k\wedge \fGN \label{eq:FktrR2} \, ,   \\
&&    {\rm AdS}_{2k+7} \,\, {\rm with}\,\, \fPanom= c\, \fF^k \wedge\tr(\fR^4) \qquad\,\,: \qquad  \nonumber \\
 &&\quad\SeeCS \nonumber\\
 &&\quad =2 (2\pi)(4c)  \int_{\Sigma}
\fF^k\wedge\Biggl[
\fGN\wedge \fRN^2
+[-2\fRN +3 (\fKbar \cdot \fK)]\wedge \left[(\fK\cdot \Dhat \fKbar) + (\fKbar\cdot \Dhat \fK)\right]
\nonumber \\
&&\qquad\qquad\qquad\qquad\qquad \qquad\qquad\qquad\qquad\qquad
+
(\Dhat \fK \cdot \fr \cdot  \fKbar)-
(\Dhat \fKbar \cdot\fr \cdot  \fK)
\Biggr]\,, \label{eq:FktrR4}\nonumber\\ \\ 
&&    {\rm AdS}_7 \,\, {\rm with}\,\,\fPanom= c\, \fF^k \wedge\tr(\fR^2)\wedge \tr(\fR^2)\,\,\,\,: \qquad  \nonumber \\
&& \quad \SeeCS \nonumber \\
&&\quad = 2^2(2\pi)(4c) \int_{\Sigma} 
\fF^k \wedge\Biggl[
\fGN\wedge \fRN^2
+\frac{1}{2}\fGN\wedge{{\rm tr}\left({\fr}^2\right)}\nonumber\\
&&\qquad\qquad\qquad\qquad\qquad\qquad\qquad
+2\left[-\fRN+(\fKbar\cdot \fK)\right]\wedge \left[
(\fK\cdot \Dhat \fKbar)
+(\fKbar \cdot \Dhat \fK)\right]
\Biggr]\,.\label{eq:FtrR2trR2}
\nonumber \\
 \end{eqnarray} 
We note that the generalization to the mixed case with non-Abelian gauge field is straightforward:
we can get the holographic entanglement entropy formula by multiplying 
the purely gravitational results with a wedge product of $\tr(\fF^k)$'s.


\subsection{Consistency with Chern-Simons Contribution to Black Hole Entropy}
For general stationary black holes, the extrinsic curvature vanishes at the bifurcation surface, ${\bf K}_i=0$ and $\bar{\bf K}_i=0$. 
We can then easily see that our holographic entanglement entropy for Chern-Simons terms evaluated there
indeed reproduces the Tachikawa entropy formula 
\begin{equation}\label{eq:tachikawa-entropy}
\begin{split}
 S_{\text{Tachikawa}} &= \int_{Bif}  \sum_{k=1}^\infty 8\pi k\ \fGamma_N (d\fGamma_N)^{2k-2} \frac{\partial \fPanom}{\partial\ \text{tr} \fR^{2k} }\, . 
 \end{split}
\end{equation} 
We note that this formula was first proposed in \cite{Tachikawa:2006sz} and 
then various covariance issues in its derivation  were pointed out for 5d and higher dimensions in \cite{Bonora:2011gz}. 
Recently,  by using a manifestly covariant formulation of differential Noether charge, 
\cite{Azeyanagi:2014sna} gave a covariant derivation of this formula. 
In particular for the anomaly polynomials $\fPanom^{3d}=\tr(\fR^2)$, $\fPanom^{7d,single}=\tr(\fR^4)$ and 
$\fPanom^{7d,double}=\tr(\fR^2)\wedge \tr(\fR^2)$, the explicit forms of the entropy formulae are given by\footnote{In 3d, the black hole entropy formula below was in fact first derived in \cite{Solodukhin:2005ah}.} 
\begin{eqnarray}
&&S_{\text{Tachikawa}}^{3d} =  8\pi \int_{Bif} \fGamma_N\, ,    \\
&& S_{\text{Tachikawa}}^{7d, single} =  2(8 \pi) \int_{Bif} \fGamma_N\wedge \fR_N^2 \, , 
\qquad S_{\text{Tachikawa}}^{7d, double} 
=  2(8\pi) \int_{Bif} \fGamma_N\wedge \tr(\fR^2) \, . 
\end{eqnarray}
We note that $\tr(\fR^2) = 2[\fR_N^2+(1/2)\tr({\fr}^2)]$ which follows from Eq.~\eqref{eq:panom3d0th} evaluated at the bifurcation surface. 
Therefore, the holographic entanglement entropy formulae with ${\bf K}_i=0, \bar{\bf K}_i=0$ at the 
bifurcation horizon indeed reproduce the Tachikawa entropy formula given as above at the bifurcation horizon. 
We note that the holographic entanglement entropy formulae 
for the mixed Chern-Simons terms derived in \S\S\ref{sec:mixanom}
are  also consistent with the Tachikawa entropy formula.

\subsection{Lorentz Boost and Frame-Dependence} 
  One of the interesting properties of the Chern-Simons contribution to entanglement entropy 
  is that it depends on the choice of the Lorentz-frame.  
  This reflects non-covariant transformation property of the Chern-Simons terms and 
  quantum anomalies of Lorentz symmetry in the dual CFT side.   
  A general local Lorentz boost can be decomposed into the ones tangent to the surface $\Sigma$ and 
  the ones normal to it. The  holographic entanglement entropy formulae in Eq.~(\ref{eq:trR2})-(\ref{eq:trR2trR2}) are obviously invariant under
  the Lorentz boosts tangent to $\Sigma$. We thus consider the normal one, written as    a  normal-bundle gauge transformation 
  $\fGN \rightarrow\fGN+d\Lambda$.
  Under this transformation, 
  the variation $\Delta\SeeCS$ of 
  the holographic entanglement entropy functionals is summarized as follows: 
 \begin{eqnarray}
&& {\rm AdS}_3 \,\, {\rm with}\,\, \fPanom= c\, \tr(\fR^2) \qquad\,\,\quad: \qquad  
  \Delta\SeeCS=2(2\pi)  (2c)  \int_{\partial \Sigma }  
 \Lambda\,, \label{eq:trR2frame} \\
&&  {\rm AdS}_5 \,\, {\rm with}\,\, \fPanom= c\, {\fF}\wedge \tr(\fR^2) \,\,\,\quad: \qquad 
  \Delta\SeeCS =2(2\pi)  (2 c)  \int_{\partial \Sigma }\Lambda  \fF  
 \,,  \\
&&   {\rm AdS}_7 \,\, {\rm with}\,\, \fPanom= c\, {\fF}^2\wedge \tr(\fR^2) \,\quad: \qquad 
   \Delta\SeeCS =2(2\pi)  (2 c)  \int_{\partial \Sigma }\Lambda  \fF^2
\,,  \\
&&    {\rm AdS}_7 \,\, {\rm with}\,\, \fPanom= c\, \tr(\fR^4) \qquad\,\,\,\,\quad: \qquad  
\Delta\SeeCS =2(2\pi) (4  c ) \int_{\partial \Sigma } \Lambda  \fRN^2\, ,  \\
&& {\rm AdS}_7 \,\, {\rm with}\,\,  \fPanom= {c}\,  \tr (\fR^2) \wedge \tr (\fR^2) : \quad 
\Delta\SeeCS =2^2 (2\pi)  (4 c)\int_{\partial \Sigma }\Lambda   \left[
\fRN^2+\half \tr(\fr^2)
\right]\,,\label{eq:trR2trR2frame} \nonumber\\
 \end{eqnarray}
 where $\partial \Sigma$ is the boundary of the bulk entangling surface $\Sigma$, 
 which is identical 
 to the entangling surface of the dual CFT. We note that we can trivially generalize 
 this to more general mixed Chern-Simons terms discussed in \S\S\ref{sec:mixanom}. 
 
Here is a comment. For AdS${}_3$ dual to CFT$_2$, 
there is only one spatial direction normal to $\Sigma$. Therefore it is straightforward to 
generalize the above analysis to disconnected entangling regions\footnote{See \cite{Wall:2011kb} for a nice geometric 
argument in the single interval case.} (i.e. multi-interval cases).
For a $m$-interval vacuum entanglement entropy in CFT$_2$, by doing 
the Lorentz boost with constant $\Lambda$, we obtain 
$\Delta\SeeCS = (8\pi c) \Lambda  (2m) $, 
where each interval contributes to a `2' to the integral on the RHS of AdS$_3$ case above. 
This indeed reproduces the result obtained
in \cite{Wall:2011kb,Castro:2014tta} for $m=1$ as well as the the Lorentz-boost-dependence in the 
multi-interval answer of \cite{Hartman:2013mia} for CFTs in the large central charge limit.

As stated before, the main focus of the current paper is the derivation of holographic entanglement entropy formulae 
in the gravity side, but here we would like to comment on its relation to the dual CFT side.
More generally, let us start with $(2k)$-dimensional QFT (QFT$_{2k}$ where $k$ is a positive integer) with anomalies associated to some global continuous symmetries. The anomaly in this even-dimensional QFT 
can be understood systematically via the anomaly inflow mechanism 
 \cite{Callan:1984sa} (see also \cite{Jensen:2013kka} for finite temperature case). 
 In the anomaly inflow mechanism, the anomalies observed in QFT$_{2k}$ 
 can be interpreted as follows: we regard that this even-dimensional 
 QFT is living on the boundary of one-dimensional higher bulk (note that this bulk is irrelevant to 
 gauge/gravity dualities) 
 where an appropriate Chern-Simons term is turned on. This Chern-Simons term 
 then generates a flow of current from the bulk to the boundary, explaining the 
 breakdown of the conservation law due to the anomalies observed in this QFT$_{2k}$.   
 
 Once we have this description based on the anomaly-inflow mechanism, 
 in order to compute anomalous contribution to entanglement entropy from QFT, 
 what we need to do is to study the regularized cone  geometry for this bulk Chern-Simons system 
 of the anomaly-inflow. 
 Since the Chern-Simons term in the anomaly inflow mechanism is exactly the 
 same as the Chern-Simons term on the gravity side of gauge/gravity duality, the formal computation 
 is exactly the same on both cases.  
 Thus, in the same way as 
we have computed the holographic entanglement entropy formula from Chern-Simons term in  \S\ref{sec:directchernsimons}, we can derive the anomalous contribution to entanglement entropy on QFT side. 
Then, by doing the $U(1)$-gauge transformation along the normal bundle direction, we obtain 
the same result as we have obtained from the dual gravity side. 
Although we have used holography to obtain Eq.~(\ref{eq:trR2frame})-(\ref{eq:trR2trR2frame}), the above argument suggests that the validity of these formulae for the frame-dependence part of entanglement entropy goes beyond CFTs with a gravity dual, and hence they are applicable for a generic QFT$_{2k}$ with anomalies.

\subsection{Application to Anomaly Polynomials of 6D $(2,0)$ theories} 
As an application, we derive the holographic entanglement entropy functional associated with anomaly polynomials of $(2,0)$ theories with gauge group $G$ (see \cite{Intriligator:2000eq,Alday:2009qq,Bah:2012dg} for some detailed discussions):
\be
\fPanom=-\frac{2\pi r_G}{48} \left[p_2(\fF)-p_2(\fR)+\frac{1}{4}\left(p_1(\fR)-p_1(\fF)\right)^2\right]
-\frac{2\pi d_G h_G}{24}p_2(\fF)\, , 
\ee where $r_G, d_G$ and $h_G$ are the rank, the dimension and the Coxeter number of the gauge group $G$ respectively while the Pontryagin classes are defined with the conventions:
\be
p_1(\fF)=\half \left(\frac{i}{2\pi}\right)^2 \tr (\fF^2),\quad 
p_2(\fF) = \frac{1}{8} \left(\frac{i}{2\pi}\right)^4\left[
(\tr \fF^2)\wedge ( \tr \fF^2)
-2 \tr (\fF^4)
\right]\, , 
\ee and accordingly the same definitions for $p_1 (\fR)$ and $p_2(\fR)$ upon sending $\fF\rightarrow \fR$. The field strength $\fF$ here is the $SO(5)_R$ field strength. Applying the results in Eqs.~(\ref{eq:FktrR2})-(\ref{eq:FtrR2trR2}), we obtain
\bea
&&\SeeCS \nonumber \\
&&=\frac{r_G}{2\times (48)\times (2\pi)^2}
\int_{\Sigma}
\Biggl[
-2\,\fGN \wedge\fRN^2
+\fGN \wedge\left[\tr (\fr^2)+\tr (\fF^2)\right]
\nonumber\\
&&\qquad\qquad\qquad\qquad\qquad\qquad\qquad
+4 \left[\fRN-2(\KK\right)] \wedge\left[(\fKbar\cdot \Dhat \fK)+(\fK\cdot \Dhat \fKbar)\right] 
\nonumber \\
&&\qquad\qquad\qquad \qquad\qquad\qquad\qquad
+4\left[
(\Dhat \fKbar\cdot \fr \cdot \fK)
-(\Dhat \fK\cdot \fr \cdot \fKbar)
\right]
\Biggr]\,. 
\eea

\section{Conclusion and Discussion} \label{sec:conclusion}
In this work, we have taken the first step towards understanding how extrinsic corrections to entanglement entropy can be reliably derived
for gravitational theories with CS terms. While the Christoffel connection on the holographic replica background should be carefully dealt with,
once this is done, one gets an answer that matches with indirect methods of computation using anomaly polynomial. 

The simplest extension of our work would be to compute the extrinsic corrections for an arbitrary CS term and give a closed form answer. While this is in 
principle a straightforward exercise using the methods described in this paper, in practice, the computations become more and more tedious.
Perhaps an intelligent reader can organise this computation into  a nice closed form answer which uncovers the structure of these  
corrections.

Given the explicit form of corrections in the case of AdS$_7$, there is a clear cut challenge to reproduce it from the field theory side by using CFT$_6$
with Lorentz anomalies. There is a good reason to assume that the corrections we have computed do not depend on the coupling (since they are 
tied to anomalies) and hence should be visible even in free theory computations. Both the free theory of chiral fermions and that of free chiral 2-form fields in 6d 
contribute to Lorentz anomalies and it is an interesting question as to how  these extrinsic corrections show up in entanglement entropy computations of these theories. 
We hope that our work will motivate entanglement studies of these free theories which would shed light on the physics behind the Dong-type corrections.

Although we focused on the derivation of holographic  entanglement entropy functional, 
in order to evaluate the Chern-Simons contribution to entanglement entropy for 
an entangling region for a given state,  another step is involved---identification of the bulk entangling surface in the dual gravitational background. 
For Einstein gravity,  the location of the surface is identified in \cite{Lewkowycz:2013nqa}  
by requiring that the singular terms in the bulk equation of motion vanish  on the regularized 
cone background.\footnote{See \cite{Camps:2014voa} for a more detailed analysis of the bulk replicated geometry.} 
This proves the 
prescription by Ryu and Takayanagi that the bulk entangling surface minimizes the holographic entanglement entropy functional. 

For a general higher derivative theory,  it is still unclear if the approach by  \cite{Lewkowycz:2013nqa} , which becomes in general complicated, 
is equivalent to the minimization of the holographic entanglement 
functional  (see  \cite{Dong:2013qoa,Bhattacharyya:2013gra,Erdmenger:2014tba, Camps:2014voa} for a related discussion in the context of Lovelock gravity). 
By using a relatively simple structure of the holographic entanglement entropy functional for Chern-Simons terms and bulk equation of motion, 
we hope that our analysis will lead to a deeper understanding of this second step.  In  \cite{Castro:2014tta}, a geodesic equation for spinning particle 
was analyzed to identify the bulk entangling surface for 3d gravitational Chern-Simons term. It would be nice
to generalize this approach to higher dimensions  by considering spinning membranes.\footnote{We thank  A.~Castro for interesting 
discussions on this issue.}

We note that the non-static situations might be necessary in order to get a nontrivial contribution to holographic entanglement entropy from the Chern-Simons terms. 
For the simplest case with the spherical entangling region for CFT vacuum state, 
the terms examined in this paper vanish. This follows from the fact that both the 
usual BH entropy part (given by Tachikawa formula) and the  extrinsic curvature
corrections vanish in this case.\footnote{We would like to thank Mukund Rangamani for emphasizing this point.}  It would be worthwhile to consider a gravitational background dual to an excited state or a more general entangling surface to obtain non-trivial contribution.

Another future direction is a generalization to time-dependent background to construct the CS analogue of HRT
proposal \cite{Hubeny:2007xt} of Einstein gravity. Hydrodynamics for anomalous systems and its holographic description
are of great interests both from theoretical and experimental point of view. 
As a result of recent developments,  we now have a systematic understanding 
of the leading-order  anomaly-induced transport both from quantum field theory side \cite{Jensen:2013kka,Jensen:2013rga} 
as well as dual gravity side \cite{Azeyanagi:2013xea,Azeyanagi:2014sna,Azeyanagi:2015gqa}. 
It would be worthwhile to add entanglement entropy to the list 
of items to probe and characterize these anomalous systems. 
 
\section*{Acknowledgements}
The authors would like to thank J.~Camps, A.~Castro, S.~Detournay, X.~Dong, N.~Halmagyi, N.~Iqbal, K.~Jensen, F.~L. Lin, E.~Perlmutter, M.~Rangamani, S.~N.~Solodukhin and A.~C.~Wall, A.~Yarom for valuable discussions.  
T.~A. would like to thank the Galileo Galilei Institute and Okinawa Institute of Science and Technology 
for hospitality. 
T.~A. was in part supported by INFN during his stay in the Galileo Galilei Institute for the workshop ``Holographic Methods for Strongly Coupled System." 
T.~A. was financially supported by the LabEx ENS-ICFP: ANR-10-LABX-0010/ANR-10-IDEX-0001-02 PSL*. 
R.~L. was supported by Institute for Advanced Study, Princeton. R.~L. would like to thank Institute for Mathematical Sciences Chennai and
Tata Institute for Fundamental Research Mumbai for their hospitality during the final stages of this work. 
G.~N. was supported by an NSERC Discovery Grant.

\appendix

\section{Some Useful Quantities}\label{sec:useful}
This Appendix summarizes the connection one-form $\fGamma$, 
curvature two-form $\fR$ and their wedge products evaluated on the regularized cone background  \eqref{eq:metric_conical}.   
We will use them in the evaluation of anomaly polynomials as well as gravitational Chern-Simons terms.
It is noted that, as we mentioned in \S\ref{sec:buildingblocks}, we dropped 
many terms that are irrelevant to the evaluation of the holographic entanglement 
entropy formulas for Chern-Simons terms in this paper.   

Here is one remark on the notation. In this Appendix, 
after evaluating $\fR$ etc. on the regularized cone geometry, 
we will decompose them into pieces and  use the following notation and terminology:
By superscript ${}^{(0)}$ (for example, ${}^{(0)}\fR$), we label the contributions of 
order-$\epsilon^0$ and at the same time zeroth order in $(z,\bar{z})$-expansion, 
while the superscript ${}^{(d)}$ labels the order-$\epsilon$ term of the from 
$\partial Adz$ or $\bar{\partial}Ad\bar{z}$ multiplied by zeroth order terms in $(z, \bar{z})$-expansion. 
By superscript ${}^{(d^2)}$, we denote the order-$\epsilon$ term of the form
$\partial\bar{\partial}Adz\wedge d\bar{z}$ multiplied by zeroth order terms in $(z, \bar{z})$-expansion. 
Finally the superscript ${}^{(dd)}$ is used for  the order-$\epsilon^2$ term of the form
$\partial A \bar{\partial}Adz\wedge d\bar{z}$ multiplied by zeroth order terms in $(z, \bar{z})$-expansion. 

\subsection{Christoffel Symbol}
On the regularized cone geometry \eqref{eq:metric_conical}, 
the Christoffel symbol up to order $\epsilon^1$ and 
up to first order in $(z,\bar{z})$-expansion is computed as 
\begin{eqnarray}\label{eq:christoffel}
&&\Gamma^{z}{}_{zz} = 2\,\partial A -4e^{2A}\,T\,\bar{z}\, , \qquad 
\Gamma^{z}{}_{z\bar{z}} = 2 e^{2A}\, T\, z\, \qquad 
\Gamma^{z}{}_{\bar{z}\bar{z}} =0\, , \nonumber \\ 
&&\Gamma^{z}{}_{zj} =
{+2i\epsilon U_j-i(zV_{zj} + \bar{z}V_{\bar{z}j})}
-2\,i\,(U_j+V_{aj} x^a) + 2\, i\, U_m \,g^{ml}K_{zlj}\,z\, , \nonumber \\ 
&& \Gamma^{z}{}_{\bar{z}j} = 2\,i\, U_m\, g^{ml}K_{\bar{z}lj}\,z\, , \nonumber \\
&& \Gamma^{z}{}_{jk} =
 -i\,z\,(\partial_j U_k+\partial_k U_j) 
-2\, e^{-2A}(K_{\bar{z}jk}+Q_{\bar{z}ajk}\,x^a) + 2\,i\,U_m \gamma^m{}_{jk}\, z\, ,  \nonumber\\
&&\Gamma^{\bar{z}}{}_{\bar{z}\bar{z}} = 2\,\bar{\partial} A -4e^{2A}\,T\,z\, , \qquad 
\Gamma^{\bar{z}}{}_{z\bar{z}} = 2 e^{2A}\, T\, \bar{z}\, ,\qquad 
\Gamma^{\bar{z}}{}_{zz} =0\, , \nonumber \\ 
&&\Gamma^{\bar{z}}{}_{\bar{z}j} = 
{-2i\epsilon U_j+i(zV_{zj}+\bar{z}V_{\bar{z}j})}
+ 2\,i\,(U_j+V_{aj} x^a) - 2\, i\, U_m \,g^{ml}K_{\bar{z}lj}\,\bar{z}\, , \nonumber \\ 
&&  \Gamma^{\bar{z}}{}_{zj} = -2\,i\, U_m\, g^{ml}K_{zlj}\,\bar{z}\, , \nonumber \\
&& \Gamma^{\bar{z}}{}_{jk} = i\,\bar{z}\,(\partial_j U_k+\partial_k U_j) 
-2\, e^{-2A}(K_{zjk}+Q_{zajk}\,x^a) - 2\,i\,U_m \gamma^m{}_{jk}\, \bar{z}\, ,  \nonumber \\
&&
\Gamma^i{}_{zz}= 
 i e^{2A} g^{il}V_{zl}\, \bar{z}\,, 
\qquad 
\Gamma^i{}_{\bar{z}\bar{z}}= 
- i e^{2A} g^{il}V_{\bar{z}l}\, z\,, 
\nonumber \\
&&
\Gamma^i{}_{z\bar{z}} 
= -\frac{i}{2}e^{2A}g^{il}(V_{zl}\,z- 
V_{\bar{z}l}\,\bar{z}) 
\,, \\
&& 
\Gamma^i{}_{jz} = 
-2e^{2A}\bar{z}g^{il}U_l U_j +g^{il}K_{zjl}-2g^{im}g^{ln}K_{amn}x^a K_{zjl}
\nonumber\\
&&\qquad\qquad\qquad
+\frac{i}{2}e^{2A}\bar{z}g^{il}(\partial_{j} U_l -\partial_{l} U_j)+g^{il}Q_{zajl}x^a
\,, \nonumber \\
&& 
\Gamma^i{}_{j\bar{z}} = 
-2e^{2A}zg^{il}U_l U_j +g^{il}K_{\bar{z}jl}-2g^{im}g^{ln}K_{amn}x^a K_{\bar{z}jl}
\nonumber\\
&&\qquad\qquad\qquad
-\frac{i}{2}e^{2A}zg^{il}(\partial_{j} U_l -\partial_{l} U_j)+g^{il}Q_{\bar{z}ajl}x^a
\,, \nonumber \\
&&
\Gamma^{i}{}_{jk} = -2g^{il}U_l (K_{zjk} z -K_{\bar{z}jk}\bar{z})
+\gamma^{i}{}_{jk} -2g^{in}K_{a nm} x^a \gamma^m{}_{jk}
\nonumber\\
&&\qquad\qquad\qquad 
+g^{il}\left(\partial_j K_{alk}+\partial_k K_{ajl}-\partial_l K_{ajk} \right)x^a\, . \nonumber
\end{eqnarray}
Here we have defined the Christoffel symbol for $g_{ij}$ as 
$\gamma^{i}{}_{jk}\equiv 1/2g^{il}(\partial_{j}g_{lk}+\partial_{k}g_{jl}- \partial_{l}g_{jk})$.

It is also convenient for our purpose to define the
normal-bundle connection $\GN_i$ and its field strength as 
\be
\GN_i \equiv(-2i U_i) \, , \quad
\RN_{ij}\equiv 2 \partial_{[i} \GN_{j]}\, , 
\ee with their differential forms being
$\fGamma_N \equiv \GN_i dx^i$ and $\fR_N \equiv d\fGamma_N$.

\subsection{Connection One-Form}
The connection one-form is defined by 
$\fGamma^\mu{}_{\nu} \equiv \Gamma^\mu{}_{\nu\rho}\,dx^\rho$. 
From the expression of the Christoffel symbol in Eq.~\eqref{eq:christoffel},
it is obvious that, in the connection one-form,   
$\partial A$ and $\bar{\partial} A$ are always accompanied by $dz$ and $d\bar{z}$, respectively. 
Up to zeroth order in $(z,\bar{z})$-expansion
and up to order $\epsilon^1$, the connection one-form is computed as 
\begin{eqnarray} \label{eq:connectionzeroth}
&&\fGamma^z{}_z = 2\partial A dz +\fGN+\ldots \, , \quad 
\fGamma^{\bar{z}}{}_{\bar{z}} = 2\bar{\partial} A d\bar{z}
-\fGN+\ldots\, , \quad \fGamma^{z}{}_{\bar{z}} = \fGamma^{\bar{z}}{}_z = 0+\ldots \, , \nonumber \\
&& \fGamma^z{}_j = -2 e^{-2A} \bar{\bf K}_j+\ldots\, , \qquad 
\fGamma^{\bar{z}}{}_j =  -2 e^{-2A} {\bf K}_j+\ldots\, , \nonumber \\
&&
\fGamma^i{}_{z} = {\bf K}^i+\ldots\,, \qquad 
\fGamma^i{}_{\bar{z}} = \bar{\bf K}^i+\ldots\,, \qquad 
\fGamma^i{}_j = {\bf \gamma}^{i}{}_{j}+\ldots\, , 
\end{eqnarray} 
where we have defined extrinsic curvature one-forms ${\bf K}_i\equiv{ K}_{zij}dy^j$, 
$\bar{\bf K}_i\equiv \bar{K}_{\bar{z}ij}dy^j$, 
${\bf K}^i\equiv g^{il} K_{zlj}dy^j$, 
$\bar{{\bf K}}^i\equiv g^{il}{K}_{\bar{z}lj}dy^j$ and the connection one-form 
${\bf \gamma}^{i}{}_{j}\equiv \gamma^{i}{}_{jk}dy^k$ for $\gamma^i{}_{jk}$. 
For later convenience (in the direct evaluation from Chern-Simons terms in \S\ref{sec:directchernsimons}), here we decompose
the above expression of the connection one form as follows: 
\begin{eqnarray} \label{eq:fgammaexpanded}
&&{}^{(d)}\fGamma^z{}_z = 2\partial A dz  \, , \qquad 
{}^{(d)}\fGamma^{\bar{z}}{}_{\bar{z}} = 2 \bar{\partial} A d\bar{z}\, , 
\qquad {}^{(d)}\fGamma^{z}{}_{\bar{z}} =0\,, \qquad {}^{(d)}\fGamma^{\bar{z}}{}_z = 0 \, , \nonumber \\
&& {}^{(d)}\fGamma^z{}_j = 0\, , \qquad 
{}^{(d)}\fGamma^{\bar{z}}{}_j =  0\, , \qquad
{}^{(d)}\fGamma^i{}_{z} = 0\,, \qquad 
{}^{(d)}\fGamma^i{}_{\bar{z}} = 0\,, \qquad 
{}^{(d)}\fGamma^i{}_j = 0\, , \\
&&{}^{(0)}\fGamma^z{}_z = \fGN \, , \qquad 
{}^{(0)}\fGamma^{\bar{z}}{}_{\bar{z}} = 
-\fGN\, , \qquad {}^{(0)}\fGamma^{z}{}_{\bar{z}} = \fGamma^{\bar{z}}{}_z = 0 \, , \nonumber \\
&& {}^{(0)}\fGamma^z{}_j = (-2 e^{-2A}) \bar{\bf K}_j\, , \qquad 
{}^{(0)}\fGamma^{\bar{z}}{}_j =  (-2 e^{-2A} ){\bf K}_j\, , \qquad
{}^{(0)}\fGamma^i{}_{z} = {\bf K}^i\,, \qquad 
{}^{(0)}\fGamma^i{}_{\bar{z}} = \bar{\bf K}^i\,, \nonumber\\
&& 
{}^{(0)}\fGamma^i{}_j = {\bf \gamma}^{i}{}_{j}\, . \nonumber
\end{eqnarray}
We note that we have dropped the terms proportional to $dz$ or $d\bar{z}$ without
 $\partial A$ or $\bar{\partial} A$ accompanied, since they are irrelevant 
 for the purpose of this paper as mentioned in \S\ref{sec:buildingblocks}. 
 In the rest of this Appendix, we will also drop these terms. 

\subsection{Derivative of Connection One-Form}
We next evaluate the derivative of the connection one-form, $d\fGamma^\mu{}_\nu$ 
up to order $\epsilon^1$ and 
up to zeroth order in $(z,\bar{z})$-expansion:
\begin{eqnarray} \label{eq:derivativeconnection1form}
&&d\fGamma^{z}{}_z  = -2\partial \bar{\partial} A dz\wedge d\bar{z} 
+\fRN+\ldots\, , \quad 
d\fGamma^{\bar{z}}{}_{\bar{z}}  = 2\partial \bar{\partial} A dz\wedge d\bar{z} 
-\fRN+\ldots\, , \nonumber  \\
&& d\fGamma^{z}{}_{\bar{z}} =0\, , \qquad  d\fGamma^{\bar{z}}{}_z =0+\ldots\, , \quad 
d\fGamma^i{}_z =d {\bf K}^i+\ldots\,, \qquad 
d\fGamma^i{}_{\bar{z}} =d\bar{\bf K}^i+\ldots\,, \nonumber \\
&&
d\fGamma^z{}_j = (-2e^{-2A})(-2dA)\wedge \bar{\bf K}_j+(-2e^{-2A}) d\bar{\bf K}_j+\ldots\, , \nonumber \\
&&
d\fGamma^{\bar{z}}{}_j =(-2e^{-2A})(-2dA)\wedge {\bf K}_j+( -2e^{-2A} )d{\bf K}_j+\ldots\, ,  \nonumber \\
&&d\fGamma^i{}_j = d{\boldsymbol \gamma}^{i}{}_j+\ldots\, . 
\end{eqnarray}

\subsection{Curvature Two-Form}
By using the above results, the curvature two-form 
${\bf R}^\mu{}_\nu \equiv d{\bf \Gamma}^\mu{}_\nu+{\bf \Gamma}^\mu{}_\rho\wedge {\bf \Gamma}^\rho{}_\nu$
is calculated as 
{
\begin{eqnarray}\label{eq:curvature2form}
&&{\fR}^z{}_z = -2\partial \bar{\partial}A dz\wedge d\bar{z}
+\fRN +(-2e^{-2A})(\bar{\bf K}\cdot {\bf K}) +\ldots\, , \nonumber \\
&&{\fR}^{\bar{z}}{}_{\bar{z}} = 2\partial \bar{\partial}A dz\wedge d\bar{z}
-\fRN-(-2e^{-2A})(\bar{\bf K}\cdot {\bf K}) +\ldots\, , \nonumber \\
&&{\fR}^z{}_{\bar{z}}={\fR}^{\bar{z}}{}_{z}=0 \nonumber \, , \\
&& 
{\fR}^i{}_{z} = -2 {\partial}Adz\wedge{\bf K}^i
+\Dhat{\fK}^i+\ldots\, , 
\nonumber \\
&& 
{\fR}^i{}_{\bar{z}} = -2 \bar{\partial} A d\bar{z}\wedge\bar{\bf K}^i
+\Dhat \bar{\bf K}^i+\ldots\, , 
\nonumber \\
&&
{\fR}^{z}{}_j = (-2e^{-2A})(-2\bar{\partial} A d\bar{z})\wedge\bar {\bf K}_j 
+(-2e^{-2A})\Dhat\bar{\bf K}_j+\ldots\, , 
\nonumber \\
&&
{\fR}^{\bar{z}}{}_j = (-2e^{-2A})(-2 \partial A dz)\wedge {\bf K}_j 
+(-2e^{-2A})\Dhat{\bf K}_j+\ldots\, , 
\nonumber \\
&&{\fR}^i{}_j = {\fr}^i{}_j - 2e^{-2A}({\bf K}^i \wedge \bar{\bf K}_j+ \bar{\bf K}^i \wedge {\bf K}_j)+\ldots \, ,
\end{eqnarray}
}
where $\bar{\bf K}\cdot {\bf K}= \bar{\bf K}^i\wedge{\bf K}_i$.  
Here the covariant derivative $\Dhat$ compatible with the normal bundle gauge field $\fGamma_N$ 
is defined such that 
\begin{eqnarray}
&& \Dhat{\bf K}_i \equiv  D{\bf K}_i - \fGamma_N\wedge {\bf K}_i \, ,
\qquad
\Dhat\bar{\bf K}_i \equiv  D\bar{\bf K}_i + \fGamma_N\wedge \bar{\bf K}_i \, , \nonumber \\
&& \Dhat{\bf K}^i \equiv  D{\bf K}^i - \fGamma_N \wedge{\bf K}^i\, ,
\qquad
\Dhat \bar{\bf K}^i \equiv  D\bar{\bf K}^i + \fGamma_N\wedge\bar{\bf K}^i\, ,
\end{eqnarray}
where the covariant derivative $D$
with respect to $g_{ij}$ is defined as
\begin{eqnarray}
&& D{\bf K}_i \equiv  d{\bf K}_i -{\bf \gamma}^j{}_i \wedge {\bf K}_j  \, ,
\qquad
D\bar{\bf K}_i \equiv  d\bar{\bf K}_i - {\bf \gamma}^j{}_i\wedge \bar{\bf K}_j \, , \nonumber \\
&& D{\bf K}^i \equiv  d{\bf K}^i + {\bf \gamma}^i{}_j\wedge{\bf K}^j\, ,
\qquad
D\bar{\bf K}^i \equiv  d\bar{\bf K}^i + {\bf \gamma}^i{}_j\wedge\bar{\bf K}^j\, . 
\end{eqnarray}
We also note that ${\fr}^i{}_j \equiv (1/2)r^i{}_{jkl}dy^k\wedge dy^l
\equiv d {\bf \gamma}^i{}_j +{\bf  \gamma}^i{}_k \wedge {\bf \gamma}^k{}_j$
is the curvature two-form for $g_{ij}$.

In Eq.~\eqref{eq:curvature2form}, 
we have written down the explicit form of the terms in the curvature two-form 
that are needed for our purpose.  
By using the notation introduced at the beginning of this Appendix, 
it is convenient to decompose the above 
expression in the following way: 
{
\begin{eqnarray}\label{eq:Rdecomposed}
&&{}^{(d^2)}{\fR}^z{}_z = -2\partial \bar{\partial} A dz\wedge d\bar{z}\,, \qquad 
{}^{(d^2)}{\fR}^{\bar{z}}{}_{\bar{z}} = 2\partial \bar{\partial}A dz\wedge d\bar{z}\,,\nonumber \\
&&
{}^{(d^2)}{\fR}^z{}_{\bar{z}} = 0\, , \qquad  {}^{(d^2)}{\fR}^{\bar{z}}{}_{{z}} = 0\, , 
 \nonumber \\
&&{}^{(d^2)}{\fR}^i{}_z=0\,, \quad {}^{(d^2)}{\fR}^i{}_{\bar{z}} =0\, , \quad {}^{(d^2)}{\fR}^z{}_{j}=0\, , \quad 
{}^{(d^2)}{\fR}^{\bar{z}}{}_{j}=0\, ,\quad {}^{(d^2)}{\fR}^i{}_j=0\, ,  \nonumber \\
&& {}^{(dd)}{\fR}^{\mu}{}_{\nu}=0\, , \nonumber \\
&&{}^{(d)}{\fR}^z{}_z =0\, , \quad  
{}^{(d)}{\fR}^{\bar{z}}{}_{\bar{z}} = 0\, ,\quad  {}^{(d)}{\fR}^z{}_{\bar{z}}=0\, , \quad 
{}^{(d)}{\fR}^{\bar{z}}{}_{z}=0\, , \nonumber \\
&& {}^{(d)}{\fR}^i{}_{z} = -2\partial Adz\wedge{\bf K}^i\,,\qquad 
{}^{(d)}{\fR}^i{}_{\bar{z}} = -2 \bar{\partial} A d\bar{z}\wedge\bar{\bf K}^i\, , \nonumber \\
&&{}^{(d)}{\fR}^{z}{}_j = (-2e^{-2A})(-2\bar{\partial} A d\bar{z})\wedge\bar {\bf K}_j \,,\qquad 
{}^{(d)}{\fR}^{\bar{z}}{}_j = (-2e^{-2A})(-2\partial A dz)\wedge {\bf K}_j \, , \nonumber \\
&&{}^{(d)}{\fR}^i{}_j = 0\,, \nonumber \\
&&{}^{(0)}{\fR}^z{}_z = 
\fRN+(-2e^{-2A})(\bar{\bf K}\cdot {\bf K}) \, , \qquad
{}^{(0)}{\fR}^{\bar{z}}{}_{\bar{z}} = 
-\fRN-(-2e^{-2A})(\bar{\bf K}\cdot {\bf K}) \, , \nonumber \\
&&{}^{(0)}{\fR}^z{}_{\bar{z}}=0\, , \quad {}^{(0)}{\fR}^{\bar{z}}{}_{z}=0\, , \nonumber \\
&& {}^{(0)}{\fR}^i{}_{z} = \Dhat{\fK}^i\, , \qquad 
{}^{(0)}{\fR}^i{}_{\bar{z}} =
\Dhat\bar{\bf K}^i \, , \nonumber \\
&&{}^{(0)}{\fR}^{z}{}_j = 
(-2e^{-2A})\Dhat\bar{\bf K}_j\, , \qquad
{}^{(0)}{\fR}^{\bar{z}}{}_j = 
(-2e^{-2A})\Dhat{\bf K}_j\, , \nonumber \\
&&{}^{(0)}{\fR}^i{}_j = {\fr}^i{}_j +(- 2e^{-2A})({\bf K}^i \wedge \bar{\bf K}_j+ \bar{\bf K}^i \wedge {\bf K}_j) \, . 
\end{eqnarray}
}

\subsection{Structures of Products of Connections and Curvatures}

\subsubsection{\fixform{${\fGamma}^2$}}
For the wedge product $({\bf \Gamma}^2)^\mu{}_\nu=\fGamma^\mu{}_\rho\wedge\fGamma^\rho{}_\nu$, 
after some computation, we can obtain the following result: 
\begin{eqnarray}
{}^{(d^2)}(\fGamma^2)^\mu{}_\nu=0 \, , \qquad {}^{(dd)}(\fGamma^2)^\mu{}_\nu=0\, , 
\end{eqnarray}
\begin{eqnarray}
&&{}^{(d)}(\fGamma^2)^{z}{}_z = 0 , \qquad 
{}^{(d)}(\fGamma^2)^{\bar{z}}{}_{\bar{z}} = 0\, , \qquad 
{}^{(d)}(\fGamma^2)^{z}{}_{\bar{z}}=0\,, \qquad {}^{(d)}(\fGamma^2)^{\bar{z}}{}_z=0\, , \nonumber \\
&& {}^{(d)}(\fGamma^2)^{z}{}_j =(-2e^{-2A})(2\partial A dz)\wedge \bar{\bf K}_j \, , 
\qquad {}^{(d)}(\fGamma^2)^{\bar{z}}{}_j =(-2e^{-2A})(2\bar{\partial} A d\bar{z})\wedge {\bf K}_j\, , 
\nonumber \\
&&{}^{(d)}(\fGamma^2)^{i}{}_z = {\bf K}^i \wedge(2\partial A dz)\,,  \qquad 
{}^{(d)}(\fGamma^2)^{i}{}_{\bar{z}} = \bar{\bf K}^i \wedge(2\bar{\partial} A d\bar{z})\, ,  \nonumber \\
&& {}^{(d)}(\fGamma^2)^{i}{}_j = 0\, ,   
\end{eqnarray} 
and
\begin{eqnarray}
&&{}^{(0)}(\fGamma^2)^{z}{}_z =(-2e^{-2A})(\bar{\bf K}\cdot {\bf K})\, , \qquad 
{}^{(0)}(\fGamma^2)^{\bar{z}}{}_{\bar{z}} =-(-2e^{-2A})(\bar{\bf K}\cdot {\bf K})\, , \nonumber \\
&&
{}^{(0)}(\fGamma^2)^{z}{}_{\bar{z}}=0\,, \qquad {}^{(0)}(\fGamma^2)^{\bar{z}}{}_z=0\, , \nonumber \\
&& {}^{(0)}(\fGamma^2)^{z}{}_j =(-2e^{-2A})\left[\fGN\wedge\bar{\bf K}_j + (\bar{\bf K}\cdot {\bf \gamma})_j \right] \, , \nonumber\\
&&
 {}^{(0)}(\fGamma^2)^{\bar{z}}{}_j =(-2e^{-2A})\left[-\fGN\wedge{\bf K}_j + ({\bf K}\cdot {\bf \gamma})_j \right] \, , 
\nonumber \\
&&{}^{(0)}(\fGamma^2)^{i}{}_z = {\bf K}^i \wedge\fGN+(\gamma\cdot {\bf K})^i\, , \qquad 
{}^{(0)}(\fGamma^2)^{i}{}_{\bar{z}} = -\bar{\bf K}^i \wedge\fGN+(\gamma\cdot \bar{\bf K})^i\, , \nonumber \\
&& {}^{(0)}(\fGamma^2)^{i}{}_j = (-2e^{-2A})({\bf K}^i\wedge\bar{\bf K}_j+\bar{\bf K}^i\wedge{\bf K}_j) + (\gamma^2)^{i}{}_j\, .  
\end{eqnarray}
Here $(\bar{\bf K}\cdot {\bf \gamma})_j = \bar{\bf K}_i \wedge {\bf \gamma}^i{}_j$, 
$(\gamma^2)^{i}{}_j =  \gamma^{i}{}_k \wedge \gamma^k{}_j$ etc. 

\subsubsection{${\bf \Gamma}^3$}
The wedge product ${\bf \Gamma}^3$ is decomposed as follows:
\begin{eqnarray}
{}^{(d^2)}(\fGamma^3)^{\mu}{}_\nu=0\, , \qquad  {}^{(dd)}(\fGamma^3)^{\mu}{}_\nu=0\, , 
\end{eqnarray} 
\begin{eqnarray}
&&{}^{(d)}(\fGamma^3)^{z}{}_z = 2(2\partial A dz)(-2e^{-2A})(\bar{\bf K}\cdot {\bf K})\, , \qquad 
{}^{(d)}(\fGamma^3)^{\bar{z}}{}_{\bar{z}} = (-2)(2\bar{\partial} A d\bar{z})(-2e^{-2A})(\bar{\bf K}\cdot {\bf K})\, , \nonumber\\
&&{}^{(d)}(\fGamma^3)^{z}{}_{\bar{z}} =0\, , \qquad 
{}^{(d)}(\fGamma^3)^{\bar{z}}{}_{z} = 0\, , \nonumber\\ 
&& {}^{(d)}(\fGamma^3)^{z}{}_j =(-2e^{-2A})(2\partial A dz)\wedge (\bar{\bf K}\cdot\gamma)_j \, , 
\qquad {}^{(d)}(\fGamma^3)^{\bar{z}}{}_j =(-2e^{-2A})(2\bar{\partial} A d\bar{z})\wedge ({\bf K}\cdot\gamma)_j\, , 
\nonumber \\
&&{}^{(d)}(\fGamma^3)^{i}{}_z = (2\partial A dz)\wedge (\gamma\cdot {\bf K})^i\, , 
\nonumber \\
&& 
{}^{(d)}(\fGamma^3)^{i}{}_{\bar{z}} = (2\bar{\partial} A d\bar{z})\wedge (\gamma\cdot \bar{\bf K})^i 
\, , \qquad 
  \nonumber \\  
&& {}^{(d)}(\fGamma^3)^{i}{}_j =  
(2\partial A dz)(-2e^{-2A})(-1)({\bf K}^i \wedge\bar{\bf K}_j) + 
(2\bar{\partial} A d\bar{z})(-2e^{-2A})(-1)(\bar{\bf K}^i \wedge {\bf K}_j) \, , 
\end{eqnarray}
and
\begin{eqnarray}
&&{}^{(0)}(\fGamma^3)^{z}{}_z = 
(-2e^{-2A}) \left[ 2\fGN\wedge (\bar{\bf K}\cdot {\bf K})+(\bar{\bf K}\cdot\gamma\cdot {\bf K})\right]\, , \nonumber \\
&&{}^{(0)}(\fGamma^3)^{\bar{z}}{}_{\bar{z}}= 
(-2e^{-2A}) \left[ 2\fGN\wedge (\bar{\bf K}\cdot {\bf K}){+({\bf K}\cdot\gamma\cdot \bar{\bf K})}\right]\, , \nonumber \\
&&{{}^{(0)}(\fGamma^3)^{z}{}_{\bar{z}} =  (-2e^{-2A})(\bar{\bf K}\cdot \gamma \cdot \bar{\bf K})\, , \qquad {}^{(0)}(\fGamma^3)^{\bar{z}}{}_{z} = (-2e^{-2A})({\bf K}\cdot \gamma \cdot {\bf K}) }\, , \nonumber  \\
&& {}^{(0)}(\fGamma^3)^{z}{}_j = (-2e^{-2A})^2(\bar{\bf K}\cdot {\bf K})\wedge \bar{\bf K}_j 
+ (-2e^{-2A})\left[\fGN\wedge(\bar{\bf K}\cdot\gamma)_j +(\bar{\bf K}\cdot\gamma\cdot\gamma)_j\right]\, , \nonumber \\
&& {}^{(0)}(\fGamma^3)^{\bar{z}}{}_j = -(-2e^{-2A})^2(\bar{\bf K}\cdot {\bf K})\wedge{\bf K}_j 
+ (-2e^{-2A})\left[-\fGN\wedge({\bf K}\cdot\gamma)_j +({\bf K}\cdot\gamma\cdot\gamma)_j\right]\, , \nonumber \\
&&{}^{(0)}(\fGamma^3)^{i}{}_z = 
\left[(\gamma\cdot {\bf K})^i\wedge\fGN+(\gamma\cdot\gamma\cdot {\bf K})^i\right] + (-2e^{-2A}){\bf K}^i\wedge(\bar{\bf K}\cdot {\bf K})\, , 
\nonumber \\
&&{}^{(0)}(\fGamma^3)^{i}{}_{\bar{z}} = 
\left[-(\gamma\cdot \bar{\bf K})^i\wedge\fGN+(\gamma\cdot\gamma\cdot \bar{\bf K})^i\right] - (-2e^{-2A})\bar{\bf K}^i\wedge(\bar{\bf K}\cdot {\bf K})\, , 
\nonumber \\
&& {}^{(0)}(\fGamma^3)^{i}{}_j =   (\gamma^3)^i{}_j \nonumber \\
&&\qquad\qquad \quad
+(-2e^{-2A})\left[
-\fGN\wedge({\bf K}^i\wedge\bar{\bf K}_j - \bar{\bf K}^i\wedge{\bf K}_j )
\right.\nonumber\\
&&\left.\qquad\qquad\qquad\qquad\qquad\,\,
+(\gamma\cdot {\bf K})^i\wedge \bar{\bf K}_j +(\gamma\cdot \bar{\bf K})^i \wedge{\bf K}_j 
+{\bf K}^i \wedge (\bar{\bf K}\cdot\gamma)_j +\bar{\bf K}^i\wedge ({\bf K}\cdot\gamma)_j 
\right]\,. \nonumber \\
\end{eqnarray}

\subsubsection{\fixform{${\bf \Gamma}{\fR}$}}
The wedge product $\fGamma\fR$ can be computed as follows: 
\begin{eqnarray}
&&{}^{(d^2)}(\fGamma \fR)^{z}{}_z 
=  \fGN \wedge (-2\partial\bar{\partial}Adz\wedge d\bar{z})\, , \qquad 
{}^{(d^2)}(\fGamma \fR)^{\bar{z}}{}_{\bar{z}} 
=  \fGN \wedge (-2\partial\bar{\partial}Adz\wedge d\bar{z})\, , \nonumber\\
&&{}^{(d^2)}(\fGamma \fR)^{z}{}_{\bar{z}} =0 \, , \qquad 
 {}^{(d^2)}(\fGamma \fR)^{\bar{z}}{}_z =0\, , \qquad 
{}^{(d^2)}(\fGamma \fR)^{z}{}_{j} =0 \, , \qquad 
 {}^{(d^2)}(\fGamma \fR)^{\bar{z}}{}_j =0\, , \nonumber \\ 
&& {}^{(d^2)}(\fGamma \fR)^{i}{}_z 
=  {\bf K}^i \wedge (-2\partial\bar{\partial}Adz\wedge d\bar{z})\, , \qquad 
{}^{(d^2)}(\fGamma \fR)^{i}{}_{\bar{z}} 
=  (-1)\bar{\bf K}^i \wedge (-2\partial\bar{\partial}Adz\wedge d\bar{z})\, , \nonumber\\
&&
{}^{(d^2)}(\fGamma \fR)^{i}{}_{j} 
= 0\, , \nonumber \\
&&{}^{(dd)}(\fGamma \fR)^{z}{}_{z}=0\, ,   \qquad {}^{(dd)}(\fGamma \fR)^{\bar{z}}{}_{\bar{z}}=0\, , 
\qquad {}^{(dd)}(\fGamma \fR)^{z}{}_{\bar{z}}=0\, ,   \qquad {}^{(dd)}(\fGamma \fR)^{\bar{z}}{}_{{z}}=0\, , \nonumber \\
&&{}^{(dd)}(\fGamma \fR)^{z}{}_{j}=(-2e^{-2A})(-4\partial A\bar{\partial}A dz\wedge d\bar{z})\wedge \bar{\bf K}_j\, ,   
\qquad 
{}^{(dd)}(\fGamma \fR)^{\bar{z}}{}_{j}=(-2e^{-2A})(4\partial A\bar{\partial}A dz\wedge d\bar{z})\wedge {\bf K}_j\, ,   
\nonumber \\
&& {}^{(dd)}(\fGamma \fR)^{i}{}_{z}=0\, ,\qquad  {}^{(dd)}(\fGamma \fR)^{i}{}_{\bar{z}}=0\, , \nonumber \\
&&{}^{(dd)}(\fGamma \fR)^{i}{}_{j} =0 \, ,
\end{eqnarray}
and
{
\begin{eqnarray}
&&{}^{(d)}({\fGamma \fR})^{z}{}_z = (2\partial A dz) \wedge\left[
\fRN+2(-2e^{-2A})(\bar{\bf K}\cdot {\bf K})
\right]\, , \nonumber \\
&&{}^{(d)}({\fGamma \fR})^{\bar{z}}{}_{\bar{z}} = -(2\bar{\partial} A d \bar{z}) \wedge\left[
\fRN+2(-2e^{-2A})(\bar{\bf K}\cdot {\bf K})
\right]\, , \nonumber \\
&&{}^{(d)}({\fGamma \fR})^{z}{}_{\bar{z}} =0\, , \qquad {}^{(d)}({\fGamma \fR})^{\bar{z}}{}_z =0\, , \nonumber \\
&&{}^{(d)}({\fGamma \fR})^{z}{}_{j} 
=(2\partial A dz) \wedge (-2e^{-2A})
\Dhat \bar{\bf K}_j
+(-2e^{-2A})(2\bar{\partial} A d{\bar z}) \wedge\fGN\wedge \bar{\bf K}_j \,, \nonumber \\
&&{}^{(d)}({\fGamma \fR})^{\bar{z}}{}_{j} 
=(2 \bar{\partial} A d\bar{z}) \wedge (-2e^{-2A})
\Dhat{\bf K}_j
 -(-2e^{-2A})(2{\partial} A d{z}) \wedge\fGN\wedge {\bf K}_j \,, \nonumber \\
&& {
{}^{(d)}(\fGamma \fR)^i{}_z  = (2\partial A dz)\wedge(\gamma\cdot {\bf K})^i\, , \qquad 
{}^{(d)}(\fGamma \fR)^i{}_{\bar{z}}  = (2\bar{\partial} A d\bar{z})\wedge(\gamma\cdot \bar{\bf K})^i\, , 
}
\nonumber \\ 
&&{}^{(d)}({\fGamma \fR})^{i}{}_{j} 
= (2\partial A dz)\wedge (-2e^{-2A})(\bar{\bf K}^i\wedge {\bf K}_j) +  (2\bar{\partial} A d\bar{z}) \wedge
(-2e^{-2A})({\bf K}^i \wedge\bar{\bf K}_j)\, ,  
\end{eqnarray}
\begin{eqnarray}
&&{}^{(0)}({\fGamma \fR})^{z}{}_z  =\fGN\wedge \fRN+ (-2e^{-2A}) 
\left[\fGN\wedge(\bar{\bf K}\cdot {\bf K}) + \bar{\bf K}\cdot \Dhat{\bf K}\right]\, , \nonumber \\
&&{}^{(0)}({\fGamma \fR})^{\bar{z}}{}_{\bar{z}}  =\fGN\wedge \fRN+ (-2e^{-2A}) 
\left[\fGN\wedge(\bar{\bf K}\cdot {\bf K}) + {\bf K}\cdot \Dhat\bar{\bf K}\right]\, , \nonumber \\
&&{}^{(0)}({\fGamma \fR})^{z}{}_{\bar{z}} = (-2e^{-2A})(\bar{\bf K}\cdot \Dhat \bar{\bf K})\, , \nonumber \\
&&
{}^{(0)}({\fGamma \fR})^{\bar{z}}{}_{{z}} = (-2e^{-2A})({\bf K}\cdot \Dhat{\bf K})\, , \nonumber \\
&&{}^{(0)}({\fGamma \fR})^{z}{}_{j} = (-2e^{-2A})\left[
\fGN\wedge \Dhat \bar{\bf K}_j +(\bar{\bf K}\cdot{\fr})_j
\right] +(-2e^{-2A})^2 (\bar{\bf K}\cdot {\bf K})\wedge\bar{\bf K}_j
\, ,\\
&&{}^{(0)}({\fGamma \fR})^{\bar{z}}{}_{j} = (-2e^{-2A})\left[
-\fGN\wedge \Dhat{\bf K}_j +({\bf K}\cdot{\fr})_j
\right] +(-1)(-2e^{-2A})^2 (\bar{\bf K}\cdot {\bf K})\wedge{\bf K}_j
\, , \nonumber \\
&&{}^{(0)}({\fGamma \fR})^{i}{}_{z} = 
\left[
\fRN\wedge{\bf K}^i + (\gamma\cdot \Dhat{\bf K})^i \right] +(-2e^{-2A}){\bf K}^i \wedge
(\bar{\bf K}\cdot {\bf K})
\, , \nonumber \\
&&{}^{(0)}({\fGamma \fR})^{i}{}_{\bar{z}} = 
\left[
-\fRN\wedge\bar{\bf K}^i + (\gamma\cdot \Dhat\bar{\bf K})^i \right] +(-1)(-2e^{-2A})\bar{\bf K}^i \wedge
(\bar{\bf K}\cdot {\bf K})
\, , \nonumber \\
&&{}^{(0)}({\fGamma \fR})^{i}{}_{j} 
= (\gamma {\fr})^i{}_j \nonumber \\
&&\qquad \qquad\qquad +
(-2e^{-2A})\left[
{\bf K}^i \wedge \Dhat\bar{\bf K}_j +\bar{\bf K}^i \wedge \Dhat {\bf K}_j  
+(\gamma\cdot {\bf K})^i \wedge\bar{\bf K}_j + (\gamma\cdot\bar{\bf K})^i\wedge {\bf K}_j
\right] .\nonumber
\end{eqnarray}
}

{
 \subsubsection{\fixform{$\fR\fGamma$}}
 After some straightforward computation, the wedge product $\fR\fGamma$ is obtained as 
\begin{eqnarray}
&&{}^{(d^2)}(\fR\fGamma)^{z}{}_{z} = (-2\partial\bar{\partial} A dz\wedge d\bar{z})\wedge \fGamma_N\, , 
\qquad {}^{(d^2)}(\fR\fGamma)^{\bar{z}}{}_{\bar{z}} = (-2\partial\bar{\partial} A dz\wedge d\bar{z})\wedge \fGamma_N\, ,  
\nonumber \\
&& {}^{(d^2)}(\fR\fGamma)^{z}{}_{\bar{z}} =0\, , \qquad {}^{(d^2)}(\fR\fGamma)^{\bar{z}}{}_{z}=0\, , \nonumber \\
&& {}^{(d^2)}(\fR\fGamma)^{z}{}_{j}  = (-2\partial\bar{\partial}A dz\wedge d \bar{z})\wedge(-2e^{-2A})\bar{\bf K}_j\, , 
\qquad  {}^{(d^2)}(\fR\fGamma)^{\bar{z}}{}_{j}  = (-1)(-2\partial\bar{\partial}A dz\wedge d \bar{z})\wedge(-2e^{-2A}){\bf K}_j\, ,
\nonumber \\
&& {}^{(d^2)}(\fR\fGamma)^{i}{}_{z}=0\, , \qquad {}^{(d^2)}(\fR\fGamma)^{i}{}_{\bar{z}} =0\, , \qquad 
{}^{(d^2)}(\fR\fGamma)^{i}{}_{j} =0\, , \nonumber  \\
&&
{}^{(dd)}(\fR\fGamma)^{\mu}{}_{\nu} =0\, , 
\end{eqnarray}
and 
\begin{eqnarray}
&& 
{}^{(d)}(\fR\fGamma)^{z}{}_{{z}} = (2\partial A dz)\wedge[\fR_N+(-2e^{-2A})(\bar{\bf K}\cdot{\bf K})] 
+  (-2\bar{\partial}A d\bar{z}) \wedge (-2e^{-2A})(\bar{\bf K}\cdot {\bf K})\, , \nonumber \\
&& 
{}^{(d)}(\fR\fGamma)^{\bar{z}}{}_{{\bar{z}}} = (-2\bar{\partial} A d\bar{z})\wedge[\fR_N+(-2e^{-2A})(\bar{\bf K}\cdot{\bf K})] 
+  (2 {\partial}A d{z}) \wedge (-2e^{-2A})(\bar{\bf K}\cdot {\bf K})\, , \nonumber \\
&& 
{}^{(d)}(\fR\fGamma)^{z}{}_{\bar{z}} =0 \, , \qquad  {}^{(d)}(\fR\fGamma)^{\bar{z}}{}_{{z}} =0 \, ,\nonumber \\
&&
{}^{(d)}(\fR\fGamma)^{{z}}{}_{j} =(-2\bar{\partial}A d\bar{z})\wedge(-2e^{-2A}) (\bar{\bf K}\cdot\gamma)_j \, , 
\qquad 
{}^{(d)}(\fR\fGamma)^{\bar{z}}{}_{j} =(-2 {\partial}A d{z})\wedge(-2e^{-2A}) ({\bf K}\cdot\gamma)_j \, , \nonumber \\
&& 
{}^{(d)}(\fR\fGamma)^{{i}}{}_{z} =(2\partial A dz)\wedge [\hat{D} {\bf K}^i-{\bf K}^i\wedge \fGamma_N] \, , 
\qquad 
{}^{(d)}(\fR\fGamma)^{{i}}{}_{\bar{z}} =(2\bar{\partial} A d\bar{z})\wedge [\hat{D} \bar{\bf K}^i+\bar{\bf K}^i\wedge \fGamma_N] \, ,  \nonumber \\
&& 
{}^{(d)}(\fR\fGamma)^{{i}}{}_{j} = (-2\partial A dz)\wedge(-2e^{-2A})({\bf K}^i\wedge\bar{\bf K}_j) 
+ (-2\bar{\partial} A d\bar{z})\wedge(-2e^{-2A})(\bar{\bf K}^i\wedge{\bf K}_j) \, , 
\end{eqnarray}
\begin{eqnarray}
&& 
{}^{(0)}(\fR\fGamma)^{z}{}_{{z}}  = \fGamma_N\wedge \fR_N +(-2e^{-2A})[\fGamma_N\wedge(\bar{\bf K}\cdot {\bf K})
+(\hat{D}\bar{\bf K}\cdot {\bf K})]\, , \nonumber \\
&&
{}^{(0)}(\fR\fGamma)^{\bar{z}}{}_{\bar{z}}  = \fGamma_N\wedge \fR_N +(-2e^{-2A})[\fGamma_N\wedge(\bar{\bf K}\cdot {\bf K})
+(\hat{D}{\bf K}\cdot \bar{\bf K})]\, , \nonumber \\
&&
{}^{(0)}(\fR\fGamma)^{z}{}_{\bar{z}} = (-2e^{-2A})(\hat{D}\bar{\bf K}\cdot \bar{\bf K})  \, , \qquad 
{}^{(0)}(\fR\fGamma)^{\bar{z}}{}_{{z}} = (-2e^{-2A})(\hat{D}{\bf K}\cdot {\bf K})  \, , \nonumber \\
&&
{}^{(0)}(\fR\fGamma)^{z}{}_{j} 
=(-2e^{-2A})\left[\fR_N \wedge \bar{\bf K}_j + (\hat{D}\bar{\bf K}\cdot \gamma)_j \right]
+(-2e^{-2A})^2 (\bar{\bf K}\cdot {\bf K})\wedge \bar{\bf K}_j \, , \nonumber \\
&&
{}^{(0)}(\fR\fGamma)^{\bar{z}}{}_{j} 
=(-2e^{-2A})\left[-\fR_N \wedge {\bf K}_j + (\hat{D}{\bf K}\cdot \gamma)_j \right]
-(-2e^{-2A})^2 (\bar{\bf K}\cdot {\bf K})\wedge {\bf K}_j \, , \nonumber \\
&&
{}^{(0)}(\fR\fGamma)^{i}{}_{z}
=[\hat{D}{\bf K}^i\wedge \fGamma_N +(\fr \cdot {\bf K})^i] +(-2e^{-2A}){\bf K}^i \wedge(\bar{\bf K}\cdot {\bf K})\, , \nonumber\\
&&
{}^{(0)}(\fR\fGamma)^{i}{}_{\bar{z}}
=[-\hat{D}\bar{\bf K}^i\wedge \fGamma_N +(\fr \cdot \bar{\bf K})^i] -(-2e^{-2A})\bar{\bf K}^i \wedge(\bar{\bf K}\cdot {\bf K})\, , 
\nonumber \\
&& 
{}^{(0)}(\fR\fGamma)^{i}{}_{j} 
= (\fr \cdot \gamma)^i{}_j +(-2e^{-2A})[
\Dhat{\bf K}^i \wedge \bar{\bf K}_j + \Dhat\bar{\bf K}^i\wedge {\bf K}_j
 +{\bf K}^i\wedge(\bar{\bf K}\cdot \gamma)_j +\bar{\bf K}^i\wedge({\bf K}\cdot \gamma)_j]\,. \nonumber \\
 &&
\end{eqnarray}

 \subsubsection{\fixform{$\fGamma^2 \fR$}}
 \label{sec:gamma2R}
 We can also compute the wedge product $\fGamma^2\fR$ as 
 \begin{eqnarray}
 && {}^{(d^2)}(\fGamma^2 \fR)^{z}{}_z = (-2\partial \bar{\partial}A dz\wedge d\bar{z})\wedge 
 (-2e^{-2A})(\bar{\bf K}\cdot {\bf K})\, , \nonumber \\
 && {}^{(d^2)}(\fGamma^2 \fR)^{\bar{z}}{}_{\bar{z}} = (-2\partial \bar{\partial}A dz\wedge d\bar{z})\wedge 
 (-2e^{-2A})(\bar{\bf K}\cdot {\bf K})\, , \nonumber \\
 &&  {}^{(d^2)}(\fGamma^2 \fR)^{z}{}_{\bar{z}} =0\, , \qquad  {}^{(d^2)}(\fGamma^2 \fR)^{\bar{z}}{}_z=0\, ,  
 \qquad {}^{(d^2)}(\fGamma^2 \fR)^{z}{}_{j} =0\, , \qquad  {}^{(d^2)}(\fGamma^2 \fR)^{\bar{z}}{}_j=0\, ,  
 \nonumber \\
 && {}^{(d^2)}(\fGamma^2 \fR)^{i}{}_z = 
(-2\partial\bar{\partial}Adz\wedge d\bar{z})\wedge \left[{\bf K}^i\wedge \fGamma_N +(\gamma \cdot {\bf K})^i\right]\, , 
\nonumber \\
 && {}^{(d^2)}(\fGamma^2 \fR)^{i}{}_{\bar{z}} = 
(-2\partial\bar{\partial}Adz\wedge d\bar{z})\wedge \left[\bar{\bf K}^i\wedge \fGamma_N -(\gamma \cdot \bar{\bf K})^i\right]\, , 
\nonumber \\
 && {}^{(d^2)}(\fGamma^2 \fR)^{i}{}_{j} = 0\, , \nonumber \\
 &&  {}^{(dd)}(\fGamma^2 \fR)^{z}{}_z=0\, , \qquad {}^{(dd)}(\fGamma^2 \fR)^{\bar{z}}{}_{\bar{z}} = 0\, , 
 \qquad   {}^{(dd)}(\fGamma^2 \fR)^{z}{}_{\bar{z}} =0\, , \qquad  {}^{(dd)}(\fGamma^2 \fR)^{\bar{z}}{}_z=0\, , 
 \nonumber \\
 &&   {}^{(dd)}(\fGamma^2 \fR)^{z}{}_{j} =0\, , \qquad  {}^{(dd)}(\fGamma^2 \fR)^{\bar{z}}{}_j=0\, ,  
 \qquad {}^{(dd)}(\fGamma^2 \fR)^{i}{}_z =0\, , \qquad {}^{(dd)}(\fGamma^2 \fR)^{i}{}_{\bar{z}}=0\, , \nonumber \\
 && 
 {}^{(dd)}(\fGamma^2 \fR)^{i}{}_{j}  = (-4\partial A \bar{\partial}A dz\wedge  d\bar{z})\wedge
 (-2e^{-2A})({\bf K}^i\wedge\bar{\bf K}_j- \bar{\bf K}^i\wedge{\bf K}_j)\, , 
 \end{eqnarray}
 and 
 \begin{eqnarray}
&& {}^{(d)}(\fGamma^2 \fR)^{z}{}_z = (2\partial A dz)\wedge (-2e^{-2A})
\left[
(\bar{\bf K}\cdot \Dhat {\bf K}) -\fGamma_N \wedge(\bar{\bf K}\cdot {\bf K})-(\bar{\bf K}\cdot \gamma \cdot {\bf K})
\right]\, , \nonumber \\
&& {}^{(d)}(\fGamma^2 \fR)^{\bar{z}}{}_{\bar{z}} = (2\bar{\partial} A d\bar{z})\wedge (-2e^{-2A})
\left[
({\bf K}\cdot \Dhat \bar{\bf K}) -\fGamma_N \wedge(\bar{\bf K}\cdot {\bf K})-({\bf K}\cdot \gamma \cdot \bar{\bf K})
\right]\, , \nonumber \\
&& 
{}^{(d)}(\fGamma^2 \fR)^{z}{}_{\bar{z}} =(2\partial A dz)\wedge(-2e^{-2A})(\bar{\bf K}\cdot \Dhat\bar{\bf K})
+(-2\bar{\partial}A d\bar{z})\wedge(-2e^{-2A})(\bar{\bf K}\cdot \gamma \cdot \bar{\bf K})\, , \nonumber \\ 
&& 
{}^{(d)}(\fGamma^2 \fR)^{\bar{z}}{}_{{z}} =(2\bar{\partial} A d\bar{z})\wedge(-2e^{-2A})({\bf K}\cdot \Dhat{\bf K})
+(-2{\partial}A d{z})\wedge(-2e^{-2A})({\bf K}\cdot \gamma \cdot {\bf K})\, , \nonumber \\ 
&& 
{}^{(d)}(\fGamma^2 \fR)^{z}{}_{j}
= (2\partial A dz)\wedge[(-2e^{-2A})(\bar{\bf K}\cdot \fr)_j + (-2e^{-2A})^2(\bar{\bf K}\cdot {\bf K})\wedge\bar{\bf K}_j ] 
\nonumber \\
&&\qquad\qquad \qquad \qquad 
+ (-2\bar{\partial}Ad\bar{z})\wedge(-2e^{-2A})^2(\bar{\bf K}\cdot {\bf K})\wedge \bar{\bf K}_j \, , \nonumber  \\
&& 
{}^{(d)}(\fGamma^2 \fR)^{\bar{z}}{}_{j}
= (2\bar{\partial} A d\bar{z})\wedge[(-2e^{-2A})({\bf K}\cdot \fr)_j - (-2e^{-2A})^2(\bar{\bf K}\cdot {\bf K})\wedge{\bf K}_j ] 
\nonumber \\
&&\qquad\qquad \qquad \qquad 
+ (2{\partial}Ad{z})\wedge(-2e^{-2A})^2(\bar{\bf K}\cdot {\bf K})\wedge {\bf K}_j \, , \nonumber  \\
&& 
{}^{(d)}(\fGamma^2 \fR)^{i}{}_{z} 
=(-2\partial A dz)\wedge[\fR_N\wedge {\bf K}^i+(\gamma\cdot \gamma\cdot {\bf K})^i +2(-2e^{-2A})(\bar{\bf K}\cdot {\bf K})\wedge {\bf K}^i]\, , \nonumber \\
&& 
{}^{(d)}(\fGamma^2 \fR)^{i}{}_{\bar{z}} 
=(-2\bar{\partial} A d\bar{z})\wedge[-\fR_N\wedge \bar{\bf K}^i+(\gamma\cdot \gamma\cdot \bar{\bf K})^i 
-2(-2e^{-2A})(\bar{\bf K}\cdot {\bf K})\wedge \bar{\bf K}^i]\, , \nonumber \\
&& 
{}^{(d)}(\fGamma^2 \fR)^{i}{}_{j} 
= (-2\partial A dz)\wedge (-2e^{-2A})\left[
({\bf K}^i\wedge \Dhat\bar{\bf K}_j) +[-\bar{\bf K}^i\wedge \fGamma_N+(\gamma\cdot \bar{\bf K})^i]\wedge {\bf K}_j
\right]  \nonumber \\
&&\qquad\qquad \qquad \qquad 
+(-2\bar{\partial} A d\bar{z})\wedge (-2e^{-2A})\left[
(\bar{\bf K}^i\wedge \Dhat{\bf K}_j) +[{\bf K}^i\wedge \fGamma_N+(\gamma\cdot {\bf K})^i]\wedge \bar{\bf K}_j
\right]\, , \nonumber \\
&&
\end{eqnarray}
 \begin{eqnarray}
 && {}^{(0)}(\fGamma^2 \fR)^{z}{}_z 
 =(-2e^{-2A})[\fR_N\wedge(\bar{\bf K}\cdot{\bf K})+\fGamma_N\wedge(\bar{\bf K}\cdot \Dhat{\bf K})+(\bar{\bf K}\cdot 
 \gamma \cdot \Dhat{\bf K})] +(-2e^{-2A})^2(\bar{\bf K}\cdot {\bf K})^2\, , \nonumber \\
 && {}^{(0)}(\fGamma^2 \fR)^{\bar{z}}{}_{\bar{z}} 
 =(-2e^{-2A})[\fR_N\wedge(\bar{\bf K}\cdot{\bf K})-\fGamma_N\wedge({\bf K}\cdot \Dhat\bar{\bf K})+({\bf K}\cdot 
 \gamma \cdot \Dhat\bar{\bf K})] +(-2e^{-2A})^2(\bar{\bf K}\cdot {\bf K})^2\, , \nonumber \\
 && {}^{(0)}(\fGamma^2 \fR)^{z}{}_{\bar{z}} 
=(-2e^{-2A})\left[\fGamma_N\wedge(\bar{\bf K}\cdot \Dhat \bar{\bf K})+(\bar{\bf K}\cdot\gamma\cdot
\Dhat\bar{\bf K})\right]\, , \nonumber \\
 && {}^{(0)}(\fGamma^2 \fR)^{\bar{z}}{}_{{z}} 
=(-2e^{-2A})\left[-\fGamma_N\wedge({\bf K}\cdot \Dhat {\bf K})+({\bf K}\cdot\gamma\cdot
\Dhat{\bf K})\right]\, , \nonumber \\
&&
{}^{(0)}(\fGamma^2 \fR)^{z}{}_j 
= (-2e^{-2A})[\fGamma_N\wedge(\bar{\bf K}\cdot \fr)_j+(\bar{\bf K}\cdot \gamma\cdot \fr)_j] \nonumber \\
&&\qquad\qquad\qquad\qquad
+ (-2e^{-2A})^2 
\left[
(\bar{\bf K}\cdot {\bf K})\wedge[\Dhat \bar{\bf K}_j+\fGamma_N\wedge\bar{\bf K}_j] 
+(\bar{\bf K}\cdot \gamma\cdot {\bf K})\wedge \bar{\bf K}_j
+(\bar{\bf K}\cdot \gamma\cdot \bar{\bf K})\wedge{\bf K}_j
\right]\, , \nonumber \\
&&
{}^{(0)}(\fGamma^2 \fR)^{\bar{z}}{}_j 
= (-2e^{-2A})[-\fGamma_N\wedge({\bf K}\cdot \fr)_j+({\bf K}\cdot \gamma\cdot \fr)_j] \nonumber \\
&&\qquad\qquad\qquad\qquad
+ (-2e^{-2A})^2 
\left[-
(\bar{\bf K}\cdot {\bf K})\wedge[\Dhat {\bf K}_j-\fGamma_N\wedge{\bf K}_j] 
+({\bf K}\cdot \gamma\cdot \bar{\bf K})\wedge {\bf K}_j
+({\bf K}\cdot \gamma\cdot {\bf K})\wedge\bar{\bf K}_j
\right]\, , \nonumber \\
&& {}^{(0)}(\fGamma^2 \fR)^{i}{}_z
= [{\bf K}^i\wedge \fGamma_N +(\gamma\cdot {\bf K})^i]\wedge \fR_N + (\gamma\cdot \gamma\cdot \Dhat{\bf K})^i 
\nonumber \\
&&\qquad\qquad\qquad\qquad
+(-2e^{-2A})
\left[
[{\bf K}^i\wedge \fGamma_N+(\gamma \cdot {\bf K})^i]\wedge(\bar{\bf K}\cdot {\bf K})
+{\bf K}^i\wedge (\bar{\bf K}\cdot \Dhat{\bf K}) +\bar{\bf K}^i \wedge({\bf K}\cdot \Dhat{\bf K})
\right]\, , \nonumber \\
&& {}^{(0)}(\fGamma^2 \fR)^{i}{}_{\bar{z}}
= [\bar{\bf K}^i\wedge \fGamma_N -(\gamma\cdot \bar{\bf K})^i]\wedge \fR_N + (\gamma\cdot \gamma\cdot \Dhat\bar{\bf K})^i 
\nonumber \\
&&\qquad\qquad\qquad\qquad
+(-2e^{-2A})
\left[
[\bar{\bf K}^i\wedge \fGamma_N-(\gamma \cdot \bar{\bf K})^i]\wedge(\bar{\bf K}\cdot {\bf K})
+\bar{\bf K}^i\wedge ({\bf K}\cdot \Dhat\bar{\bf K}) +{\bf K}^i \wedge(\bar{\bf K}\cdot \Dhat\bar{\bf K})
\right]\, , \nonumber \\
&& {}^{(0)}(\fGamma^2 \fR)^{i}{}_{j}
=(\gamma\cdot\gamma\cdot \fr)^i{}_j \nonumber \\
&&\qquad\qquad\qquad\qquad 
+(-2e^{-2A})\Bigl[
[{\bf K}^i\wedge \fGamma_N+(\gamma\cdot {\bf K})^i]\wedge \Dhat\bar{\bf K}_j
+[-\bar{\bf K}^i\wedge \fGamma_N +(\gamma\cdot \bar{\bf K})^i]\wedge \Dhat{\bf K}_j
\nonumber \\
&&\qquad\qquad\qquad\qquad\qquad\qquad\qquad
+{\bf K}^i\wedge (\bar{\bf K}\cdot \fr)_j + \bar{\bf K}^i\wedge({\bf K}\cdot \fr)_j
+(\gamma\cdot\gamma\cdot {\bf K})^i\wedge \bar{\bf K}_j 
+(\gamma\cdot\gamma\cdot \bar{\bf K})^i\wedge {\bf K}_j 
\Bigr]\nonumber \\
&& 
\qquad\qquad\qquad\qquad
+(-2e^{-2A})^2 (\bar{\bf K}\cdot {\bf K})\wedge ({\bf K}^i\wedge\bar{\bf K}_j-\bar{\bf K}^i\wedge{\bf K}_j)\, . \nonumber \\
&&
\end{eqnarray}

}

\subsubsection{\fixform{${\fR}^2$}}
For $\fR^2$, after some computation, we obtain the following result: 
{
\begin{eqnarray}
&&{}^{(d^2)}({\fR}^2)^z{}_z = 2(-2\partial\bar{\partial}A dz\wedge d\bar{z}) \wedge\left[
\fRN+ (-2e^{-2A})(\bar{\bf K}\cdot {\bf K})
\right]\, , \nonumber \\
&&{}^{(d^2)}({\fR}^2)^{\bar{z}}{}_{\bar{z}} = 2(-2\partial\bar{\partial}A dz\wedge d\bar{z}) \wedge\left[
\fRN+ (-2e^{-2A})(\bar{\bf K}\cdot {\bf K})
\right]\, , \nonumber \\
&&{}^{(d^2)}({\fR}^2)^z{}_{\bar{z}}=0\, , \qquad  {}^{(d^2)}({\fR}^2)^{\bar{z}}{}_{{z}} =0\, , \nonumber \\
&&{}^{(d^2)}({\fR}^2)^z{}_j =(-2\partial\bar{\partial}A dz\wedge d\bar{z})\wedge (-2e^{-2A})
\Dhat\bar{\bf K}_j
\nonumber \\
&&{}^{(d^2)}({\fR}^2)^{\bar{z}}{}_j =(-1)(-2\partial\bar{\partial}A dz\wedge d\bar{z})\wedge(-2e^{-2A})
\Dhat{\bf K}_j \, , 
\nonumber \\
&&{}^{(d^2)}({\fR}^2)^i{}_z =(-2\partial\bar{\partial}A dz\wedge d\bar{z})\wedge
\Dhat{\bf K}^i \, , \nonumber \\
&&{}^{(d^2)}({\fR}^2)^i{}_{\bar{z}} =(-1)(-2\partial\bar{\partial}A dz\wedge d\bar{z})\wedge
\Dhat\bar{\bf K}^i \, , \nonumber \\
&&{}^{(d^2)}({\fR}^2)^i{}_{j} = 0\, ,  
\end{eqnarray} 
}
and
\begin{eqnarray}
&&{}^{(dd)}({\fR}^2)^z{}_z = (4\partial A \bar{\partial}A dz\wedge d\bar{z})\wedge(-2e^{-2A})(\bar{\bf K}\cdot {\bf K})\, , \nonumber\\
&&{}^{(dd)}({\fR}^2)^{\bar{z}}{}_{\bar{z}} = (4\partial A \bar{\partial}A dz\wedge d\bar{z})\wedge
(-2e^{-2A})(\bar{\bf K}\cdot {\bf K})\, , \nonumber\\
&&{}^{(dd)}({\fR}^2)^{z}{}_{\bar{z}}=0\, , \qquad {}^{(dd)}({\fR}^2)^{\bar{z}}{}_{{z}}=0\, , \qquad 
{}^{(dd)}({\fR}^2)^{z}{}_{j}=0\, , \qquad {}^{(dd)}({\fR}^2)^{\bar{z}}{}_j=0\, , \nonumber \\
&&{}^{(dd)}({\fR}^2)^{i}{}_{z}=0\, , \qquad {}^{(dd)}({\fR}^2)^{i}{}_{\bar{z}}=0\, , \nonumber \\ 
&&{}^{(dd)}({\fR}^2)^{i}{}_{j}=
(-1)(-2e^{-2A})(4\partial A \bar{\partial}A dz\wedge d\bar{z})\wedge({\bf K}^i\wedge \bar{\bf K}_j- \bar{\bf K}^i\wedge{\bf K}_j)
\, , 
\end{eqnarray}
as well as
{
\begin{eqnarray}
&&{}^{(d)}(\fR^2)^{z}{}_z = 
(-2e^{-2A})(-2{\partial A}d{z})\wedge({\bf K}\cdot \Dhat\bar{\bf K}) \nonumber \\
&&\qquad\qquad\qquad
+(-2e^{-2A})(-2\bar{\partial} Ad\bar{z})\wedge(\bar{\bf K}\cdot \Dhat{\bf K})\, , \nonumber \\
&&{}^{(d)}(\fR^2)^{\bar{z}}{}_{\bar{z}} = 
(-2e^{-2A})(-2{\partial A}d{z})\wedge({\bf K}\cdot \Dhat\bar{\bf K}) \nonumber \\
&&\qquad\qquad\qquad
+(-2e^{-2A})(-2\bar{\partial} Ad\bar{z})\wedge(\bar{\bf K}\cdot \Dhat{\bf K})\, , \nonumber \\
&&{}^{(d)}(\fR^2)^{z}{}_{\bar{z}} = 2(-2e^{-2A})(-2\bar{\partial} Ad\bar{z})(\bar{\bf K}\cdot \Dhat\bar{\bf K})\, , \nonumber \\
&&{}^{(d)}(\fR^2)^{\bar{z}}{}_{{z}} = 2(-2e^{-2A})(-2{\partial} Ad{z})({\bf K}\cdot \Dhat{\bf K})\, , \nonumber \\
&&{}^{(d)}(\fR^2)^{z}{}_j = 
(-2\bar{\partial}A d\bar{z})\wedge
\left[
(-2e^{-2A})[(\bar{\bf K}\cdot {\fr})_j +\fRN\wedge\bar{\bf K}_j]
+2(-2e^{-2A})^2(\bar{\bf K}\cdot {\bf K})\wedge\bar{\bf K}_j
\right] \, ,  \nonumber \\
&&{}^{(d)}(\fR^2)^{\bar{z}}{}_j = 
(-2{\partial}A d{z})\wedge
\left[
(-2e^{-2A})[({\bf K}\cdot{\fr})_j -\fRN\wedge{\bf K}_j]
-2(-2e^{-2A})^2(\bar{\bf K}\cdot {\bf K})\wedge{\bf K}_j
\right] \, , \nonumber \\
&&{}^{(d)}(\fR^2)^{i}{}_z = 
(-2\partial A dz) \wedge \left[
{\bf K}^i \wedge \fRN+ (\fr\cdot{\bf K})^i +2(-2e^{-2A}){\bf K}^i\wedge(\bar{\bf K}\cdot {\bf K})
\right] \, , \nonumber \\
&&{}^{(d)}(\fR^2)^{i}{}_{\bar{z}} = 
(-2\bar{\partial} A d\bar{z}) \wedge \left[
-\bar{\bf K}^i \wedge\fRN+ (\fr\cdot\bar{\bf K})^i -2(-2e^{-2A})\bar{\bf K}^i\wedge(\bar{\bf K}\cdot {\bf K})
\right] \, , \nonumber \\
&&{}^{(d)}(\fR^2)^{i}{}_{j} = 
(-2e^{-2A})(-2\partial A dz)\wedge\left[
{\bf K}^i \wedge \Dhat \bar{\bf K}_j
+  \Dhat\bar{\bf K}^i \wedge{\bf K}_j
\right]  \nonumber \\
&&\qquad\qquad \qquad
+ (-2e^{-2A})(-2\bar{\partial} A d\bar{z})\wedge \left[
\bar{\bf K}^i \wedge \Dhat{\bf K}_j 
+  \Dhat{\bf K}^i \wedge\bar{\bf K}_j
\right] \, , \nonumber\\ 
\end{eqnarray} 
}
and
{
\begin{eqnarray}
&&{}^{(0)}(\fR^2)^{z}{}_z = \fRN^2 
+ (-2e^{-2A})\left[2\fRN\wedge(\bar{\bf K}\cdot {\bf K}) 
+(\Dhat\bar{\bf K}\cdot \Dhat{\bf K})
\right] + (-2e^{-2A})^2 (\bar{\bf K}\cdot {\bf K})^2\, , \nonumber \\
&&{}^{(0)}(\fR^2)^{\bar{z}}{}_{\bar{z}} = \fRN^2 
+ (-2e^{-2A})\left[2\fRN\wedge(\bar{\bf K}\cdot {\bf K}) 
+(\Dhat\bar{\bf K}\cdot \Dhat{\bf K})
\right]+ (-2e^{-2A})^2 (\bar{\bf K}\cdot {\bf K})^2\, , \nonumber \\
&&{}^{(0)}(\fR^2)^{z}{}_{\bar{z}} = (-2e^{-2A})(\Dhat\bar{\bf K}\cdot \Dhat\bar{\bf K})\, ,\qquad 
{}^{(0)}(\fR^2)^{\bar{z}}{}_{{z}} = (-2e^{-2A})(\Dhat{\bf K}\cdot \Dhat{\bf K})\, ,\nonumber \\
&&
{}^{(0)}(\fR^2)^{z}{}_{j} = (-2e^{-2A})\left[
\fRN\wedge \Dhat\bar{\bf K}_j
 +(\Dhat\bar{\bf K}\cdot{\fr})_j 
 \right] \nonumber \\
 &&\qquad\qquad \quad\,\,
 +(-2e^{-2A})^2 \left[
 (\bar{\bf K}\cdot {\bf K})\wedge \Dhat\bar{\bf K}_j +(\Dhat\bar{\bf K}\cdot {\bf K})\wedge\bar{\bf K}_j
 +(\Dhat\bar{\bf K}\cdot \bar{\bf K})\wedge{\bf K}_j
 \right]\, , \nonumber \\
&&
{}^{(0)}(\fR^2)^{\bar{z}}{}_{j} = (-2e^{-2A})\left[
-\fRN\wedge \Dhat{\bf K}_j
 +(\Dhat{\bf K}\cdot{\fr})_j 
 \right] \nonumber \\
 &&\qquad\qquad\quad\,\, 
 +(-2e^{-2A})^2 \left[
 -(\bar{\bf K}\cdot {\bf K})\wedge \Dhat{\bf K}_j +(\Dhat{\bf K}\cdot \bar{\bf K})\wedge{\bf K}_j
 +(\Dhat{\bf K}\cdot {\bf K})\wedge\bar{\bf K}_j
 \right]\, , \nonumber \\ 
 &&
{}^{(0)}(\fR^2)^{i}{}_{z} = 
\left[
\fRN\wedge \Dhat{\bf K}^i+ ({\fr}\cdot \Dhat{\bf K})^i 
\right] \nonumber \\
&&\qquad\qquad\quad\,\, 
+ (-2e^{-2A}) \left[
(\bar{\bf K}\cdot {\bf K})\wedge \Dhat{\bf K}^i +{\bf K}^i \wedge(\Dhat{\bf K}\cdot \bar{\bf K}) + \bar{\bf K}^i \wedge(\Dhat{\bf K}\cdot {\bf K})
\right]\, , \nonumber \\
 &&
{}^{(0)}(\fR^2)^{i}{}_{\bar{z}} = 
\left[
-\fRN\wedge \Dhat\bar{\bf K}^i + ({\fr}\cdot \Dhat\bar{\bf K})^i 
\right] \nonumber \\
&&\qquad\qquad\quad \,\,
+ (-2e^{-2A}) \left[
-(\bar{\bf K}\cdot {\bf K})\wedge \Dhat\bar{\bf K}^i +\bar{\bf K}^i \wedge(\Dhat\bar{\bf K}\cdot {\bf K}) + {\bf K}^i \wedge(\Dhat\bar{\bf K}\cdot \bar{\bf K})
\right]\, , \nonumber \\
 \end{eqnarray}
 along with 
\begin{eqnarray}
{}^{(0)}(\fR^2)^{i}{}_{j} &&=
({\fr}^2)^{i}{}_j + (-2e^{-2A})\biggl[
\Dhat{\bf K}^i \wedge \Dhat\bar{\bf K}_j +\Dhat\bar{\bf K}^i \wedge \Dhat{\bf K}_j  \nonumber \\
&&\qquad\qquad
+{\bf K}^i\wedge (\bar{\bf K}\cdot{\fr})_j+\bar{\bf K}^i \wedge({\bf K}\cdot{\fr})_j +({\fr}\cdot{\bf K})^i \wedge\bar{\bf K}_j 
+({\fr}\cdot\bar{\bf K})^i\wedge {\bf K}_j 
\biggr] \nonumber \\
&&
+(-2e^{-2A})^2 (\bar{\bf K}\cdot {\bf K})\wedge({\bf K}^i\wedge\bar{\bf K}_j - \bar{\bf K}^i \wedge{\bf K}_j)\, . 
\end{eqnarray}
}

\subsubsection{\fixform{$\ICS^{3d}$}}
\label{sec:ics3dcomp}
By using the above results, we can 
compute 3d Chern-Simons terms $\ICS^{3d}$ on the regularized cone geometry. 
For the contribution of the form $\partial\bar{\partial}A dz\wedge d\bar{z}$ or 
$\partial A \bar{\partial}A dz\wedge d\bar{z}$
multiplied by zeroth order terms in $(z,\bar{z})$-expansion, we obtain 
\begin{eqnarray}
&&\ICS^{3d}|_{\partial\bar{\partial}A} =
(-4)(\partial\bar{\partial}Adz\wedge d\bar{z}) \wedge\fGN\, ,   \\
&&\ICS^{3d}|_{\partial A\bar{\partial}A} = 0 \, . 
\end{eqnarray}
On the other hand, the contribution of the form 
 $\partial A$ (or $\bar{\partial}A$) times zeroth order terms in $(z,\bar{z})$-expansion 
as well as that of order-$\epsilon^0$ and zeroth order in $(z,\bar{z})$-expansion are
{
\begin{eqnarray}
&&\ICS^{3d}|_{\partial A, \bar{\partial}A} = (2\partial A dz)\wedge\left[
\fRN + 2(-2e^{-2A})(\bar{\bf K}\cdot {\bf K}) \right] \nonumber \\
&&\qquad\qquad\qquad+ (-1) (2\bar{\partial} A d\bar{z}) \wedge \left[
\fRN+ 2(-2e^{-2A})(\bar{\bf K}\cdot {\bf K})
\right]\, ,  \\
&&\ICS^{3d}|_{0th}  = 2\fGN\wedge\fRN+\tr\left({\bf \gamma }\fr - \frac{1}{3}\gamma^3\right) 
 +2(-2e^{-2A})\left[
({\bf K}\cdot \Dhat\bar{\bf K}) +(\bar{\bf K}\cdot \Dhat{\bf K})
\right]\, . 
\end{eqnarray}
}

\subsubsection{\fixform{$\fPanom^{3d}$}}
From the above expression for ${\fR}^2$, we can also straightforwardly compute 
the anomaly polynomial $\fPanom^{3d}=\tr({\fR}^2)$ corresponding to the 3d Chern-Simons term. 
By using the same notation as the 3d Chern-Simons term case just above, we have
\begin{eqnarray}
&&\fPanom^{3d}|_{\partial\bar{\partial}A} =  (-8)(\partial\bar{\partial}A dz\wedge d\bar{z})
\wedge\left[\fRN+(-2e^{-2A})(\bar{\bf K}\cdot {\bf K}) \right]\, , \\
&&\fPanom^{3d}|_{\partial A \bar{\partial}A} =  16 (\partial A \bar{\partial}A dz\wedge d\bar{z})
\wedge (-2e^{-2A})(\bar{\bf K}\cdot {\bf K})\, , 
\end{eqnarray}
and 
{
\begin{eqnarray}
&&\fPanom^{3d}|_{\partial A,  \bar{\partial}A} 
= (-8)(\partial A dz)\wedge(-2e^{-2A})({\bf K}\cdot \Dhat\bar {\bf K})
\nonumber \\
&&\qquad\qquad \qquad\qquad  
+(-8)(\bar{\partial} A d\bar{z})\wedge(-2e^{-2A})(\bar{\bf K}\cdot \Dhat{\bf K})
\, ,
\end{eqnarray}
\begin{eqnarray} \label{eq:panom3d0th}
&&\fPanom^{3d}|_{0th} 
= 4(-2e^{-2A})\left[
\fRN\wedge(\bar{\bf K}\cdot{\bf K})+(\Dhat\bar{\bf K}\cdot \Dhat{\bf K})
- (\bar{\bf K}\cdot {\fr}\cdot {\bf K})
\right] 
+2\fRN^2 + \tr({\fr}^2)\, .  \nonumber \\
\end{eqnarray}
}

\section{Details of Computation in  Anomaly Polynomial Method}
\label{sec:anompolydetail}
In this Appendix, we will explain in detail the derivation of holographic entanglement entropy formulas 
for purely gravitational Chern-Simons terms from the anomaly polynomials we have 
briefly summarized the results in \S\ref{sec:directchernsimons}.

\subsection{\fixform{$\tr({\fR}^2)$}}
Let us first start with the anomaly polynomial $\fPanom^{3d}=\tr({\fR}^2)$ corresponding to 
3d gravitational Chern-Simons term. We will evaluate this polynomial on the regularized cone
geometry. As reviewed in \S\S\ref{sec:anompolystrategy}, there are two types of potential contributions 
to the holographic entanglement entropy: the contribution proportional to $\partial A \bar{\partial}A$ and 
that proportional to $\partial \bar{\partial}A $. We will first evaluate them separately and then
combine to get the holographic entanglement entropy formula from the anomaly polynomial. 
\subsubsection{$\partial \bar{\partial} A$ Term}
Among the anomaly polynomial $\tr(\fR^2)$, there are two possibilities to have 
the contribution proportional to $\partial\bar{\partial} A$ :
\begin{eqnarray}
{}^{(d^2)}\fR^{z}{}_{z}{}^{(0)}\fR^{z}{}_{z}\times 2\, , \qquad 
{}^{(d^2)}\fR^{\bar{z}}{}_{\bar{z}}{}^{(0)}\fR^{\bar{z}}{}_{\bar{z}}\times 2\, , 
\end{eqnarray} 
where the factor $2$ comes from the choice for the location of  ${}^{(d^2)}\fR\,$  in $\tr(\fR^2)$\,. 
By using the expression of $\fR$ summarized in Eq.~\eqref{eq:Rdecomposed} and 
summing up the contribution from the above two possibilities, we obtain after integration 
\begin{eqnarray}
\int_{\onCone} \fPanom^{3d}|_{\partial\bar{\partial}A} &=& \int_{\onCone} 
(-8\,\partial \bar{\partial}A dz\wedge d\bar{z})
\wedge \left[\fRN+(-2e^{-2A})(\bar{\bf K}\cdot {\bf K})\right]\,  \nonumber \\
&=& 8\pi\,\epsilon \int_{\Sigmatilde} \left[\fRN- 2 (\bar{\bf K}\cdot {\bf K})\right]\, . 
\label{eq:3danompolyppbarA}
\end{eqnarray}

\subsubsection{$\partial A \bar{\partial} A$ Term}
As for the terms proportional to $\partial A \bar{\partial} A$, 
there are two possibilities in $\tr(\fR^2)$ :
\begin{eqnarray}
{}^{(d)}\fR^{i}{}_{z}{}^{(d)}\fR^{z}{}_{i}\times 2\, , \qquad 
{}^{(d)}\fR^{i}{}_{\bar{z}}{}^{(d)}\fR^{\bar{z}}{}_{i} \times 2\, .
\end{eqnarray}
Here the factor $2$ comes from the choice for the location of  ${}^{(d)}\fR^i{}_z$ and ${}^{(d)}\fR^i{}_{\bar{z}}$\,. 
By summing up and integrating these two contributions, we obtain 
\begin{eqnarray}
\int_{\onCone} \fPanom^{3d}|_{\partial A\bar{\partial}A} &=& \int_{\onCone} 
16(-2\,e^{-2A})(\partial A \bar{\partial} A dz\wedge d\bar{z})\,\wedge (\bar{\bf K}\cdot {\bf K})\, \nonumber \\
&=& 16\pi\epsilon \int (\bar{\bf K}\cdot {\bf K})\, .\label{eq:3danompolypApbarA}
\end{eqnarray}

\subsubsection{Final Result}
By summing Eqs.~\eqref{eq:3danompolyppbarA} and \eqref{eq:3danompolypApbarA}, 
we obtain the total of the order $\epsilon$ contribution to the anomaly polynomial as follows:
\begin{eqnarray}\label{eq:finaltrR2}
8\pi\,\epsilon \int_{\Sigmatilde} \fRN = 8\pi\,\epsilon \int_{\Sigma} \fGN\, . 
\end{eqnarray}
Therefore, the holographic entanglement entropy formula for 3d gravitational Chern-Simons term 
is computed from the anomaly polynomial as 
\begin{eqnarray}
  \SeeCS^{3d}
 = 8\pi\, \int_{\Sigma} \fGN\, , 
 \end{eqnarray}
 which reproduces the result obtained in \cite{Castro:2014tta}.

\subsection{\fixform{$\tr(\fR^4)$}}
As a next example, we proceed to 7d and consider the single-trace type of 
gravitational Chern-Simons term $\ICS^{7d, \, single}$ and the corresponding 
anomaly polynomial $\fPanom^{7d,\, single}= \tr({\fR}^4)$. 
In the same ways as 3d case, we first evaluate the terms proportional to 
$\partial\bar{\partial} A$ and $\partial A\bar{\partial}A$ separately and then 
combine to get the holographic entanglement entropy formula.  
\subsubsection{$\partial \bar{\partial}A$ Term}
We first evaluate the term proportional to $\partial \bar{\partial} A$ in $\tr({\fR}^4)$. 
There are two possibilities in $\tr(\fR^4)$:
\begin{eqnarray}
{}^{(d^2)}\fR^{z}{}_{z}{}^{(0)}(\fR\fR\fR)^{z}{}_{z}\times 4\, , \qquad 
{}^{(d^2)}\fR^{\bar{z}}{}_{\bar{z}}{}^{(0)}(\fR\fR\fR)^{\bar{z}}{}_{\bar{z}} \times 4\, . 
\end{eqnarray}
Here ${}^{(0)}(\fR\fR\fR)^{z}{}_{z}={}^{(0)}\fR^z{}_\mu{}^{(0)}\fR^\mu{}_\nu{}^{(0)}\fR^\nu{}_z$\, 
and the factor 4 comes from the choice for the location of ${}^{(d^2)}\fR$ in $\tr(\fR^4)$.
By summing up these contributions, we obtain after integration
{
\begin{eqnarray}
&&\int_{\onCone} \fPanom^{7d, single}|_{\partial\bar{\partial}A} \nonumber \\
&&=\int_{\onCone} (-16\partial \bar{\partial} Adz\wedge d\bar{z}) \nonumber \\
&&\qquad \wedge \Biggl[ 
\left[\fRN +(-2e^{-2A})(\bar{\bf K}\cdot {\bf K})\right]^3 \nonumber \\
&&\qquad\qquad 
+2(-2\,e^{-2A})[\fRN+(-2e^{-2A})(\bar{\bf K}\cdot {\bf K})]\wedge
(\Dhat\bar{\bf K}\cdot \Dhat{\bf K} )
+(-2e^{-2A})(\Dhat \bar{\bf K}\cdot \fr \cdot \Dhat{\bf K})
\nonumber \\
&&\qquad\qquad 
+(-2e^{-2A})^2 \left[
(\Dhat\bar{\bf K}\cdot {\bf K})\wedge (\Dhat{\bf K}\cdot \bar{\bf K})
+(\Dhat\bar{\bf K}\cdot \bar{\bf K})\wedge (\Dhat{\bf K}\cdot {\bf K})
\right]
\Biggr] \nonumber \\
&&= 16\pi\epsilon\int_{\Sigmatilde} \Biggl[
\fRN^3 -6\fRN^2\wedge(\bar{\bf K}\cdot {\bf K}) 
- 4\fRN\wedge (\Dhat \fKbar\cdot \Dhat \fK)
+12 \fRN\wedge(\bar{\bf K}\cdot {\bf K})^2
\nonumber \\
&&\qquad\qquad\qquad\qquad 
+4\,(\Dhat\bar{\bf K}\cdot {\bf K})\wedge(\Dhat{\bf K}\cdot \bar{\bf K}) 
+4(\Dhat\bar{\bf K}\cdot \bar{\bf K})\wedge(\Dhat{\bf K}\cdot {\bf K})
-2\,(\Dhat\bar{\bf K}\cdot {\fr}\cdot \Dhat{\bf K}) \nonumber \\
&&\qquad\qquad \qquad\qquad
+8\,(\fKbar\cdot \fK)\wedge(\Dhat \fKbar\cdot \Dhat \fK)
-8\,(\bar{\bf K}\cdot {\bf K})^3
\Biggr]\, . 
  \label{eq:7danompolyppbarA}
\end{eqnarray}
}

\subsubsection{$\partial A \bar{\partial} A$ Term}
For the terms proportional to $\partial A  \bar{\partial} A$, 
there are eight possibilities in $\tr(\fR^4)$ :
\begin{eqnarray}
&&{}^{(d)}{\fR}^i{}_z{}^{(d)}{\fR}^z{}_j{}^{(0)}(\fR\fR)^j{}_i\times 4\, , 
\qquad 
{}^{(d)}{\fR}^i{}_{\bar{z}}{}^{(d)}{\fR}^{\bar{z}}{}_j{}^{(0)}(\fR\fR)^j{}_i\times 4\, , \nonumber \\
&&{}^{(d)}{\fR}^z{}_{j}{}^{(d)}{\fR}^{j}{}_z{}^{(0)}(\fR\fR)^z{}_z\times 4\, , \qquad 
{}^{(d)}{\fR}^{\bar{z}}{}_j{}^{(d)}{\fR}^j{}_{\bar{z}}{}^{(0)}(\fR\fR)^{\bar{z}}{}_{\bar{z}}\times 4\, , 
\nonumber\\
&&{}^{(d)}{\fR}^i{}_z{}^{(0)}{\fR}^z{}_z {}^{(d)}{\fR}^z{}_j{}^{(0)}{\fR}^j{}_i\times 4\,,
\qquad 
{}^{(d)}{\fR}^i{}_{\bar{z}}{}^{(0)}{\fR}^{\bar{z}}{}_{\bar{z}} {}^{(d)}{\fR}^{\bar{z}}{}_j{}^{(0)}{\fR}^j{}_i\times 4\,,  \\
&&{}^{(d)}{\fR}^i{}_z{}^{(0)}{\fR}^z{}_j {}^{(d)}{\fR}^j{}_{\bar{z}}{}^{(0)}{\fR}^{\bar{z}}{}_i\times 4\,,
\qquad 
{}^{(d)}{\fR}^z{}_j{}^{(0)}{\fR}^{j}{}_{\bar{z}} {}^{(d)}{\fR}^{\bar{z}}{}_i{}^{(0)}{\fR}^i{}_z\times 4\,. \nonumber 
\end{eqnarray}
Here the factor 4 comes from the locations of  ${}^{(d)}\fR$'s in $\tr(\fR^4)$\,. 
Each term in this list is then computed as follows : 
{
\begin{eqnarray}
&&{}^{(d)}{\fR}^i{}_z{}^{(d)}{\fR}^z{}_j{}^{(0)}(\fR\fR)^j{}_i\times 4 \nonumber \\
&&\quad = 16 (-2e^{-2A})(\partial A \bar{\partial} Adz\wedge d\bar{z})  \nonumber \\
&&\qquad \wedge \Biggl[
-(-2e^{-2A})(\Dhat\bar{\bf K}\cdot {\bf K})\wedge(\Dhat{\bf K}\cdot \bar{\bf K}) 
+(\bar{\bf K}\cdot {\fr}\cdot {\fr}\cdot{\bf K})
+(-2e^{-2A})(\Dhat\bar{\bf K}\cdot \bar{\bf K})\wedge (\Dhat{\bf K}\cdot {\bf K})
\nonumber \\
&&\qquad\qquad  
+2(-2e^{-2A})(\bar{\bf K}\cdot {\bf K})\wedge (\bar{\bf K}\cdot{\fr}\cdot {\bf K})
+(-2e^{-2A})^2(\bar{\bf K}\cdot {\bf K})^3\Biggr]\, ,\nonumber \\
&&{}^{(d)}{\fR}^i{}_z{}^{(d)}{\fR}^z{}_j{}^{(0)}(\fR\fR)^j{}_i\times 4
= {}^{(d)}{\fR}^i{}_{\bar{z}}{}^{(d)}{\fR}^{\bar{z}}{}_j{}^{(0)}(\fR\fR)^j{}_i\times 4\, , \nonumber\\
&&{}^{(d)}{\fR}^z{}_{j}{}^{(d)}{\fR}^{j}{}_z{}^{(0)}(\fR\fR)^z{}_z\times 4 \nonumber \\
&&\quad = 16(-2e^{-2A})(\partial A \bar{\partial} Adz\wedge d\bar{z})\wedge(\bar{\bf K}\cdot {\bf K})
\nonumber \\
&&\qquad\qquad 
 \wedge \Biggl[
[\fRN+(-2e^{-2A})(\bar{\bf K}\cdot {\bf K})]^2+(-2e^{-2A})
(\Dhat\bar{\bf K}\cdot \Dhat{\bf K}) 
\Biggr]\, , 
 \nonumber \\
 &&
{}^{(d)}{\fR}^{\bar{z}}{}_j{}^{(d)}{\fR}^j{}_{\bar{z}}{}^{(0)}(\fR\fR)^{\bar{z}}{}_{\bar{z}}\times 4 = 
{}^{(d)}{\fR}^z{}_{j}{}^{(d)}{\fR}^{j}{}_z{}^{(0)}(\fR\fR)^z{}_z\times 4\,, \nonumber \\
&&{}^{(d)}{\fR}^i{}_z{}^{(0)}{\fR}^z{}_z {}^{(d)}{\fR}^z{}_j{}^{(0}{\fR}^j{}_i\times 4
=16(-2\,e^{-2A})(\partial A \bar{\partial} Adz\wedge d\bar{z}) 
\wedge
[\fRN+(-2e^{-2A})(\bar{\bf K}\cdot {\bf K})]
\nonumber \\
&& \qquad \qquad\qquad\qquad\qquad\qquad\qquad \wedge 
[(\bar{\bf K}\cdot {\fr}\cdot {\bf K})+(-2e^{-2A})(\bar{\bf K}\cdot {\bf K})^2]\,, \nonumber \\
&&
{}^{(d)}{\fR}^i{}_{\bar{z}}{}^{(0)}{\fR}^{\bar{z}}{}_{\bar{z}} {}^{(d)}{\fR}^{\bar{z}}{}_j{}^{(0)}{\fR}^j{}_i\times 4
={}^{(d)}{\fR}^i{}_z{}^{(0)}{\fR}^z{}_z {}^{(d)}{\fR}^z{}_j{}^{(0)}{\fR}^j{}_i\times 4\, , \nonumber \\
&&
{}^{(d)}{\fR}^i{}_z{}^{(0)}{\fR}^z{}_j {}^{(d)}{\fR}^j{}_{\bar{z}}{}^{(0)}{\fR}^{\bar{z}}{}_i\times 4
= 16(-2e^{-2A})^2(\partial A \bar{\partial} Adz\wedge d\bar{z}) \wedge
(\Dhat\bar{\bf K}\cdot \bar{\bf K})\wedge (\Dhat{\bf K}\cdot {\bf K})\, , \nonumber \\
&&
{}^{(d)}{\fR}^z{}_j{}^{(0)}{\fR}^{j}{}_{\bar{z}} {}^{(d)}{\fR}^{\bar{z}}{}_i{}^{(0)}{\fR}^i{}_z\times 4
= 16(-2e^{-2A})^2(\partial A \bar{\partial} Adz\wedge d\bar{z} )\wedge
(\Dhat\bar{\bf K}\cdot \bar{\bf K})\wedge (\Dhat{\bf K}\cdot {\bf K})\, . \nonumber \\
\end{eqnarray}
By summing them up and doing integration, we obtain 
\begin{eqnarray}
&&\int_{\onCone}\fPanom^{7d, single}|_{\partial A\bar{\partial }A} \nonumber \\
&&=\int_{\onCone} 32(-2\,e^{-2A})(\partial A \bar{\partial} Adz\wedge d\bar{z})  \\
&&\qquad \wedge 
\Biggl[
\fRN^2\wedge(\bar{\bf K}\cdot {\bf K})
+\fRN\wedge [3(-2e^{-2A})(\bar{\bf K}\cdot{\bf K})^2+(\bar{\bf K}\cdot{\fr}\cdot{\bf K} )]
\nonumber \\
&&\qquad \qquad
+2 (-2 e^{-2A})(\Dhat\bar{\bf K}\cdot \bar{\bf K})\wedge (\Dhat{\bf K}\cdot {\bf K})
-(-2e^{-2A})(\Dhat\bar{\bf K}\cdot {\bf K})\wedge (\Dhat{\bf K}\cdot \bar{\bf K})
+(\bar{\bf K}\cdot {\fr}\cdot {\fr}\cdot {\bf K})\nonumber \\
&&\qquad\qquad 
+(-2e^{-2A})(\bar{\bf K}\cdot{\bf K})\wedge(\Dhat\bar{\bf K}\cdot \Dhat{\bf K})
+3(-2e^{-2A})(\bar{\bf K}\cdot {\bf K})\wedge (\bar{\bf K}\cdot {\fr}\cdot {\bf K})\nonumber \\
&&\qquad\qquad 
-3(-2e^{-2A})^2(\bar{\bf K}\cdot {\bf K})^3
\Biggr]\, \nonumber \\
&&=16\pi\epsilon\int_{\Sigmatilde} \Biggl[
2\fRN^2\wedge(\fKbar\cdot \fK)
+\fRN\wedge[-6\,(\bar{\bf K}\cdot {\bf K})^2
+2(\bar{\bf K}\cdot{\fr}\cdot {\bf K})] \nonumber \\
&&\qquad\qquad\qquad
\qquad+2(\Dhat\bar{\bf K}\cdot {\bf K})\wedge(\Dhat{\bf K}\cdot \bar{\bf K})
-4(\Dhat\bar{\bf K}\cdot \bar{\bf K})\wedge(\Dhat{\bf K}\cdot {\bf K})
+2(\bar{\bf K}\cdot{\fr}\cdot{\fr }\cdot {\bf K}) \nonumber \\
&&\qquad\qquad\qquad
\qquad -2(\bar{\bf K}\cdot {\bf K})\wedge(\Dhat\bar{\bf K}\cdot \Dhat{\bf K})
-6(\bar{\bf K}\cdot {\bf K})\wedge(\bar{\bf K}\cdot{\fr}\cdot {\bf K})
+8(\bar{\bf K}\cdot {\bf K})^3
\Biggr]\, . \label{eq:7danompolypApbarA}
\end{eqnarray}
}

\subsubsection{Final Result}
Summing these two contributions in Eqs.~\eqref{eq:7danompolyppbarA} and \eqref{eq:7danompolypApbarA}, 
we obtain 
\begin{eqnarray}
&&16\pi \epsilon \int _{\Sigmatilde} 
\Biggl[
\fR_N^3 -4\fR_N^2\wedge (\bar{\bf K}\cdot {\bf K}) 
+\fR_N\wedge [ -4(\Dhat \bar{\bf K}\cdot \Dhat {\bf K})+6(\bar{\bf K}\cdot {\bf K})^2 + 2(\bar{\bf K}\cdot {\fr}\cdot {\bf K})]
\nonumber  \\
&&\qquad\qquad\qquad 
+6(\Dhat \bar{\bf K}\cdot{\bf K})\wedge ( \Dhat{\bf K}\cdot \bar{\bf K})
+6(\bar{\bf K}\cdot {\bf K})\wedge (\Dhat\bar{\bf K}\cdot \Dhat{\bf K}) -6(\bar{\bf K}\cdot {\bf K})\wedge (\bar{\bf K}\cdot 
{\fr}\cdot {\bf K}) \nonumber \\
&&\qquad\qquad\qquad 
-2(\Dhat\bar{\bf K}\cdot {\fr}\cdot \Dhat{\bf K}) + 2(\bar{\bf K}\cdot {\fr}\cdot {\fr}\cdot {\bf K})
\Biggr] \label{eq:sumtrr4}\, . 
\end{eqnarray}
Now we rewrite this into the form $d(\ldots)$. 
For this purpose, we note the following identities: $d[\fGamma_N \wedge \fR_N^2] = \fR_N^3\,$ along with 
\begin{equation}
\begin{split}
d &\left[(-2\fR_N)\wedge[({\bf K}\cdot \Dhat \bar{\bf K}) + (\bar{\bf K}\cdot \Dhat {\bf K}) ]\right] \\
&\qquad= 
4\left[
\fR_N \wedge(\bar{\bf K} \cdot \fr\cdot {\bf K} ) -\fR_N^2 \wedge (\bar{\bf K}\cdot {\bf K}) -\fR_N\wedge (\Dhat \bar{\bf K}\cdot \Dhat{\bf K}) 
\right]\,,   \\
d&\left[
3(\bar{\bf K}\cdot {\bf K})\wedge[({\bf K}\cdot \Dhat \bar{\bf K}) + (\bar{\bf K}\cdot \Dhat{\bf K})]
\right] \\
&\qquad= 
 6\left[
\fR_N\wedge (\bar{\bf K}\cdot {\bf K})^2 +(\bar{\bf K}\cdot {\bf K})\wedge (\Dhat \bar{\bf K}\cdot \Dhat{\bf K})
-(\bar{\bf K}\cdot {\bf K})\wedge (\bar{\bf K}\cdot {\fr} \cdot {\bf K}) - (\bar{\bf K}\cdot \Dhat{\bf K})\wedge({\bf K}\cdot \Dhat\bar{\bf K})
\right]\, , \\
d&\left[
(\Dhat{\bf K}\cdot {\fr} \cdot  \bar{\bf K}) - (\Dhat \bar{\bf K}\cdot {\fr} \cdot {\bf K}) 
\right]\\
&\qquad=  2\left[
(\bar{\bf K}\cdot {\fr}\cdot {\fr}\cdot {\bf K}) -\fR_N\wedge (\bar{\bf K}\cdot {\fr}\cdot {\bf K})
-(\Dhat\bar{\bf K}\cdot {\fr} \cdot \Dhat{\bf K} ) 
\right]\, , 
\end{split}
\end{equation}
where we have used 
\begin{eqnarray}
&&\Dhat^2 {\bf K}^i = ({\fr}\cdot {\bf K})^i -\fR_N \wedge {\bf K}^i\, , \qquad 
\Dhat^2 \bar{\bf K}^i = ({\fr}\cdot \bar{\bf K})^i +\fR_N \wedge \bar{\bf K}^i\, .
\end{eqnarray}
By using the above identities, we can rewrite Eq.~\eqref{eq:sumtrr4} as 
\begin{eqnarray}\label{eq:trR4final}
&&16\pi\epsilon \int_{\Sigmatilde} d
\Biggl[
\fGN\wedge \fRN^2
+[-2\fRN +3 (\fKbar \cdot \fK)]\wedge \left[(\fK\cdot \Dhat \fKbar) + (\fKbar\cdot \Dhat \fK)\right]
\nonumber \\
&&\qquad\qquad\qquad\qquad\qquad 
+
(\Dhat \fK \cdot \fr \cdot  \fKbar)-
(\Dhat \fKbar \cdot\fr \cdot  \fK)
\Biggr]\, . 
\end{eqnarray}
Therefore, the holographic entanglement entropy formula for the 7d single-trace 
gravitational Chern-Simons term is obtained from the anomaly polynomial as 
\begin{eqnarray}
&&  \SeeCS^{7d,\, single} \nonumber  \\
 &&\quad =2 (2\pi)(4)  \int_{\Sigma} 
\Biggl[
\fGN\wedge \fRN^2
+[-2\fRN +3 (\fKbar \cdot \fK)]\wedge \left[(\fK\cdot \Dhat \fKbar) + (\fKbar\cdot \Dhat \fK)\right]
\nonumber \\
&&\qquad\qquad\qquad\qquad\qquad +
(\Dhat \fK \cdot \fr \cdot  \fKbar)-
(\Dhat \fKbar \cdot\fr \cdot  \fK)
\Biggr]\,.  \nonumber\\
\end{eqnarray}

\subsection{\fixform{$\tr(\fR^2)\wedge \tr(\fR^2)$}}
As a final example, here we derive the holographic entanglement entropy 
formula for 7d double-trace gravitational Chern-Simons terms by 
evaluating the anomaly polynomial $\fPanom^{7d, double} = \tr({\fR}^2)\wedge \tr({\fR}^2)$ 
on the regularized cone background.  
\subsubsection{$\partial \bar{\partial} A$ Term}
There are two possibilities for this type of contribution in $\tr(\fR^2)\wedge\tr(\fR^2)$:
\begin{eqnarray}
{}^{(d^2)}\fR^z{}_z {}^{(0)}\fR^z{}_z \wedge \fPanom^{3d}|_{0th}\times 4\, , 
\qquad 
{}^{(d^2)}\fR^{\bar{z}}{}_{\bar{z}} {}^{(0)}\fR^{\bar{z}}{}_{\bar{z}} \wedge \fPanom^{3d}|_{0th} \times 4\, , 
\end{eqnarray}
where the factor 4 comes from the choice for the location of ${}^{(d^2)}\fR$\,. 
By summing over these contributions and doing integration, we obtain 
{
\begin{eqnarray}
&&\int_{\onCone} \fPanom^{7d, double}|_{\partial\bar{\partial} A} \nonumber \\
&&\qquad = \int_{\onCone}
(-16\,\partial \bar{\partial} Adz\wedge d\bar{z})
\wedge[\fRN+(-2e^{-2A})(\bar{\bf K}\cdot {\bf K})]\wedge \fPanom^{3d}|_{0th} \nonumber \\
&&\qquad =16\pi\, \epsilon\,  [\fRN+(-2)(\bar{\bf K}\cdot {\bf K})] \nonumber \\
&&\qquad\qquad  \wedge
\Biggl[
2\fRN^2-8\fRN\wedge(\bar{\bf K}\cdot {\bf K})-8(\Dhat\bar{\bf K}\cdot \Dhat{\bf K})
+ \tr({\fr}^2)+8(\bar{\bf K}\cdot {\fr}\cdot {\bf K})
\Biggr]\, . \nonumber \\
\label{eq:7danomdoubleppbarA} 
\end{eqnarray}
}
Here we have used the expression of $\fPanom^{3d}|_{0th}$ given in Eq.~\eqref{eq:panom3d0th}. 

\subsubsection{$\partial A \bar{\partial} A$ Term}
As a next step, we evaluate the terms proportional to $\partial A \bar{\partial} A$. 
There are four possibilities in $\tr(\fR^2)\wedge\tr(\fR^2)$:
\begin{eqnarray}
&&{}^{(d)}\fR^i{}_z{}^{(d)}\fR^z{}_i\wedge \fPanom^{3d}|_{0th} \times 4\, ,\qquad  
{}^{(d)}\fR^i{}_{\bar{z}}{}^{(d)}\fR^{\bar{z}}{}_i\wedge \fPanom^{3d}|_{0th}\times 4\, , \nonumber \\
&&{}^{(d)}\fR^i{}_z{}^{(0)}\fR^z{}_i\wedge {}^{(d)}\fR^z{}_j{}^{(0)}\fR^j{}_z \times 8\, ,\qquad  
{}^{(d)}\fR^i{}_{\bar{z}}{}^{(0)}\fR^{\bar{z}}{}_i\wedge 
{}^{(d)}\fR^{\bar{z}}{}_j{}^{(0)}\fR^j{}_{\bar{z}} \times 8\, ,  \\
&&{
{}^{(d)}\fR^i{}_z{}^{(0)}\fR^z{}_i\wedge {}^{(d)}\fR^j{}_{\bar{z}}{}^{(0)}\fR^{\bar{z}}{}_j \times 8\, ,\qquad
{}^{(d)}\fR^z{}_i{}^{(0)}\fR^i{}_z\wedge 
{}^{(d)}\fR^{\bar{z}}{}_j{}^{(0)}\fR^j{}_{\bar{z}} \times 8\, },  \nonumber
\end{eqnarray}
and each term is evaluated as follows :
{
\begin{eqnarray}
&&{}^{(d)}\fR^i{}_z{}^{(d)}\fR^z{}_i\wedge\fPanom^{3d}|_{0th}\times 4
= 16(-2e^{-2A})(\partial A \bar{\partial} Adz\wedge d\bar{z})\wedge(\bar{\bf K}\cdot {\bf K})
\wedge \fPanom^{3d}|_{0th}\, , \nonumber \\
&&  
{}^{(d)}\fR^i{}_{\bar{z}}{}^{(d)}\fR^{\bar{z}}{}_i\wedge \fPanom^{3d}|_{0th}\times 4
={}^{(d)}\fR^i{}_z{}^{(d)}\fR^z{}_i\wedge \fPanom^{3d}|_{0th}\times 4
\, , \\
&&
{}^{(d)}\fR^i{}_z{}^{(0)}\fR^z{}_i\wedge {}^{(d)}\fR^z{}_j{}^{(0)}\fR^j{}_z \times 8 \nonumber \\
&&\qquad = -32(-2 e^{-2A})^2(\partial A \bar{\partial} Adz\wedge d\bar{z})\wedge
(\Dhat\bar{\bf K}\cdot {\bf K})\wedge(\Dhat{\bf K}\cdot \bar{\bf K}) , \nonumber \\
&&
{}^{(d)}\fR^i{}_{\bar{z}}{}^{(0)}\fR^{\bar{z}}{}_i\wedge 
{}^{(d)}\fR^{\bar{z}}{}_j{}^{(0)}\fR^j{}_{\bar{z}} \times 8=
{}^{(d)}\fR^i{}_z{}^{(0)}\fR^z{}_i\wedge {}^{(d)}\fR^z{}_j{}^{(0)}\fR^j{}_z \times 8\, \nonumber \\
&& {\qquad = 
{}^{(d)}\fR^i{}_z{}^{(0)}\fR^z{}_i\wedge {}^{(d)}\fR^j{}_{\bar{z}}{}^{(0)}\fR^{\bar{z}}{}_j \times 8
=   {}^{(d)}\fR^z{}_i{}^{(0)}\fR^i{}_z\wedge 
{}^{(d)}\fR^{\bar{z}}{}_j{}^{(0)}\fR^j{}_{\bar{z}} \times 8\,.} \nonumber
\end{eqnarray}
By summing up these terms and doing integration, we obtain
\begin{eqnarray}
&& \int_{\onCone} \fPanom^{7d, double}|_{\partial A\bar{\partial}A} \nonumber \\
&&= \int_{\onCone}32(-2e^{-2A})(\partial A \bar{\partial} Adz\wedge d\bar{z})  \nonumber \\
&&\qquad 
\wedge \left[
(\bar{\bf K}\cdot{\bf K})\wedge \fPanom^{3d}|_{0th}
- {4}(-2e^{-2A})(\Dhat\bar{\bf K}\cdot{\bf K})\wedge (\Dhat{\bf K}\cdot\bar{\bf K})
\right]\nonumber \, \\
&&=32\pi\epsilon \int_{\Sigmatilde}
\Biggl[
2\fRN^2\wedge(\bar{\bf K}\cdot {\bf K})-4
\fRN\wedge(\bar{\bf K}\cdot {\bf K})^2
-4(\bar{\bf K}\cdot {\bf K})\wedge (\Dhat\bar{\bf K}\cdot \Dhat{\bf K}) \nonumber \\
&&\qquad\qquad\qquad\qquad 
+4(\Dhat\bar{\bf K}\cdot {\bf K})\wedge (\Dhat{\bf K}\cdot \bar{\bf K})
+4(\bar{\bf K}\cdot {\bf K})\wedge (\bar{\bf K}\cdot {\fr}\cdot {\bf K})
+(\bar{\bf K}\cdot {\bf K}) \wedge \tr({\fr}^2)
\Biggr]\, . \nonumber \\
\label{eq:7danomdoublepApbarA}
\end{eqnarray}
}

\subsubsection{Final Result}
By summing the above two contributions \eqref{eq:7danomdoubleppbarA} 
and \eqref{eq:7danomdoublepApbarA} and rewriting in the total derivative form, we obtain 
{
\begin{eqnarray}
&&32\pi\epsilon\int_{\Sigmatilde} d 
\Biggl[
\fGN\wedge \fRN^2
+\frac{1}{2}\fGN\wedge{{\rm tr}\left({\fr}^2\right)}
+2\left[-\fRN+(\fKbar\cdot \fK)\right]\wedge [
(\fK\cdot \Dhat \fKbar)
+(\fKbar \cdot \Dhat \fK)]
\Biggr]\, .\nonumber\\ 
\end{eqnarray}
}Therefore, the holographic entanglement entropy formula for 7d double-trace gravitational Chern-Simons 
term obtained from the anomaly polynomial is 
\bea
&&  \SeeCS^{7d,\, double} \nonumber \\
&&\qquad = 2^2(2\pi)(4) \int_{\Sigma} 
\Biggl[
\fGN\wedge \fRN^2
+\frac{1}{2}\fGN\wedge{{\rm tr}\left({\fr}^2\right)}
\nonumber\\
&&\qquad\qquad\qquad\qquad\qquad\qquad\qquad
+2\left[-\fRN+(\fKbar\cdot \fK)\right]\wedge [
(\fK\cdot \Dhat \fKbar)+(\fKbar \cdot \Dhat \fK)]
\Biggr]\,.\label{eq:trR2trR2v2}
\nonumber \\
\eea

\section{Alternative Derivation from Anomaly Polynomial: Application of Dong's Formula}
\label{sec:anomalymethod2}
This Appendix provides an alternative way to derive the holographic entanglement entropy formulas from 
the anomaly polynomials. The idea is that, since the anomaly polynomials depend only on 
the Riemann tensor, we directly apply Dong's formula \cite{Dong:2013qoa} to the anomaly polynomials 
and rewrite the integrand in the total derivative form, 
instead of doing $\epsilon$-expansion near the regularized cone geometry explicitly. 
We will confirm that the results in \S\ref{sec:anomalymethod1} are reproduced correctly. 

Here is one remark: Since we do not carry out $\epsilon$-expansion near the 
regularized cone geometry explicitly in this Appendix, we will take the coordinate to be 
general (for example, we do not introduce $(z, \bar{z})$ coordinate),  
except when we compare the final results with those 
obtained by using $\epsilon$-expansion in the rest part of this paper. 
For the comparison,  we summarize in Appendix \ref{sec:usefulconversion} some useful formulas 
that convert 
the notation in this Appendix into the one we used in the main text of this paper.

\subsection{Some Geometry of Co-Dimension Two Surface in Differential Form}
\label{app:codimension2geo}
Before introducing the Dong's holographic entanglement entropy formula and applying 
it to the anomaly polynomials, we start with a brief summary of geometry relevant to a
co-dimension two surface $\tilde{\Sigma}$ living in a spacetime (in Euclidean signature) with metric $G_{\mu\nu}$. 

For the co-dimension two surface $\tilde{\Sigma}$, we can introduce two normal 
vectors $n_{(a)}{}^\mu$ (where $a=1,2$) such that
\beq
n_{(1)}{}^\mu n_{(1)\mu}=
n_{(2)}{}^\mu n_{(2)\mu }=1\, ,\quad
n_{(1)}{}^\mu n_{(2)\mu}=0\, . 
\eeq 
Then the induced metric on the tangent space spanned by these normal vectors is given by 
\be
n_{ab} =n_{(a)}{}^\mu n_{(b)}{}^\nu G_{\mu\nu}\, .
\ee We also denote the binormal on $\tilde{\Sigma}$ by $\ve_{ab}$. 
Here we note that indices $a, b, c, \ldots$ denote the directions orthogonal to $\tilde{\Sigma}$, 
while we labels the ones along $\tilde{\Sigma}$ with $i, j, k, l, \ldots$ 
(we use $y^i$ for the coordinates along $\tilde{\Sigma}$).\footnote{For the geometry of the co-dimension two surface, we follow the notation and discussion in 
the Appendix B of \cite{Rosenhaus:2014woa}. Here, we only summarize some quantities 
that we need to build the differential forms on $\tilde{\Sigma}$ used for the derivation 
of the holographic entanglement entropy formula (for example, for the Riemann tensor,
we only summarize $R_{\mu\nu ij}$ components only). We also note that 
we chose coordinates as $n_{(a)}{}^\mu=\delta^\mu_a$ such that the index `$a$' also means a coordinate. 
One could equally think of all the $a,b,\ldots$ indices as projection on the normal vectors.}

By using these quantities, the extrinsic curvatures on the co-dimension two surface $\tilde{\Sigma}$ are given by
\be
K_{aij} \equiv D_i n_{aj} |_{\tilde{\Sigma}}=\half \partial_a G_{ij} \, ,\quad
K_{aij}=K_{aji}\,, 
\ee 
where $D_i$ is the covariant derivative compatible with the induced metric on $\tilde{\Sigma}$. 
It is also useful for our purpose to define 
 ``gauge field'' $\GN_i$ for the $SO(2)$  normal bundle rotation 
 and its field strength $\RN_{ij}$ as 
\be
\GN_i =\half  \iepsUUab \partial_a G_{ib}\,,\quad
\RN_{ij}\equiv 2 \partial_{[i} \GN_{j]}\,.
\ee
Then the Christoffel symbol and Riemann tensor are written as 
\be
\Gamma^\mu{}_{bc} =0\,,\quad
\Gamma^a{}_{bj} =\iepsUdab \GN_j\, ,\quad
\Gamma^a{}_{ij}=- K^a{}_{ij} \,,\quad
\Gamma^i{}_{bj} = K_{bj}{}^i\,, 
\ee 
and 
\bea
R_{abij}&=&\iepsddab\RN_{ij} +K_{bil}K_{a}{}^l{}_{j}-K_{ail} K_{b}{}^l{}_{j} \, , \nonumber\\
R_{ijkl}&=&r_{ijkl}  - K_{aik}K^a{}_{jl} +K_{ail} K^a{}_{jk}\, ,  \nonumber\\
R_{ajkl}&=&-D_k K_{ajl}+D_l K_{ajk} -2 \iepsddab\GN_{[k} K^b{}_{l]j}\, . 
\eea 
Here $r_{ijkl}$ is the Riemann tensor for the induced metric along $\tilde{\Sigma}$.

For our purpose to deal with the anomaly polynomials, it is convenient to 
introduce the differential forms for some geometric quantities defined above:
\bea
\fGamma^\mu{}_\nu&\equiv &\Gamma^\mu{}_{\nu k} dy^k \quad , \quad
\fGN \equiv \GN_i dy^i \,  , \quad
\fRN \equiv d \fGN \, , \quad
\fK_{ai} \equiv K_{aik} dy^k  \,,  \nonumber\\
\fR_{\mu\nu} &\equiv&\half R_{\mu\nu kl} dy^k \wedge dy^l\, ,\quad
\fr_{ij} \equiv\half r_{ij kl} dy^k \wedge dy^l\, . \quad
\eea
More explicitly, by substituting the above component-based expressions to these definitions, we obtain 
\bea
\fGamma^a{}_{b} &=&\iepsUdab \fGN\, ,\quad
\fGamma^a{}_{i}=- \fK^a{}_{i} \,,\quad
\fGamma^i{}_{b} = \fK_{b}{}^i\, , \\
\fR^{a}{}_{b}&=&\iepsUdab \left[\fRN +\fsigma\right]\, , \quad 
\fR^i{}_{j} = \fr^i{}_{j}  - 2\fw^i{}_j\, ,  
 \nonumber\\
\fR^{a}{}_{i}&=&-\Dhat \fK^a{}_{i}  \,  , \quad
\Dhat \fK{}_{aj} \equiv D \fK_{aj}  +\iepsddab \fGN \wedge\fK^b{}_{j}\,,
\eea 
where $\Dhat$ is the covariant derivative associated with the normal bundle gauge field $\fGN$. 
We have also defined
\bea
\fsigma_{ij}&\equiv&\half \epsilon^{ab} \fK_{ai} \wedge \fK_{bj},\quad \fsigma_{ij}=\fsigma_{ji},\quad
\fK_{al} \wedge \fK_{b}{}^l=\miepsab \fsigma\, ,\quad \fsigma\equiv \fsigma^i{}_i\,,
\nonumber\\
\fw_{ij}&\equiv &\half \fK^a{}_{i} \wedge \fK_{aj},\quad\fw_{ij}=-\fw_{ji} \,.
\eea To show the third relation in the first line, we have used 
$\epsUUab\epsddab=2$ which follows from
$\epsddab \epsddcd
=n_{ac} n_{bd}-n_{ad}n_{bc}$ and 
$n^a{}_a=2\,$.  
For later use, here we summarize some useful identities relevant to $\fw_{ij}$ 
and $\fsigma_{ij}$:\footnote{It should be clear that whenever $\fsigma$ appears inside a $\tr[\ldots]$, it is $\fsigma^i{}_j$. Otherwise $\fsigma=\fsigma^i{}_i$.}
\bea
&&\fw_{ik}\wedge \fw^{k}{}_j
=-\half \fsigma\wedge
\fsigma_{ij}
\, ,\quad
\tr[\fomega^2]
=-\half \fsigma^2 \, , \quad  
\fsigma_{ik}\wedge \fsigma^k{}_{j}
=-\frac{1}{2}
 \fsigma\wedge \fsigma_{ij}\, ,\quad
 \tr[\fsigma^2]=-\half \fsigma^2\, , \nonumber \\
&& \fw^i{}_k \wedge \fsigma^k{}_j 
=-\half \fsigma  \wedge \fw^i{}_j 
\, ,\quad
 \fsigma^i{}_k \wedge\fw^k{}_j 
 =-\frac{1}{2}\fsigma\wedge \fw^i{}_j\, . 
\eea

\subsection{Dong's Formula for Holographic Entanglement Entropy}
In \cite{Dong:2013qoa}, the holographic entanglement entropy formula is derived 
for a general theory with a Lagrangian which is a functional  of Riemann tensor and does not
involve covariant derivatives of Riemann tensor. 
The purpose of this part of Appendix is to rewrite  this formula in a way it can be readily applied  to the anomaly polynomials. 
Here we focus only on pure gravitational anomaly polynomials since  the extension
to mixed anomaly polynomials is straightforward.

In Dong's  formula, there is a parameter $q_\alpha$ which essentially counts 
the number of the regularization factor $e^{-2A}$ carried by each term (labeled by the subscript ${}_\alpha$)
in the second order derivative of the Lagrangian with respect to Riemann tensor evaluated on the regularized cone background. For our purposes,
$Q$'s and $V$'s in \cite{Dong:2013qoa} do not contribute to the final answer.
Using this, we can simplify the Dong's rule to compute $q_\alpha$ to the following: 
(1) first expand the product of $\fR$'s into all lower-indices,  
(2) then keep track of $q_\alpha$, which counts half of the total number of  $\fK_{ai}$ in each term in the sum.
To implement this,   it is convenient to introduce $\fR_t$ as 
\bea\label{eq:Rt}
(\fR_t)^a{}_{b}&\equiv &\iepsUdab (\fRN +t\,\fsigma)\,, \quad 
(\fR_t)^i{}_{j} \equiv \fr^i{}_{j}  - t\,(2\fw^i{}_{j}) \,, \quad 
(\fR_t)^a{}_{j} \equiv -t^{1/2}\,\Dhat \fK^a{}_{j}  \, .  
\eea 
We note that $\fsigma_{ij}$ and $\fomega_{ij}$ each gets a factor $t$ since they are quadratic in $\fK_{ai}$\,. 
By noticing that 
\be
\int_0^1 dt~ t^{q_\alpha} = \frac{1}{1+q_\alpha}\,, 
\ee 
we can now write a covariant form of Dong's formula derived in \cite{Dong:2013qoa} specialised to our case as: 
\bea
S_{\rm ent}&=&\SEEWald+\SEEKK \,  \qquad {\rm where} \nonumber\\
\SEEWald&\equiv &
 2\pi \int_{\tilde{\Sigma}}\left[
  -\frac{\pa L}{
  \pa R_{\m\r\n\s}
  }
   \ve_{\m\r}
    \ve_{\n\s}
    \right]
  (\hodge_{\tilde{\Sigma}} 1)\,,  \nonumber\\
\SEEKK&\equiv &
 4\pi \int_{\tilde{\Sigma}}\int_0^1 dt \lt\{ 
 \left(
 \frac{\pa^2 L}{\pa R_{\m_1\r_1\n_1\s_1} \pa R_{\m_2\r_2\n_2\s_2}}
 \right)_t K_{\l_1\r_1\s_1} K_{\l_2\r_2\s_2}  \rt. \nonumber\\
&&\lt.
\phantom{\frac12} 
\qquad\qquad\quad  \times \left[ (n_{\m_1\m_2} n_{\n_1\n_2}-\ve_{\m_1\m_2} \ve_{\n_1\n_2}) n^{\l_1\l_2} + (n_{\m_1\m_2} \ve_{\n_1\n_2}+\ve_{\m_1\m_2} n_{\n_1\n_2}) \ve^{\l_1\l_2}\right] \rt\} (\hodge_{\tilde{\Sigma}} 1)\,. \nonumber\\
\eea Here the quantities with subscript $t$ means that the curvature two-forms are evaluated by using Eq.~(\ref{eq:Rt}) 
and  $\hodge_{\tilde{\Sigma}}(\ldots)$ denotes the Hodge dual of $(\ldots)$ on $\tilde{\Sigma}$.

In the rest of this Appendix, we use the above expression to reproduce 
the holographic entanglement entropy formula for 3d and 7d gravitational 
Chern-Simons terms from the anomaly polynomial \`a la Dong's formula. For the co-dimension two surface $\tilde{\Sigma}$, 
since we are using the anomaly polynomial after uplifting to a one-dimension higher 
spacetime as explained in \S\ref{sec:buildingblocks}, we take as $\tilde{\Sigma}=\Sigmatilde
$ 
where $\Sigma$ is the bulk entangling surface for the Chern-Simons terms and 
$I=[0,\infty)$ is a half line. 
For the reader's convenience, in the end of this Appendix (see 
Appendix \ref{sec:usefulconversion}), we have also 
summarized some useful relations for converting the notation in this Appendix 
into the one we used in the computation based on $\epsilon$-expansion.

\subsection{\fixform{$\fPanom =  \tr (\fR^2)$}}
As the simplest example, let us start with the anomaly polynomial corresponding 
to 3d gravitational Chern-Simons term and apply Dong's formula.  The corresponding Euclidean Lagrangian $L$ 
is given by:\footnote{We note that our notation for the epsilon tensors deviates from 
\cite{Jensen:2013kka} by $(-i)$: $\epsilon^{ab}_{\rm JLY} = (-i)\epsilon_{{\rm here}}^{ab}$ and 
$\epsilon^{\alpha_1\alpha_2\alpha_3\alpha_4}_{\rm JLY} = (-i)\epsilon^{\alpha_1\alpha_2\alpha_3\alpha_4}_{{\rm here}}$.}
\be
L 
=\frac{1}{(2!)^2}R^\mu{}_{\rho \alpha_1 \alpha_2} R^\rho{}_{\mu \alpha_3\alpha_4} 
(-i\ve^{\alpha_1\ldots \alpha_4})\, . 
\ee
Our convention for the epsilon-tensor is that $ \ve^{ab\alpha_1\alpha_2} = \ve^{ab} \ve^{\alpha_1\alpha_2}$ where $\ve^{ab}$ in components are given in Appendix \ref{sec:usefulconversion}.
By using $\iepsUdab\fR^b{}_{a}=2( \fRN +\fsigma)$, the Wald term $\SEEWald$ 
is given by  
\be
\SEEWald=  8\pi   \int_{\Sigmatilde} (\fRN +\fsigma)\,, 
\ee
while $\SEEKK$ term is evaluated as 
\be
\SEEKK=-8\pi  \int_{\Sigmatilde} \fsigma \, .
 \ee
Combining these two contributions, we obtain
 \be
  \SeeCS^{3d}=8 \pi  \int_{\Sigmatilde}\fRN=8\pi  \int_{\Sigmatilde} d
 \fGN
 = 8\pi  \int_{\Sigma }  \fGN
 \, .
\ee 
This indeed reproduces the result in Eq.~(\ref{eq:trR2}) we have obtained 
by applying $\epsilon$-expansion directly.

\subsection{\fixform{$\fPanom=  \tr (\fR^4)$}}
As a next example, here we consider the anomaly polynomial 
corresponding to the 7d single-trace gravitational Chern-Simons term. 
First of all, the Wald term $\SEEWald$ is obtained as 
\bea
&&\SEEWald
\nonumber\\
&&= 2\pi \times 4 \int_{\Sigmatilde}
\Biggl\{
 \fRN \wedge (\tr ( \fR^2)-  \tr[\fr^2])
 +2\fRN^2\wedge \fsigma
+2\fRN\wedge \tr[  \fr \fw]
+2\tr[ \fr^2\fsigma]
+2\fRN\wedge \fsigma^2
\nonumber\\
 &&
\qquad\qquad\qquad\qquad 
+4\fsigma\wedge \tr[\fr \fw] 
+2(\iepsUdab)(\Dhat \fK^b)\cdot  \fomega \cdot( \Dhat \fK_a)
- 2\fsigma \wedge d\left[  \fK^a
\cdot(\Dhat \fK_a)\right]
+2\fsigma^3
\nonumber\\
 &&\qquad\qquad\qquad\qquad 
+ d\left[\miepsab( \fK^b\cdot  \fr \cdot \Dhat \fK^a)\right]
\Biggr\} \,, 
 \eea where the `$\cdot$' indicates contraction of the $i$-type indices. 

As a next step, let us compute $\SEEKK$ term. Since the integrand is computed as 
  \bea
&&4\pi  \left(\frac{\pa^2 L}{\pa R_{\m_1\r_1\n_1\s_1} \pa R_{\m_2\r_2\n_2\s_2}}\right)_tK_{\l_1\r_1\s_1} K_{\l_2\r_2\s_2}\nonumber\\
&&\qquad\qquad \times  \left[ (n_{\m_1\m_2} n_{\n_1\n_2}-\ve_{\m_1\m_2} \ve_{\n_1\n_2}) n^{\l_1\l_2} + (n_{\m_1\m_2} \ve_{\n_1\n_2}+\ve_{\m_1\m_2} n_{\n_1\n_2}) \ve^{\l_1\l_2}\right] (\hodge_\Sigma 1)  \nonumber\\
 &&= 2\pi\times 4\Biggl\{
-2\fRN^2\wedge \fsigma
-2 \fRN \wedge\tr[\fr \fw]
-2\tr[\fr^2\fsigma]
-4 t \fRN \wedge \fsigma^2
 -8t\fsigma\wedge \tr[\fr \fw]  
 -6 t^2 \fsigma^3 
\nonumber\\
&&
\qquad\qquad\qquad 
 +4t \miepsab(\Dhat \fK^b )\cdot \fw\cdot (\Dhat \fK^a )
+4t\fsigma\wedge d\left[  \fK^a\cdot(\Dhat \fK_a)\right]
-3 t d \left[\fsigma(\fK^c\cdot \Dhat\fK_c)\right]
 \Biggr\}\,,  \nonumber\\
\eea
after integrating over $\int_0^1 dt$, we obtain $\SEEKK$  as 
\bea
&&\SEEKK \nonumber \\
&&=
2\pi \times 4  \int_{\Sigmatilde}\left\{
-2 \fsigma^3 
-3  \fRN\wedge \fsigma^2
 +\frac{1}{2}(\Dhat \fK^a)\cdot(  \Dhat \fK_a) \wedge  \fsigma
 -3\fsigma\wedge\tr[\fr \fw]
 -2\fRN^2\wedge \fsigma
  \right.
 \nonumber\\
&&\left.\qquad\qquad\qquad\quad 
+\frac{3}{2} (\Dhat \fK^{a} )\cdot\fsigma\cdot(\Dhat \fK_{a})
+\miepsab\frac{1}{2} (\Dhat \fK^b )\cdot \fw\cdot (\Dhat \fK^a )
-2\tr[\fr^2\fsigma]
-2 \fRN\wedge \tr[\fr \fw]
 \right\}\,.\nonumber\\
\eea

Combining these two contribution, we then obtain 
\bea
  \SeeCS^{7d, \,single}
&=&
 16\pi 	  \int_{\Sigmatilde}
\left\{
\fRN^3
- \fRN \wedge d\left[ \fK^a\cdot(\Dhat \fK_a)\right]
+\frac{3}{2} \fRN\wedge \fsigma^2
-\frac{3}{4} \fsigma\wedge(\Dhat \fK^a)\cdot(\Dhat \fK_{a} )
 \right.
\nonumber\\
&&\left.\qquad\qquad\qquad 
 -\frac{3}{2}\fsigma\wedge\tr\left[\fr \fw\right]
+\half \miepsab(\Dhat \fK^b)\cdot  \fr \cdot( \Dhat \fK^a)
+ \fRN\wedge \tr\left[\fr \fw\right]
-\tr\left[\fr^2\fsigma\right]
 \right.\nonumber\\
 &&\left.\qquad\qquad\qquad
-\frac{3}{4} \miepsab(\Dhat \fK^b)\cdot  \fomega\cdot (\Dhat \fK^a)
+\frac{3}{4} (\Dhat \fK^{a} )\cdot \fsigma\cdot(\Dhat \fK_{a})
  \right\}\,. \label{eq:7dsingledongbeforeintegral}
\eea

The final step is to rewrite this integrand in the form of $d(\ldots)\,$. 
To do this, we first notice that 
\bea
\fRN^3&=&d\left[ \fGN \wedge \fRN^2\right]\, , \nonumber\\
\fRN(\Dhat \fK^a)\cdot(\Dhat \fK_{a} )
-2\fRN^2\wedge \fsigma
&=&d\left[
\fRN\wedge   \fK^a
\cdot(\Dhat \fK_a)
\right]
-2 \fRN\wedge \tr\left[\fr \fw\right]\,. 
\eea 
Here we have used\footnote{More generally
\bea
&&(\Dhat \fK^b)
\cdot(\Dhat \fK_a)
=
\Dhat\left[  \fK^b
\cdot(\Dhat \fK_a)\right]
+  (\fK^b
\cdot \fr \cdot \fK_a)
+\delta^b_a\fRN \wedge \fsigma\, ,  \nonumber \\
&&(\Dhat \fK^a{}_i)\wedge
(\Dhat \fK_a{}_j)
=
D\left[ \fK^a{}_i\wedge
(\Dhat \fK_a{}_j)
\right]
-  
2 (\fr\fw)_{ji}
  +2  \fRN\wedge \fsigma_{ji}\, ,  \nonumber \\
&& i\ve^a{}_b (\Dhat \fK^b{}_i)\wedge
(\Dhat \fK_a{}_j)
=
{D}\left[ i\ve^a{}_b (\fK^b{}_i)\wedge
(\Dhat \fK_a{}_j)
\right]
-2 (\fr\fsigma)_{ji}
+2\fRN\wedge \fw_{ji}\,.
\eea
}

  \be\label{eq:useful11}
 (\Dhat \fK^a)
\cdot(\Dhat \fK_a)
=
d\left[  \fK^a
\cdot(\Dhat \fK_a)\right]
-2\tr[
\fr \fw]
+2 \fRN\wedge \fsigma\, , 
 \ee
 which follows from
 \be
 D^2 \fK_{ai}= \fr_{i}{}^{j} \wedge \fK_{aj}\, , \quad
 \Dhat^2 \fK_{ai}
 =
 \fr_i{}^j \wedge \fK_{aj}
  +\iepsddab \fRN \wedge \fK^{b}{}_{i}\,.
\ee Furthermore, we can also have 
\bea
 d\fsigma&= &\iepsUUab
(\fK_{b}\cdot D \fK_{a} )\, , 
\nonumber\\
D\fsigma_{ij}&=&
\half \iepsUUab\left[
(\Dhat \fK_{ai} )\wedge \fK_{bj}
+(\Dhat\fK_{aj})\wedge \fK_{bi}  \right]\, , \nonumber\\
D\fw_{ij}&=&
\half \Dhat \fK^a{}_{i} \wedge \fK_{aj}
-\half \fK^a{}_{i} \wedge \Dhat \fK_{aj} \,,
\eea
and\footnote{The easiest way to show this is to take the complex $x^a=(z,\bar{z})$ coordinates (see Appendix \ref{sec:usefulconversion}).}
\bea\label{eq:useful1}
(d\fsigma)\wedge (\fK^c\cdot \Dhat\fK_c)
&=&\iepsUUab (\fK_{b}\cdot \Dhat\fK_{a} )(\fK^c\cdot \Dhat\fK_c) 
\nonumber\\
&=&
-( \Dhat\fK_c  \cdot \fsigma\cdot \Dhat\fK^c) 
- \iepsUUab
(\Dhat \fK_{b} \cdot\fw \cdot \Dhat\fK_a) \,.
\eea 

Using all these useful identities, we can rewrite Eq.~\eqref{eq:7dsingledongbeforeintegral} 
to obtain the final expression for the holographic entanglement entropy formula: 
\bea
  &&\SeeCS^{7d,\,single}\nonumber\\
  &&=2\pi \times 4  \int_{\Sigmatilde}
d\left[
2 \fGN \wedge \fRN^2
-2\fRN \wedge( \fK^a\cdot\Dhat \fK_{a} )
 -{\frac{3}{2}\fsigma\wedge (\fK^a
\cdot{\Dhat} \fK_a)}
+\iepsUUab(\fK_a
\cdot  \fr\cdot 
\Dhat \fK_b
)\right]\, \nonumber\\
&&=
2(2\pi) \times 4  \int_{\Sigma}
\left[
 \fGN \wedge \fRN^2
-\left(\fRN +\frac{3}{4}\fsigma \right)\wedge( \fK^a\cdot\Dhat \fK_{a} )
+\half\iepsUUab(\fK_a
\cdot  \fr\cdot 
\Dhat \fK_b
)\right]\, . 
\eea
It is straightforward to see that this indeed is equivalent to Eq.~(\ref{eq:trR4}) 
by using the conversion formulas summarized in  Appendix~\ref{sec:usefulconversion}. 

\subsection{\fixform{${\fPanom}=  \tr (\fR^2)\wedge \tr (\fR^2)$}}
As the final example, here we deal with the anomaly polynomial corresponding 
to 7d double-trace gravitational Chern-Simons term. 

First of all, the Wald term is evaluated to have
\bea
\SEEWald
&=&2\pi \times 8 \int_{\Sigmatilde}
\left[
 \fRN \wedge \tr( \fR^2)
-  \fsigma \wedge \left(2 \fRN^2- \tr[\fr^2]\right)
-2\fsigma\wedge d[  \fK^a\cdot(\Dhat \fK_a)]
\right] \, . \nonumber\\
\eea

Secondly, for the $\SEEKK$ term, we first compute the integrand as 
  \bea\label{eq:compute1}
&&
4\pi  \left(\frac{\pa^2 L}{\pa R_{\m_1\r_1\n_1\s_1} \pa R_{\m_2\r_2\n_2\s_2}}\right)_tK_{\l_1\r_1\s_1} K_{\l_2\r_2\s_2}\nonumber\\
&&\qquad\times  \left[ (n_{\m_1\m_2} n_{\n_1\n_2}-\ve_{\m_1\m_2} \ve_{\n_1\n_2}) n^{\l_1\l_2} + (n_{\m_1\m_2} \ve_{\n_1\n_2}+\ve_{\m_1\m_2} n_{\n_1\n_2}) \ve^{\l_1\l_2}\right] (\hodge_\Sigma 1)  \nonumber\\
 &&\quad =
2\pi \times 8 \times \left\{
 \fsigma\wedge \left(-2\fRN^2 -\tr[\fr^2] \right)
  +4 t\fsigma \wedge d\left[  \fK^a\cdot(\Dhat \fK_a)\right]
-2td\left[\fsigma\wedge (\fK^c\cdot \Dhat\fK_c)\right]
 \right\}\,. \nonumber\\
\eea
Thus, by integrating this, we obtain
\begin{eqnarray}
\SEEKK = 
2\pi \times 8 \int_{\Sigmatilde} \left[ 
 \fsigma\wedge\left(-2\fRN^2 -\tr[\fr^2] \right)
   +2 \fsigma \wedge d\left[  \fK^a\cdot(\Dhat \fK_a)\right]
- d\left[\fsigma \wedge (\fK^c\cdot \Dhat\fK_c)\right]
 \right ]\,. \nonumber \\
\end{eqnarray} 

Combining these two contributions and after some massaging the expression, we finally get 
the holographic entanglement entropy formula as 
\bea
  \SeeCS^{7d, \, double}
&=&16\pi  \times 2 \int_{\Sigmatilde}d\left[
\fGN \wedge \fRN^2
+ \half   \fGN\wedge \tr[\fr^2]
-\left(\fRN+\half \fsigma \right)\wedge (\fK^a
\cdot{\Dhat} \fK_a)
\right]\, \nonumber \\
&=&
16\pi  \times 2 \int_{\Sigma}\left[
\fGN \wedge \fRN^2
+ \half   \fGN\wedge \tr[\fr^2]
-\left(\fRN+\half \fsigma \right)\wedge (\fK^a
\cdot{\Dhat} \fK_a)
\right]\,, 
 \eea
which reproduces Eq.~(\ref{eq:trR2trR2}) 
after using the conversion formulas summarized in Appendix~\ref{sec:usefulconversion}.

\subsection{Formulas for Conversion}
\label{sec:usefulconversion}
In this part, we summarize some dictionaries for
converting the notation in this Appendix to 
that used in the rest part of this paper.

We first note that, when we take $(z,\bar{z})$-coordinate 
for the directions orthogonal to $\tilde{\Sigma}=\Sigmatilde$, 
we can explicitly write down the induced metric on this two-dimensional plane and 
the binormal at $\tilde{\Sigma}$ as 
\be
n_{z\bar{z}}=1/2,\quad n^{z\bar{z}}=2\, , \quad
\epsilon_{z\bar{z}}=\frac{i}{2}\,,\quad \epsilon^{\bar{z}{z}}=2i \,,\quad 
\epsilon^z{}_z =
-\epsilon_z{}^z= -i\,. 
\ee 
The extrinsic curvatures are:   
\be
\fK\equiv \fK_z   \,,  \quad  \fKbar \equiv  \fK_{\zbar}\, . 
\ee 
Then, by using these, we can also have 
\begin{eqnarray}
&&\Dhat \fK_{j}  =  D \fK_{j}  - \fGN \wedge \fK_{j} \, , \qquad 
\Dhat \fKbar_{j} = D \fKbar_{j}  +\fGN \wedge\fKbar_{j}\,, \nonumber \\
&& \fsigma_{ij}=\half\iepsUUab \fK_{ai} \wedge \fK_{bj}
=
\fK_{i} \wedge \fKbar_{j} 
- \fKbar_{i} \wedge \fK_{j} 
 ,\quad
 \fsigma = \fsigma^i{}_i=-2  (\KK) \,, \nonumber\\
&& \fw_{ij} =  \half \fK^a{}_{i} \wedge \fK_{aj}
=
  \bar{\fK}_{i} \wedge {\fK}_{j}
+ \fK_{i} \wedge \bar{\fK}_{j}\,.
\end{eqnarray}
In the end, we also summarize some extra conversion formulas 
that are useful for our computation: 
\bea
( \fK^a\cdot\Dhat \fK_{a} )&=&
2
(
 \fK\cdot\Dhat \fKbar
+ \fKbar\cdot\Dhat \fK
) \, , \\
\frac{1}{2} \iepsUUab (\fK_a
\cdot  \fr\cdot 
\Dhat \fK_b
)
&=&
\Dhat\fK
\cdot  \fr\cdot 
 \fKbar
-\Dhat\fKbar
\cdot  \fr\cdot 
 \fK \,.
\eea


\section{7d Single-Trace Gravitational Chern-Simons Term}
\label{app:7dCSdetails}
In this section, we will directly evaluate the 7d single-trace gravitational Chern-Simons term on the regularized cone background. We will reproduce the result obtained 
from the anomaly-polynomial-based computation in \S\ref{sec:anomalymethod1}. 

Before the evaluation, let us recall that the 7d single-trace gravitational Chern-Simons term is given  by 
\begin{eqnarray} \label{eq:cssingletrace7dtransgression}
\ICS^{7d,\, single} 
&=& 4\int^1_0 dt\,  \tr 
\left[
\fGamma \wedge (t\fR + t(t-1)\fGamma^2)^3
\right] \nonumber \\
&=& 
 4\int^1_0 dt\,  \tr 
\Bigl[
t^3(t-1)^3\fGamma^7+  3t^3(t-1)^2 \fGamma^5\wedge \fR 
\nonumber \\
&&\qquad\qquad \qquad 
+ t^3(t-1) (2\fGamma^3\wedge \fR^2 + 
\fGamma\wedge \fR\wedge \fGamma^2\wedge \fR)+ 
t^3 \fGamma\wedge \fR^3
\Bigr] \nonumber \\
&=& 
\tr \left[
-\frac{1}{35}\fGamma^7+  \frac{1}{5} \fGamma^5\wedge \fR 
-\frac{1}{5}\fGamma\wedge \fR\wedge \fGamma^2\wedge \fR
 -\frac{2}{5}\fGamma^3\wedge \fR^2+\fGamma\wedge \fR^3
 \right]\, . 
\end{eqnarray}
 Thus, what we need to evaluate is 
 $\tr(\fGamma^7)$,  $\tr( \fGamma^5\wedge \fR)$,  $\tr(\fGamma\wedge \fR\wedge \fGamma^2\wedge \fR)$,  
 $\tr(\fGamma^3\wedge \fR^2)$ and $\tr(\fGamma\wedge \fR^3)$ on the regularized cone background. 
 We will deal with each term separately  and eventually 
 combine them to obtain the holographic entanglement entropy formula. 
 
 \subsection{Term-by-term Contributions}
We now compute the contributions of each term in $\ICS^{7d,\, single}$ one by one.
 \subsubsection{\fixform{$\tr(\fGamma^7)$}}

 \noindent
 \underline{(1) $\partial\bar{\partial}A$ Term} \\
 Since the connection one-form does not contain terms of the form $\partial \bar{\partial}A$, 
 we conclude that there is no this type of contribution:
 \begin{eqnarray}\label{eq:G7ppA}
\int_{{\rm Cone}_{\Sigma, n}}\tr(\fGamma^7)|_{\partial\bar{\partial}A} =0 \, . 
 \end{eqnarray}
 
 \noindent
 \underline{(2) $\partial A \bar{\partial}A$ Term} \\
In the connection one-form, $\partial Adz$ (resp. $\bar{\partial} Ad\bar{z}$) shows up 
only in $\fGamma^z{}_z$ (resp. only in $\fGamma^{\bar{z}}{}_{\bar{z}}$). 
Therefore, the potential nontrivial contributions of this form are 
\begin{eqnarray}
&&{}^{(d)}\fGamma^{z}{}_{z}
{}^{(0)}(\fGamma^2)^{z}{}_{\bar{z}}
 {}^{(d)}\fGamma^{\bar{z}}{}_{\bar{z}} 
{}^{(0)}(\fGamma^3)^{\bar{z}}{}_{{z}} \, , \qquad 
{}^{(d)}\fGamma^{z}{}_{z}
{}^{(0)}(\fGamma^3)^{z}{}_{\bar{z}}
 {}^{(d)}\fGamma^{\bar{z}}{}_{\bar{z}} 
{}^{(0)}(\fGamma^2)^{\bar{z}}{}_{{z}} \, , \nonumber \\
&&{}^{(d)}\fGamma^{z}{}_{z}
{}^{(0)}\fGamma^{z}{}_{\bar{z}}
 {}^{(d)}\fGamma^{\bar{z}}{}_{\bar{z}} 
{}^{(0)}(\fGamma^4)^{\bar{z}}{}_{{z}} \, , \qquad \quad
{}^{(d)}\fGamma^{z}{}_{z}
{}^{(0)}(\fGamma^4)^{z}{}_{\bar{z}}
 {}^{(d)}\fGamma^{\bar{z}}{}_{\bar{z}} 
{}^{(0)}\fGamma^{\bar{z}}{}_{{z}} \, ,  
\end{eqnarray}
but all of them are zero since ${}^{(0)}\fGamma^{z}{}_{\bar{z}}={}^{(0)}\fGamma^{\bar{z}}{}_{{z}}
={}^{(0)}(\fGamma^2)^{z}{}_{\bar{z}}={}^{(0)}(\fGamma^2)^{\bar{z}}{}_{{z}}
=0$. We thus conclude that there is no contribution of this type:
\begin{eqnarray}\label{eq:G7pApA}
\int_{{\rm Cone}_{\Sigma, n}}  \tr(\fGamma^7)|_{\partial A\bar{\partial}A} =0\, . 
\end{eqnarray}

\noindent
 \underline{(3) $\partial A$ or $\bar{\partial}A$ Term} \\
 It is trivial that there is no this type of contribution since 
 $\tr(\fGamma^7)$ does not contain $d\fGamma$:
 \begin{eqnarray}\label{eq:G7pA}
\int_{{\rm Cone}_{\Sigma, n}}  \tr(\fGamma^7)|_{\partial A\,, \bar{\partial}A} =0\, .  
 \end{eqnarray}
 
 \noindent
 \underline{Summary for $\tr(\fGamma^7)$} \\ 
From the above, by combining Eq.~\eqref{eq:G7ppA}, \eqref{eq:G7pApA} and \eqref{eq:G7pA},  
we conclude that $\tr(\fGamma^7)$ does not generate 
any order-$\epsilon$ contribution: 
\begin{eqnarray}\label{eq:G7sum}
\left[\int_{{\rm Cone}_{\Sigma, n}} \tr(\fGamma^7)\right]_{\epsilon} = 0\, . 
\end{eqnarray}

 \subsubsection{\fixform{$\tr( \fGamma^5\wedge \fR)$}}
 As a next step, 
 we evaluate $\tr( \fGamma^5\wedge \fR)= (\fGamma^2\fR)^\mu{}_\nu \wedge (\fGamma^3)^\nu{}_\mu $
 on the regularized cone background. \\
 
  \noindent
 \underline{(1) $\partial\bar{\partial}A$ Term} \\
From the expression of $\fGamma^2\fR$ and $\fGamma^3$ summarized in Appendix \ref{sec:useful}, 
we first note that there are the following potential nontrivial contributions: 
\begin{eqnarray}
&&{}^{(d^2)}(\fGamma^2\fR)^z{}_z {}^{(0)}(\fGamma^3)^z{}_z\, , \qquad 
{}^{(d^2)}(\fGamma^2\fR)^{\bar{z}}{}_{\bar{z}} {}^{(0)}(\fGamma^3)^{\bar{z}}{}_{\bar{z}}\, , \nonumber \\
&&{}^{(d^2)}(\fGamma^2\fR)^i{}_z {}^{(0)}(\fGamma^3)^z{}_i\, , \qquad 
{}^{(d^2)}(\fGamma^2\fR)^{i}{}_{\bar{z}} {}^{(0)}(\fGamma^3)^{\bar{z}}{}_{i}\, . 
\end{eqnarray}
Each term is directly computed as follows:
\begin{eqnarray}
&&{}^{(d^2)}(\fGamma^2\fR)^z{}_z {}^{(0)}(\fGamma^3)^z{}_z
= (-2\partial\bar{\partial}A dz\wedge d\bar{z})\wedge(-2e^{-2A})^2
\left[2\fGamma_N\wedge(\bar{\bf K}\cdot{\bf K})^2+ (\bar{\bf K}\cdot{\bf K})\wedge (\bar{\bf K}\cdot\gamma\cdot 
{\bf K})\right]\, , \nonumber \\
&&{}^{(d^2)}(\fGamma^2\fR)^{\bar{z}}{}_{\bar{z}} {}^{(0)}(\fGamma^3)^{\bar{z}}{}_{\bar{z}}
= (-2\partial\bar{\partial}A dz\wedge d\bar{z})\wedge(-2e^{-2A})^2
\left[2\fGamma_N\wedge(\bar{\bf K}\cdot{\bf K})^2+ (\bar{\bf K}\cdot{\bf K})\wedge ({\bf K}\cdot\gamma\cdot 
\bar{\bf K})\right]\, , \nonumber \\
&&{}^{(d^2)}(\fGamma^2\fR)^i{}_z {}^{(0)}(\fGamma^3)^z{}_i
= (-2\partial\bar{\partial}A dz\wedge d\bar{z}) \wedge
\Bigl[
(-2e^{-2A})[2\fGamma_N\wedge({\bar{\bf K}}\cdot\gamma^2\cdot{\bf K})+(\bar{\bf K}\cdot\gamma^3\cdot{\bf K})]
\nonumber \\
&&\qquad\qquad\qquad\qquad\qquad\qquad\qquad\qquad\qquad \qquad 
+(-2e^{-2A})^2[\fGamma_N\wedge(\bar{\bf K}\cdot {\bf K})^2 +(\bar{\bf K}\cdot {\bf K})\wedge(\bar{\bf K}\cdot\gamma\cdot{\bf K})]
\Bigr]\, , \nonumber \\
&&{}^{(d^2)}(\fGamma^2\fR)^{i}{}_{\bar{z}} {}^{(0)}(\fGamma^3)^{\bar{z}}{}_{i}\
= (-2\partial\bar{\partial}A dz\wedge d\bar{z}) \wedge
\Bigl[
(-2e^{-2A})[2\fGamma_N\wedge({{\bf K}}\cdot\gamma^2\cdot\bar{\bf K})-({\bf K}\cdot\gamma^3\cdot\bar{\bf K})]
\nonumber \\
&&\qquad\qquad\qquad\qquad\qquad\qquad\qquad\qquad\qquad \qquad 
+(-2e^{-2A})^2[\fGamma_N\wedge(\bar{\bf K}\cdot {\bf K})^2 +(\bar{\bf K}\cdot {\bf K})\wedge({\bf K}\cdot\gamma\cdot\bar{\bf K})]
\Bigr]\, . \nonumber \\
\end{eqnarray}
 By summing them and doing the integration, we finally obtain the total of this type of contribution:
 \begin{eqnarray} \label{eq:G5RppA}
&&\int_{{\rm Cone}_{\Sigma, n}} \tr(\fGamma^5\wedge \fR)|_{\partial\bar{\partial}A} \nonumber \\
&&\qquad 
= (4\pi\epsilon)\int_\Sigma
\Biggl[ 12\fGamma_N\wedge(\bar{\bf K}\cdot {\bf K})^2 
-2\fGamma_N\wedge[(\bar{\bf K}\cdot\gamma^2\cdot{\bf K})+({\bf K}\cdot \gamma^2\cdot\bar{\bf K})] 
\nonumber \\
&&\qquad\qquad\qquad\qquad 
+ 4(\bar{\bf K}\cdot {\bf K})\wedge [(\bar{\bf K}\cdot \gamma \cdot {\bf K}) + ({\bf K}\cdot\gamma\cdot\bar{\bf K}) ]
-[
(\bar{\bf K}\cdot \gamma^3\cdot {\bf K}) - ({\bf K}\cdot \gamma^3\cdot \bar{\bf K})
] \Biggr]\, . \nonumber \\
&&
 \end{eqnarray}
 
  \noindent
 \underline{(2) $\partial A \bar{\partial}A$ Term} \\
 For this type of contribution, from the expression of $\fGamma^2\fR$ and $\fGamma^3$ summarized in 
 Appendix \ref{sec:useful},  potential nontrivial possibilities are as follows: 
 \begin{eqnarray}
 &&{}^{(dd)}(\fGamma^2\fR)^i{}_j {}^{(0)}(\fGamma^3)^j{}_i\,, \nonumber \\
 &&{}^{(d)}(\fGamma^2\fR)^z{}_z{}^{(0)}(\fGamma^3)^z{}_z\, , \qquad 
  {}^{(d)}(\fGamma^2\fR)^{\bar{z}}{}_{\bar{z}}{}^{(d)}(\fGamma^3)^{\bar{z}}{}_{\bar{z}}\, , \nonumber \\
 &&{}^{(d)}(\fGamma^2\fR)^z{}_j{}^{(d)}(\fGamma^3)^j{}_z\, , \qquad 
  {}^{(d)}(\fGamma^2\fR)^{\bar{z}}{}_{j}{}^{(d)}(\fGamma^3)^{j}{}_{\bar{z}}\, , \nonumber \\
  &&{}^{(d)}(\fGamma^2\fR)^i{}_z{}^{(d)}(\fGamma^3)^z{}_i\, , \qquad 
  {}^{(d)}(\fGamma^2\fR)^{i}{}_{\bar{z}}{}^{(d)}(\fGamma^3)^{\bar{z}}{}_{i}\, , \nonumber \\
  &&{}^{(d)}(\fGamma^2\fR)^i{}_j{}^{(d)}(\fGamma^3)^j{}_i\, . 
 \end{eqnarray}
 We can evaluate each term as 
 \begin{eqnarray}
 &&{}^{(dd)}(\fGamma^2\fR)^i{}_j {}^{(0)}(\fGamma^3)^j{}_i \nonumber \\
 &&
 = (-4\partial A\bar{\partial}A dz\wedge d\bar{z})
 \wedge\Biggl[(-2e^{-2A})
 [(\bar{\bf K}\cdot\gamma^3\cdot{\bf K})-({\bf K}\cdot\gamma^3\cdot\bar{\bf K})]
 \nonumber \\
 &&\qquad\qquad\qquad \qquad\qquad \qquad 
 +(-2e^{-2A})^2\Bigl[
 2\fGamma_N\wedge (\bar{\bf K}\cdot {\bf K})^2 +2(\bar{\bf K}\cdot {\bf K})
 \wedge[(\bar{\bf K}\cdot\gamma\cdot{\bf K})+({\bf K}\cdot\gamma\cdot\bar{\bf K})]
 \Bigr]
 \Biggr]\, , \nonumber \\
 &&
 {}^{(d)}(\fGamma^2\fR)^z{}_z{}^{(d)}(\fGamma^3)^z{}_z=0\,, \qquad 
  {}^{(d)}(\fGamma^2\fR)^{\bar{z}}{}_{\bar{z}}{}^{(d)}(\fGamma^3)^{\bar{z}}{}_{\bar{z}}=0\, , \nonumber \\
   &&{}^{(d)}(\fGamma^2\fR)^z{}_j{}^{(d)}(\fGamma^3)^j{}_z 
    = (-4\partial\bar{\partial}Adz\wedge d\bar{z})\wedge
    (-2e^{-2A})^2(\bar{\bf K}\cdot {\bf K})\wedge(\bar{\bf K}\cdot\gamma\cdot{\bf K})\, , \nonumber \\
   &&  {}^{(d)}(\fGamma^2\fR)^{\bar{z}}{}_{j}{}^{(d)}(\fGamma^3)^{j}{}_{\bar{z}}
    = (-4\partial\bar{\partial}Adz\wedge d\bar{z})\wedge
    (-2e^{-2A})^2(\bar{\bf K}\cdot {\bf K})\wedge({\bf K}\cdot\gamma\cdot\bar{\bf K})\, , \nonumber \\
    &&
    {}^{(d)}(\fGamma^2\fR)^i{}_z{}^{(d)}(\fGamma^3)^z{}_i=0\,, \qquad 
      {}^{(d)}(\fGamma^2\fR)^{i}{}_{\bar{z}}{}^{(d)}(\fGamma^3)^{\bar{z}}{}_{i}=0\, , \nonumber\\
 &&{}^{(d)}(\fGamma^2\fR)^i{}_j{}^{(d)}(\fGamma^3)^j{}_i \nonumber \\
 && =(-4\partial A \bar{\partial}A dz\wedge d\bar{z})\wedge
      (-2e^{-2A})^2\Bigl[
      -2\fGamma_N\wedge(\bar{\bf K}\cdot {\bf K})^2 - (\bar{\bf K}\cdot {\bf K})\wedge
      [(\bar{\bf K}\cdot\gamma\cdot{\bf K})+({\bf K}\cdot\gamma\cdot\bar{\bf K})]
      \Bigr]\, . 
 \end{eqnarray}
 By summing all of them and then integrating, we conclude that the total of this type of contribution is given by
 \begin{eqnarray}\label{eq:G5RpApA}
 &&\int_{{\rm Cone}_{\Sigma, n}} \tr(\fGamma^5\wedge \fR)|_{\partial A\bar{\partial}A} \nonumber \\
 && \qquad 
 = (4\pi\epsilon) \int_\Sigma
 \Biggl[
 -[(\bar{\bf K}\cdot\gamma^3\cdot {\bf K})- ({\bf K}\cdot\gamma^3\cdot\bar{\bf K})]
 +2(\bar{\bf K}\cdot{\bf K})\wedge[(\bar{\bf K}\cdot\gamma\cdot{\bf K})+({\bf K}\cdot\gamma\cdot \bar{\bf K})]
 \Biggr]\, . \nonumber \\
 &&
 \end{eqnarray}

 \noindent
 \underline{(3) $\partial A$ or $\bar{\partial}A$ Term} \\
By using Appendix \ref{sec:useful}, we can see that all of this type of terms contain
${\bf K}$ and $\bar{\bf K}$. Thus, from the argument of \S\ref{sec:integrationbypart}, it follows that there is no contribution of this type:
 \begin{eqnarray}\label{eq:G5RpA}
  &&\int_{{\rm Cone}_{\Sigma, n}} \tr(\fGamma^5\wedge \fR)|_{\partial A, \, \bar{\partial}A} =0\, . 
 \end{eqnarray}
 
  \noindent
  \underline{Summary for $\tr(\fGamma^5\wedge \fR)$} \\
  By summing up Eqs.~\eqref{eq:G5RppA}, \eqref{eq:G5RpApA} and \eqref{eq:G5RpA}, 
  we finally obtain the order-$\epsilon$ contribution from $\tr(\fGamma^5\wedge \fR)$ as 
  \begin{eqnarray}\label{eq:G5Rsum}
  &&\left[\int_{{\rm Cone}_{\Sigma, n}} \tr(\fGamma^5\wedge \fR)\right]_{\epsilon} \nonumber \\
&&\qquad 
= (4\pi\epsilon)\int_\Sigma
\Biggl[ 12\fGamma_N\wedge(\bar{\bf K}\cdot {\bf K})^2 
-2\fGamma_N\wedge[(\bar{\bf K}\cdot\gamma^2\cdot{\bf K})+({\bf K}\cdot \gamma^2\cdot\bar{\bf K})] 
\nonumber \\
&&\qquad\qquad\qquad\qquad 
+ 6(\bar{\bf K}\cdot {\bf K})\wedge [(\bar{\bf K}\cdot \gamma \cdot {\bf K}) + ({\bf K}\cdot\gamma\cdot\bar{\bf K}) ]
-2[
(\bar{\bf K}\cdot \gamma^3\cdot {\bf K}) - ({\bf K}\cdot \gamma^3\cdot \bar{\bf K})
]
\Biggr]\, . \nonumber \\
&&
  \end{eqnarray} 
 
 \subsubsection{\fixform{$\tr(\fGamma\wedge \fR\wedge \fGamma^2\wedge \fR)$}}
 Here we will evaluate $\tr(\fGamma\wedge \fR\wedge \fGamma^2\wedge \fR)= (\fGamma^2\fR)^{\mu}{}_\nu
 \wedge (\fGamma \fR)^\nu{}_\mu$ on the regularized cone background.

  \noindent
 \underline{(1) $\partial\bar{\partial}A$ Term} \\
The followings are the potential nontrivial terms of this type:
\begin{eqnarray}
&&{}^{(d^2)}(\fGamma^2\fR)^{{z}}{}_{{z}}{}^{(0)}(\fGamma\fR)^{{z}}{}_{{z}}\, ,\qquad 
{}^{(d^2)}(\fGamma^2\fR)^{\bar{z}}{}_{\bar{z}}{}^{(0)}(\fGamma\fR)^{\bar{z}}{}_{\bar{z}}\,, \nonumber \\
&&{}^{(d^2)}(\fGamma^2\fR)^{{i}}{}_{{z}}{}^{(0)}(\fGamma\fR)^{{z}}{}_{{i}}\, ,\qquad 
{}^{(d^2)}(\fGamma^2\fR)^{i}{}_{\bar{z}}{}^{(0)}(\fGamma\fR)^{\bar{z}}{}_{i}\,, \nonumber \\
&&{}^{(0)}(\fGamma^2\fR)^{{z}}{}_{{z}}{}^{(d^2)}(\fGamma\fR)^{{z}}{}_{{z}}\, ,\qquad 
{}^{(0)}(\fGamma^2\fR)^{\bar{z}}{}_{\bar{z}}{}^{(d^2)}(\fGamma\fR)^{\bar{z}}{}_{\bar{z}}\,, \nonumber \\
&&{}^{(0)}(\fGamma^2\fR)^{{z}}{}_{{i}}{}^{(d^2)}(\fGamma\fR)^{{i}}{}_{{z}}\, ,\qquad 
{}^{(0)}(\fGamma^2\fR)^{\bar{z}}{}_{i}{}^{(d^2)}(\fGamma\fR)^{i}{}_{\bar{z}}\,, 
\end{eqnarray}
where each term is computed as 
\begin{eqnarray}
&&
{}^{(d^2)}(\fGamma^2\fR)^{{z}}{}_{{z}}{}^{(0)}(\fGamma\fR)^{{z}}{}_{{z}}\nonumber \\
&& 
= (-2\partial\bar{\partial}A dz\wedge d\bar{z})\wedge
\Bigl[
(-2e^{-2A})\fGamma_N\wedge \fR_N\wedge(\bar{\bf K}\cdot{\bf K}) 
+(-2e^{-2A})^2[\fGamma_N\wedge(\bar{\bf K}\cdot{\bf K})^2+(\bar{\bf K}\cdot{\bf K})\wedge (\bar{\bf K}\cdot\Dhat{\bf K})]
\Bigr]\, , \nonumber \\
&&
{}^{(d^2)}(\fGamma^2\fR)^{\bar{z}}{}_{\bar{z}}{}^{(0)}(\fGamma\fR)^{\bar{z}}{}_{\bar{z}}\nonumber \\
&& 
= (-2\partial\bar{\partial}A dz\wedge d\bar{z})\wedge
\Bigl[
(-2e^{-2A})\fGamma_N\wedge \fR_N\wedge(\bar{\bf K}\cdot{\bf K}) 
+(-2e^{-2A})^2[\fGamma_N\wedge(\bar{\bf K}\cdot{\bf K})^2+(\bar{\bf K}\cdot{\bf K})\wedge ({\bf K}\cdot\Dhat\bar{\bf K})]
\Bigr]\, , \nonumber \\
&&{}^{(d^2)}(\fGamma^2\fR)^{{i}}{}_{{z}}{}^{(0)}(\fGamma\fR)^{{z}}{}_{{i}} \nonumber \\
&& = (-2\partial\bar{\partial}A dz\wedge d\bar{z})\wedge
\Biggl[
(-2e^{2A})\Bigl[\fGamma_N\wedge[(\bar{\bf K}\cdot\fr\cdot {\bf K})+(\Dhat\bar{\bf K}\cdot\gamma\cdot{\bf K})]
+(\bar{\bf K}\cdot \fr \cdot \gamma \cdot {\bf K}) \Bigr] \nonumber \\
&&\qquad\qquad\qquad\qquad \qquad\qquad 
+(-2e^{2A})^2
[\fGamma_N\wedge(\bar{\bf K}\cdot{\bf K})^2 + (\bar{\bf K}\cdot{\bf K})\wedge (\bar{\bf K}\cdot\gamma\cdot {\bf K})  ]
\Biggr]\, , \nonumber \\
&&{}^{(d^2)}(\fGamma^2\fR)^{{i}}{}_{\bar{z}}{}^{(0)}(\fGamma\fR)^{\bar{z}}{}_{{i}} \nonumber \\
&& = (-2\partial\bar{\partial}A dz\wedge d\bar{z})\wedge
\Biggl[
(-2e^{2A})\Bigl[\fGamma_N\wedge[({\bf K}\cdot\fr\cdot \bar{\bf K})+(\Dhat{\bf K}\cdot\gamma\cdot\bar{\bf K})]
-({\bf K}\cdot \fr \cdot \gamma \cdot \bar{\bf K}) \Bigr] \nonumber \\
&&\qquad\qquad\qquad\qquad \qquad\qquad 
+(-2e^{2A})^2
[\fGamma_N\wedge(\bar{\bf K}\cdot{\bf K})^2 + (\bar{\bf K}\cdot{\bf K})\wedge ({\bf K}\cdot\gamma\cdot \bar{\bf K})  ]
\Biggr]\, , \nonumber 
\end{eqnarray}
\begin{eqnarray}
&&{}^{(0)}(\fGamma^2\fR)^{{z}}{}_{{z}}{}^{(d^2)}(\fGamma\fR)^{{z}}{}_{{z}} \nonumber \\
&& = (-2\partial\bar{\partial}A dz\wedge d\bar{z})\wedge
\wedge \Bigl[
(-2e^{-2A})[\fGamma_N\wedge\fR_N\wedge(\bar{\bf K}\cdot{\bf K})+ \fGamma_N\wedge (\bar{\bf K}\cdot \gamma
\cdot \Dhat {\bf K})] +(-2e^{-2A})^2\fGamma_N\wedge (\bar{\bf K}\cdot{\bf K})^2
\Bigr]\,, \nonumber \\
&&{}^{(0)}(\fGamma^2\fR)^{\bar{z}}{}_{\bar{z}}{}^{(d^2)}(\fGamma\fR)^{\bar{z}}{}_{\bar{z}} \nonumber \\
&& = (-2\partial\bar{\partial}A dz\wedge d\bar{z})\wedge
\wedge \Bigl[
(-2e^{-2A})[\fGamma_N\wedge\fR_N\wedge(\bar{\bf K}\cdot{\bf K})+ \fGamma_N\wedge ({\bf K}\cdot \gamma
\cdot \Dhat \bar{\bf K})] +(-2e^{-2A})^2\fGamma_N\wedge (\bar{\bf K}\cdot{\bf K})^2
\Bigr]\,, \nonumber \\
&&{}^{(0)}(\fGamma^2\fR)^{{z}}{}_{{i}}{}^{(d^2)}(\fGamma\fR)^{{i}}{}_{{z}} \nonumber \\
&& = (-2\partial\bar{\partial}A dz\wedge d\bar{z})\wedge\Biggl[
(-2e^{-2A})[\fGamma_N\wedge(\bar{\bf K}\cdot\fr\cdot{\bf K})+(\bar{\bf K}\cdot\gamma\cdot\fr\cdot {\bf K})]
\nonumber \\
&&\qquad\qquad\qquad\qquad \qquad\qquad 
+(-2e^{-2A})^2\Bigl[
(\bar{\bf K}\cdot{\bf K})\wedge[(\Dhat\bar{\bf K}\cdot{\bf K})+(\bar{\bf K}\cdot\gamma\cdot{\bf K})]
+\fGamma_N\wedge (\bar{\bf K}\cdot{\bf K})^2
\Bigr]
\Biggr] \, , \nonumber \\
&&{}^{(0)}(\fGamma^2\fR)^{\bar{z}}{}_{{i}}{}^{(d^2)}(\fGamma\fR)^{{i}}{}_{\bar{z}} \nonumber \\
&& = (-2\partial\bar{\partial}A dz\wedge d\bar{z})\wedge\Biggl[
(-2e^{-2A})[\fGamma_N\wedge({\bf K}\cdot\fr\cdot\bar{\bf K})-({\bf K}\cdot\gamma\cdot\fr\cdot \bar{\bf K})]
\nonumber \\
&&\qquad\qquad\qquad\qquad \qquad\qquad 
+(-2e^{-2A})^2\Bigl[
(\bar{\bf K}\cdot{\bf K})\wedge[(\Dhat{\bf K}\cdot\bar{\bf K})+({\bf K}\cdot\gamma\cdot\bar{\bf K})]
+\fGamma_N\wedge (\bar{\bf K}\cdot{\bf K})^2
\Bigr]
\Biggr] \, . 
\end{eqnarray}
 Therefore, by summing all the contributions above and doing the integration, we finally obtain
   \begin{eqnarray}\label{eq:G2RGRppA}
  &&\int_{{\rm Cone}_{\Sigma, n}} \tr(\fGamma \wedge \fR\wedge \fGamma^2\wedge \fR)|_{\partial\bar{\partial}A}
   \nonumber \\
   &&\qquad 
  = (4\pi\epsilon)\int_{\Sigma}
   \Biggl[
   -4\fGamma_N\wedge \fR_N\wedge (\bar{\bf K}\cdot{\bf K})
 \nonumber \\
 &&\qquad\qquad\qquad \qquad 
 +\fGamma_N\wedge\Bigl[
 -2[(\bar{\bf K}\cdot\fr\cdot{\bf K})+({\bf K}\cdot\fr\cdot\bar{\bf K})] +16(\bar{\bf K}\cdot{\bf K})^2 \nonumber \\
 &&\qquad\qquad\qquad \qquad \qquad \qquad 
 -[(\Dhat{\bf K}\cdot\gamma\cdot {\bf K})+(\Dhat{\bf K}\cdot\gamma\cdot\bar{\bf K})
  + (\bar{\bf K}\cdot\gamma\cdot\Dhat{\bf K})+({\bf K}\cdot\gamma\cdot\Dhat\bar{\bf K})] 
 \Bigr]\nonumber \\
 &&\qquad\qquad\qquad \qquad 
 +4(\bar{\bf K}\cdot{\bf K})\wedge[(\bar{\bf K}\cdot\gamma\cdot{\bf K})+({\bf K}\cdot\gamma\cdot \bar{\bf K})] 
  +4(\bar{\bf K}\cdot{\bf K})\wedge[(\Dhat\bar{\bf K}\cdot{\bf K})+(\Dhat{\bf K}\cdot\bar{\bf K})] \nonumber \\
  &&\qquad\qquad\qquad \qquad \qquad \qquad 
  -[(\bar{\bf K}\cdot\fr\cdot\gamma\cdot {\bf K})-({\bf K}\cdot\fr\cdot\gamma\cdot\bar{\bf K})
  +(\bar{\bf K}\cdot\gamma\cdot\fr\cdot{\bf K})-({\bf K}\cdot\gamma\cdot\fr\cdot\bar{\bf K})]
   \Biggr]\, . \nonumber \\
   &&
  \end{eqnarray}
 
  \noindent
 \underline{(2) $\partial A \bar{\partial}A$ Term} \\
 The nontrivial possibilities are as follows: 
 \begin{eqnarray}
 &&{}^{(dd)}(\fGamma^2\fR)^i{}_j {}^{(0)}(\fGamma\fR)^j{}_i\, ,  \nonumber \\
 &&{}^{(0)}(\fGamma^2\fR)^j{}_z {}^{(dd)}(\fGamma\fR)^z{}_j\, ,  \qquad 
 {}^{(0)}(\fGamma^2\fR)^j{}_{\bar{z}} {}^{(dd)}(\fGamma\fR)^{\bar{z}}{}_j\, , \nonumber \\
 &&{}^{(d)}(\fGamma^2\fR)^z{}_z {}^{(d)}(\fGamma\fR)^z{}_z\, ,  \qquad 
 {}^{(d)}(\fGamma^2\fR)^{\bar{z}}{}_{\bar{z}} {}^{(d)}(\fGamma\fR)^{\bar{z}}{}_{\bar{z}}\, , \nonumber \\
 &&{}^{(d)}(\fGamma^2\fR)^z{}_j {}^{(d)}(\fGamma\fR)^j{}_z\, ,  \qquad 
 {}^{(d)}(\fGamma^2\fR)^{\bar{z}}{}_{j} {}^{(d)}(\fGamma\fR)^{j}{}_{\bar{z}}\, , \nonumber \\
  &&{}^{(d)}(\fGamma^2\fR)^i{}_z {}^{(d)}(\fGamma\fR)^z{}_i\, ,  \qquad 
 {}^{(d)}(\fGamma^2\fR)^{i}{}_{\bar{z}} {}^{(d)}(\fGamma\fR)^{\bar{z}}{}_{i}\, , \nonumber \\
   &&{}^{(d)}(\fGamma^2\fR)^i{}_j {}^{(d)}(\fGamma\fR)^j{}_i\, . 
 \end{eqnarray}
 Each term above is computed as 
 \begin{eqnarray}
  &&{}^{(dd)}(\fGamma^2\fR)^i{}_j {}^{(0)}(\fGamma\fR)^j{}_i\, \nonumber \\
  && = (-4\partial A \bar{\partial}A dz\wedge d\bar{z})
  \wedge\Biggl[
  (-2e^{-2A})[(\bar{\bf K}\cdot\gamma\cdot\fr\cdot{\bf K})-({\bf K}\cdot\gamma\cdot\fr\cdot\bar{\bf K})] 
  \nonumber  \\
    &&\qquad\qquad\qquad \qquad \qquad \qquad 
    +(-2e^{-2A})^2(\bar{\bf K}\cdot{\bf K})\wedge
    \Bigl[
    (\bar{\bf K}\cdot\gamma\cdot{\bf K}) +  ({\bf K}\cdot\gamma\cdot\bar{\bf K}) 
    +(\Dhat\bar{\bf K}\cdot{\bf K})+(\Dhat{\bf K}\cdot\bar{\bf K})
    \Bigr]
  \Biggr] \, , \nonumber \\
&&  {}^{(0)}(\fGamma^2\fR)^j{}_z {}^{(dd)}(\fGamma\fR)^z{}_j \nonumber \\
&& = 
(-4\partial A \bar{\partial}A dz\wedge d\bar{z})
\wedge\Biggl[
(-2e^{-2A})[\fGamma_N\wedge\fR_N\wedge(\bar{\bf K}\cdot{\bf K})+\fR_N\wedge(\bar{\bf K}\cdot\gamma\cdot{\bf K})
+(\bar{\bf K}\cdot\gamma^2\cdot\Dhat{\bf K})
] \nonumber \\
    &&\qquad\qquad\qquad \qquad \qquad \qquad 
  +(-2e^{-2A})^2\Bigl[
  \fGamma_N\wedge (\bar{\bf K}\cdot{\bf K})^2 + 
  (\bar{\bf K}\cdot{\bf K})\wedge[(\bar{\bf K}\cdot\gamma\cdot{\bf K})+(\bar{\bf K}\cdot \Dhat{\bf K})]
  \Bigr]  
\Biggr]\, , \nonumber \\
&&
 {}^{(0)}(\fGamma^2\fR)^j{}_{\bar{z}} {}^{(dd)}(\fGamma\fR)^{\bar{z}}{}_j
 \nonumber \\
 &&=
 (-4\partial A \bar{\partial}A dz\wedge d\bar{z})
\wedge\Biggl[
(-2e^{-2A})[\fGamma_N\wedge\fR_N\wedge(\bar{\bf K}\cdot{\bf K})+\fR_N\wedge({\bf K}\cdot\gamma\cdot\bar{\bf K})
-({\bf K}\cdot\gamma^2\cdot\Dhat\bar{\bf K})
] \nonumber \\
    &&\qquad\qquad\qquad \qquad \qquad \qquad 
  +(-2e^{-2A})^2\Bigl[
  \fGamma_N\wedge (\bar{\bf K}\cdot{\bf K})^2 + 
  (\bar{\bf K}\cdot{\bf K})\wedge[({\bf K}\cdot\gamma\cdot\bar{\bf K})+({\bf K}\cdot \Dhat\bar{\bf K})]
  \Bigr]  
\Biggr]\, , \nonumber 
\end{eqnarray}
\begin{eqnarray}
 &&{}^{(d)}(\fGamma^2\fR)^z{}_z {}^{(d)}(\fGamma\fR)^z{}_z=0\, ,  \qquad 
 {}^{(d)}(\fGamma^2\fR)^{\bar{z}}{}_{\bar{z}} {}^{(d)}(\fGamma\fR)^{\bar{z}}{}_{\bar{z}}=0\, , \nonumber \\
  &&{}^{(d)}(\fGamma^2\fR)^z{}_j {}^{(d)}(\fGamma\fR)^j{}_z
  =(-4\partial A \bar{\partial}A dz\wedge d\bar{z})\wedge(-2e^{-2A})^2(\bar{\bf K}\cdot{\bf K})\wedge 
  (\bar{\bf K}\cdot\gamma\cdot{\bf K})\, , \nonumber \\
 &&{}^{(d)}(\fGamma^2\fR)^{\bar{z}}{}_{j} {}^{(d)}(\fGamma\fR)^{j}{}_{\bar{z}}
   =(-4\partial A \bar{\partial}A dz\wedge d\bar{z})\wedge(-2e^{-2A})^2(\bar{\bf K}\cdot{\bf K})\wedge 
  ({\bf K}\cdot\gamma\cdot\bar{\bf K})\, , \nonumber \\
  &&{}^{(d)}(\fGamma^2\fR)^i{}_z {}^{(d)}(\fGamma\fR)^z{}_i\nonumber \\ 
  && =  (-4\partial A \bar{\partial}A dz\wedge d\bar{z})\wedge
  \Biggl[
  (-2e^{-2A})[-\fGamma_N\wedge\fR_N\wedge (\bar{\bf K}\cdot{\bf K})-\fGamma_N\wedge(\bar{\bf K}\cdot\gamma^2
  \cdot {\bf K})] \nonumber \\
    &&\qquad\qquad\qquad \qquad \qquad \qquad 
      +(-2e^{-2A})^2[-2\fGamma_N\wedge(\bar{\bf K}\cdot{\bf K})^2]
  \Biggr] \, , \nonumber \\
 &&{}^{(d)}(\fGamma^2\fR)^{i}{}_{\bar{z}} {}^{(d)}(\fGamma\fR)^{\bar{z}}{}_{i} \nonumber \\
   && =  (-4\partial A \bar{\partial}A dz\wedge d\bar{z})\wedge
  \Biggl[
  (-2e^{-2A})[-\fGamma_N\wedge\fR_N\wedge (\bar{\bf K}\cdot{\bf K})-\fGamma_N\wedge({\bf K}\cdot\gamma^2
  \cdot \bar{\bf K})] \nonumber \\
    &&\qquad\qquad\qquad \qquad \qquad \qquad 
      +(-2e^{-2A})^2[-2\fGamma_N\wedge(\bar{\bf K}\cdot{\bf K})^2]
  \Biggr] \, , \nonumber \\
     &&{}^{(d)}(\fGamma^2\fR)^i{}_j {}^{(d)}(\fGamma\fR)^j{}_i 
     = (-4\partial A \bar{\partial}A dz\wedge d\bar{z})\wedge (-2e^{-2A})^2\Bigl[
     -(\bar{\bf K}\cdot{\bf K})\wedge[(\Dhat{\bf K}\cdot\bar{\bf K})+(\Dhat\bar{\bf K}\cdot{\bf K})]
     \Bigr]\, . \nonumber \\
 \end{eqnarray}
 Then, by summing all of these terms, after integration, we obtain the following expression:  
    \begin{eqnarray}\label{eq:G2RGRpApA}
  &&\int_{{\rm Cone}_{\Sigma, n}} \tr(\fGamma \wedge \fR\wedge \fGamma^2\wedge \fR)|_{\partial A\bar{\partial}A}
   \nonumber \\
   &&\qquad 
  = (4\pi\epsilon)\int_{\Sigma}
   \Biggl[
   -\fR_N\wedge[(\bar{\bf K}\cdot\gamma\cdot{\bf K})+({\bf K}\cdot\gamma\cdot \bar{\bf K})]
   +\fGamma_N\wedge [(\bar{\bf K}\cdot\gamma^2\cdot{\bf K})+({\bf K}\cdot\gamma^2\cdot\bar{\bf K})-2(\bar{\bf K}\cdot{\bf K})^2]\nonumber \\
     &&\qquad \qquad \qquad\qquad 
     +3(\bar{\bf K}\cdot{\bf K})\wedge [(\bar{\bf K}\cdot\gamma\cdot{\bf K})+({\bf K}\cdot\gamma\cdot\bar{\bf K}) ]
      +(\bar{\bf K}\cdot{\bf K})\wedge[(\Dhat{\bf K}\cdot\bar{\bf K})+(\Dhat\bar{\bf K}\cdot{\bf K})]\nonumber \\
           &&\qquad \qquad \qquad\qquad 
       -[
       (\bar{\bf K}\cdot\gamma\cdot \fr\cdot {\bf K}) -({\bf K}\cdot\gamma\cdot\fr \cdot\bar{\bf K})
       + (\bar{\bf K}\cdot\gamma^2\cdot \Dhat{\bf K}) - ({\bf K}\cdot\gamma^2\cdot\Dhat\bar{\bf K}) 
       ]    
   \Biggr] \, . 
   \end{eqnarray}
 
 \noindent
 \underline{(3) $\partial A$ or $\bar{\partial}A$ Term} \\
 All the terms of this form contain ${\bf K}$ and $\bar{\bf K}$. Therefore, following the argument in \S\ref{sec:integrationbypart}, we conclude that there is no contribution of this type: 
  \begin{eqnarray}\label{eq:G2RGRpA}
  &&\int_{{\rm Cone}_{\Sigma, n}} \tr(\fGamma \wedge \fR\wedge \fGamma^2\wedge \fR)|_{\partial A\, , \bar{\partial}A}
  =0\, . 
   \end{eqnarray}
 
  \noindent
  \underline{Summary for $\tr(\fGamma\wedge \fR\wedge \fGamma^2\wedge \fR)$} \\
    By summing up Eqs.~\eqref{eq:G2RGRppA}, \eqref{eq:G2RGRpApA} and \eqref{eq:G2RGRpA}, 
  the order-$\epsilon$ contribution from $\tr(\fGamma\wedge\fR\wedge\fGamma^2\wedge \fR)$ turns out to be 
  \begin{eqnarray}\label{eq:G2RGRsum}
    &&\left[\int_{{\rm Cone}_{\Sigma, n}} \tr(\fGamma \wedge \fR\wedge \fGamma^2\wedge \fR)\right]_{\epsilon}
    \nonumber \\
    &&\qquad 
    =(4\pi\epsilon)\int_\Sigma
    \Biggl[
   -4\fGamma_N\wedge \fR_N\wedge(\bar{\bf K}\cdot{\bf K}) -\fR_N\wedge[(\bar{\bf K}\cdot\gamma\cdot {\bf K})
   +({\bf K}\cdot\gamma\cdot\bar{\bf K})] \nonumber \\
    &&\qquad \qquad\qquad\qquad 
    +\fGamma_N\wedge 
    \Bigl[
    -2[(\bar{\bf K}\cdot\fr\cdot{\bf K})+({\bf K}\cdot\fr\cdot\bar{\bf K})]
    +[(\bar{\bf K}\cdot\gamma^2\cdot{\bf K})+({\bf K}\cdot\gamma^2\cdot \bar{\bf K})] + 14(\bar{\bf K}\cdot{\bf K})^2
    \nonumber \\ 
    &&\qquad \qquad\qquad\qquad \qquad\qquad 
    -[(\Dhat\bar{\bf K}\cdot\gamma\cdot{\bf K})+(\Dhat{\bf K}\cdot\gamma\cdot\bar{\bf K}) 
     +(\bar{\bf K}\cdot \gamma\cdot \Dhat{\bf K})+({\bf K}\cdot\gamma\cdot\Dhat\bar{\bf K})]
    \Bigr]  \nonumber \\
      &&\qquad \qquad\qquad\qquad 
    +7(\bar{\bf K}\cdot{\bf K})\wedge[(\bar{\bf K}\cdot \gamma\cdot {\bf K})+({\bf K}\cdot\gamma\cdot\bar{\bf K})]
    +5(\bar{\bf K}\cdot{\bf K})\wedge[(\Dhat\bar{\bf K}\cdot{\bf K})+(\Dhat{\bf K}\cdot\bar{\bf K})] \nonumber \\
        &&\qquad \qquad\qquad\qquad 
        -(\bar{\bf K}\cdot \fr \cdot \gamma\cdot {\bf K}) +({\bf K}\cdot \fr \cdot \gamma\cdot \bar{\bf K})
        -2(\bar{\bf K}\cdot\gamma\cdot \fr \cdot {\bf K}) + 2({\bf K}\cdot\gamma\cdot \fr \cdot \bar{\bf K})  \nonumber \\
        &&\qquad \qquad\qquad\qquad 
        -(\bar{\bf K}\cdot \gamma^2\cdot\Dhat{\bf K}) +({\bf K}\cdot\gamma^2\cdot \Dhat\bar{\bf K})  
    \Biggr]\, . \nonumber \\
    &&
  \end{eqnarray} 
 
 \subsubsection{\fixform{$\tr(\fGamma^3\wedge \fR^2)$}}
We next consider $\tr(\fGamma^3\wedge \fR^2)=(\fGamma^2\fR)^\mu{}_\nu\wedge(\fR\fGamma)^\nu{}_\mu$. 

   \noindent
 \underline{(1) $\partial\bar{\partial}A$ Term} \\
 In this case, the followings are potential nontrivial terms:
 \begin{eqnarray}
 &&{}^{(d^2)}(\fGamma^2\fR)^{{z}}{}_{{z}}{}^{(0)}(\fR\fGamma)^{{z}}{}_{{z}}\, , \qquad  
 {}^{(d^2)}(\fGamma^2\fR)^{\bar{z}}{}_{\bar{z}}{}^{(0)}(\fR\fGamma)^{\bar{z}}{}_{\bar{z}}\, , \nonumber \\ 
 &&{}^{(d^2)}(\fGamma^2\fR)^{{i}}{}_{{z}}{}^{(0)}(\fR\fGamma)^{{z}}{}_{{i}}\, , \qquad  
 {}^{(d^2)}(\fGamma^2\fR)^{i}{}_{\bar{z}}{}^{(0)}(\fR\fGamma)^{\bar{z}}{}_{i}\, , \nonumber \\ 
 &&{}^{(0)}(\fGamma^2\fR)^{{z}}{}_{{z}}{}^{(d^2)}(\fR\fGamma)^{{z}}{}_{{z}}\, , \qquad  
 {}^{(0)}(\fGamma^2\fR)^{\bar{z}}{}_{\bar{z}}{}^{(d^2)}(\fR\fGamma)^{\bar{z}}{}_{\bar{z}}\, , \nonumber \\ 
 &&{}^{(0)}(\fGamma^2\fR)^{{j}}{}_{{z}}{}^{(d^2)}(\fR\fGamma)^{{z}}{}_{{j}}\, , \qquad  
 {}^{(0)}(\fGamma^2\fR)^{j}{}_{\bar{z}}{}^{(d^2)}(\fR\fGamma)^{\bar{z}}{}_{j}\, . 
 \end{eqnarray}
 We can evaluate these terms one by one as follows: 
 \begin{eqnarray}
 &&{}^{(d^2)}(\fGamma^2\fR)^{{z}}{}_{{z}}{}^{(0)}(\fR\fGamma)^{{z}}{}_{{z}}\nonumber \\
 && =(-2\partial\bar{\partial}A dz\wedge d\bar{z})\wedge 
 \Biggl[
 (-2e^{-2A})\fGamma_N\wedge \fR_N\wedge(\bar{\bf K}\cdot{\bf K})
 \nonumber \\
    &&\qquad \qquad\qquad\qquad \qquad\qquad 
    +(-2e^{-2A})^2[\fGamma_N\wedge(\bar{\bf K}\cdot{\bf K})^2+(\bar{\bf K}\cdot{\bf K})\wedge(\Dhat\bar{\bf K}\cdot{\bf K})]
 \Biggr]\, , \nonumber \\
 && {}^{(d^2)}(\fGamma^2\fR)^{\bar{z}}{}_{\bar{z}}{}^{(0)}(\fR\fGamma)^{\bar{z}}{}_{\bar{z}} \nonumber \\
   && =(-2\partial\bar{\partial}A dz\wedge d\bar{z})\wedge 
    \Biggl[
 (-2e^{-2A})\fGamma_N\wedge \fR_N\wedge(\bar{\bf K}\cdot{\bf K})
 \nonumber \\
    &&\qquad \qquad\qquad\qquad \qquad\qquad 
    +(-2e^{-2A})^2[\fGamma_N\wedge(\bar{\bf K}\cdot{\bf K})^2+(\bar{\bf K}\cdot{\bf K})\wedge(\Dhat{\bf K}\cdot\bar{\bf K})]
 \Biggr]\, , \nonumber \\
  &&{}^{(d^2)}(\fGamma^2\fR)^{{i}}{}_{{z}}{}^{(0)}(\fR\fGamma)^{{z}}{}_{{i}}  \nonumber \\
 &&= (-2\partial\bar{\partial}A dz\wedge d\bar{z})
 \nonumber \\
     &&\qquad \qquad
 \wedge \Biggl[
  (-2 e^{-2A})[\fGamma_N\wedge \fR_N\wedge(\bar{\bf K}\cdot{\bf K})+\fR_N\wedge(\bar{\bf K}\cdot\gamma\cdot{\bf K})
  +\fGamma_N\wedge(\Dhat\bar{\bf K}\cdot\gamma\cdot{\bf K}) +(\Dhat\bar{\bf K}\cdot\gamma^2\cdot{\bf K})
  ] \nonumber \\
      &&\qquad \qquad\qquad\qquad
    +(-2e^{-2A})^2[\fGamma_N\wedge(\bar{\bf K}\cdot{\bf K})^2 +(\bar{\bf K}\cdot{\bf K})\wedge 
    (\bar{\bf K}\cdot\gamma\cdot {\bf K})]  
  \Biggr]\, , \nonumber \\
    &&{}^{(d^2)}(\fGamma^2\fR)^{i}{}_{\bar{z}}{}^{(0)}(\fR\fGamma)^{\bar{z}}{}_{i} \nonumber \\
 &&= (-2\partial\bar{\partial}A dz\wedge d\bar{z})
 \nonumber \\
     &&\qquad \qquad
 \wedge \Biggl[
  (-2 e^{-2A})[\fGamma_N\wedge \fR_N\wedge(\bar{\bf K}\cdot{\bf K})+\fR_N\wedge({\bf K}\cdot\gamma\cdot\bar{\bf K})
  +\fGamma_N\wedge(\Dhat{\bf K}\cdot\gamma\cdot\bar{\bf K}) -(\Dhat{\bf K}\cdot\gamma^2\cdot\bar{\bf K})
  ] \nonumber \\
      &&\qquad \qquad\qquad\qquad
    +(-2e^{-2A})^2[\fGamma_N\wedge(\bar{\bf K}\cdot{\bf K})^2 +(\bar{\bf K}\cdot{\bf K})\wedge 
    ({\bf K}\cdot\gamma\cdot \bar{\bf K})]  
  \Biggr]\, , \nonumber 
  \end{eqnarray}
  \begin{eqnarray}
   &&
 {}^{(0)}(\fGamma^2\fR)^{{z}}{}_{{z}}{}^{(d^2)}(\fR\fGamma)^{{z}}{}_{{z}} \nonumber \\
&& =  (-2\partial\bar{\partial}A dz\wedge d\bar{z})
 \wedge\Biggl[
 (-2e^{-2A})[\fGamma_N\wedge\fR_N\wedge(\bar{\bf K}\cdot{\bf K})+\fGamma_N\wedge(\bar{\bf K}\cdot\gamma\cdot
 \Dhat{\bf K})] + (-2e^{-2A})^2\fGamma_N\wedge(\bar{\bf K}\cdot{\bf K})^2
 \Biggr]\, , \nonumber \\
   &&
 {}^{(0)}(\fGamma^2\fR)^{\bar{z}}{}_{\bar{z}}{}^{(d^2)}(\fR\fGamma)^{\bar{z}}{}_{\bar{z}} \nonumber \\
&& =  (-2\partial\bar{\partial}A dz\wedge d\bar{z})
 \wedge\Biggl[
 (-2e^{-2A})[\fGamma_N\wedge\fR_N\wedge(\bar{\bf K}\cdot{\bf K})+\fGamma_N\wedge({\bf K}\cdot\gamma\cdot
 \Dhat\bar{\bf K})] + (-2e^{-2A})^2\fGamma_N\wedge(\bar{\bf K}\cdot{\bf K})^2
 \Biggr]\, , \nonumber \\
  &&{}^{(0)}(\fGamma^2\fR)^{{j}}{}_{{z}}{}^{(d^2)}(\fR\fGamma)^{{z}}{}_{{j}} \nonumber \\
  &&
  =  (-2\partial\bar{\partial}A dz\wedge d\bar{z})
  \wedge \Biggl[
  (-2e^{-2A})[\fGamma_N\wedge\fR_N\wedge(\bar{\bf K}\cdot{\bf K}) +\fR_N\wedge(\bar{\bf K}\cdot\gamma\cdot{\bf K}) 
 +(\bar{\bf K}\cdot\gamma^2\cdot\Dhat{\bf K}) ]  \nonumber \\
  &&\qquad \qquad\qquad\qquad\qquad\qquad 
 +(-2e^{-2A})^2\Bigl[\fGamma_N\wedge (\bar{\bf K}\cdot{\bf K})^2 +
 (\bar{\bf K}\cdot{\bf K})\wedge[(\bar{\bf K}\cdot\gamma\cdot{\bf K})+(\bar{\bf K}\cdot\Dhat{\bf K})]
 \Bigr]
 \Biggr]\, , \nonumber \\
   &&{}^{(0)}(\fGamma^2\fR)^{{j}}{}_{\bar{z}}{}^{(d^2)}(\fR\fGamma)^{\bar{z}}{}_{{j}} \nonumber \\
  &&
  =  (-2\partial\bar{\partial}A dz\wedge d\bar{z})
  \wedge \Biggl[
  (-2e^{-2A})[\fGamma_N\wedge\fR_N\wedge(\bar{\bf K}\cdot{\bf K}) +\fR_N\wedge({\bf K}\cdot\gamma\cdot\bar{\bf K}) 
 -({\bf K}\cdot\gamma^2\cdot\Dhat\bar{\bf K}) ]  \nonumber \\
  &&\qquad \qquad\qquad\qquad\qquad\qquad 
 +(-2e^{-2A})^2\Bigl[\fGamma_N\wedge (\bar{\bf K}\cdot{\bf K})^2 +
 (\bar{\bf K}\cdot{\bf K})\wedge[({\bf K}\cdot\gamma\cdot\bar{\bf K})+({\bf K}\cdot\Dhat\bar{\bf K})]
 \Bigr]
 \Biggr]\, . \nonumber \\
 \end{eqnarray}
 By summing them and doing the integration, we finally obtain the total of this type of contribution as  
   \begin{eqnarray}\label{eq:G3R2ppA}
  &&\int_{{\rm Cone}_{\Sigma, n}} \tr(\fGamma^3\wedge \fR^2)|_{\partial\bar{\partial}A} \nonumber \\
&&\qquad 
= (4\pi\epsilon)\int_\Sigma
 \Biggl[
 -8\fGamma_N\wedge\fR_N\wedge(\bar{\bf K}\cdot {\bf K}) 
 -2\fR_N\wedge[(\bar{\bf K}\cdot\gamma\cdot{\bf K})+({\bf K}\cdot\gamma\cdot\bar{\bf K})] 
 \nonumber \\
  &&\qquad \qquad\qquad\qquad 
 +\fGamma_N\wedge\Bigl[
 16(\bar{\bf K}\cdot{\bf K})^2 -
 [(\Dhat\bar{\bf K}\cdot \gamma\cdot {\bf K})+(\Dhat{\bf K}\cdot\gamma\cdot\bar{\bf K})
 +(\bar{\bf K}\cdot\gamma\cdot\Dhat{\bf K})+({\bf K}\cdot\gamma\cdot\Dhat\bar{\bf K})]
 \Bigr]        \nonumber \\ 
 &&\qquad \qquad\qquad\qquad 
 +4(\bar{\bf K}\cdot{\bf K})\wedge[(\Dhat\bar{\bf K}\cdot{\bf K})+(\Dhat{\bf K}\cdot\bar{\bf K})]
 +4(\bar{\bf K}\cdot{\bf K})\wedge[(\bar{\bf K}\cdot\gamma\cdot{\bf K})+({\bf K}\cdot\gamma\cdot\bar{\bf K})]
 \nonumber \\
  &&\qquad \qquad\qquad\qquad 
  -[(\Dhat\bar{\bf K}\cdot\gamma^2\cdot{\bf K})-(\Dhat{\bf K}\cdot\gamma^2\cdot\bar{\bf K})
  +(\bar{\bf K}\cdot\gamma^2\cdot\Dhat{\bf K})-({\bf K}\cdot\gamma^2\cdot\Dhat\bar{\bf K})
  ]
 \Biggr]\, .\nonumber \\
  &&
\end{eqnarray}

  \noindent
\underline{(2) $\partial A \bar{\partial}A$ Term} \\
 The potential nontrivial terms of this type are as follows:
 \begin{eqnarray}
 &&{}^{(dd)}(\fGamma^2\fR)^i{}_j {}^{(0)}(\fR\fGamma)^j{}_i\, , \nonumber \\
 &&{}^{(d)}(\fGamma^2\fR)^z{}_z {}^{(d)}(\fR\fGamma)^z{}_z\, , \qquad 
 {}^{(d)}(\fGamma^2\fR)^{\bar{z}}{}_{\bar{z}} {}^{(d)}(\fR\fGamma)^{\bar{z}}{}_{\bar{z}}\, , \nonumber \\
 &&{}^{(d)}(\fGamma^2\fR)^z{}_j {}^{(d)}(\fR\fGamma)^j{}_z\, , \qquad 
 {}^{(d)}(\fGamma^2\fR)^{\bar{z}}{}_{j} {}^{(d)}(\fR\fGamma)^{j}{}_{\bar{z}}\, , \nonumber \\
 &&{}^{(d)}(\fGamma^2\fR)^i{}_z {}^{(d)}(\fR\fGamma)^z{}_i\, , \qquad 
 {}^{(d)}(\fGamma^2\fR)^{i}{}_{\bar{z}} {}^{(d)}(\fR\fGamma)^{\bar{z}}{}_{i}\, , \nonumber \\
 &&{}^{(d)}(\fGamma^2\fR)^i{}_j {}^{(d)}(\fR\fGamma)^j{}_i\, .
 \end{eqnarray}
 We can compute these quantities directly as 
 \begin{eqnarray}
 &&{}^{(dd)}(\fGamma^2\fR)^i{}_j {}^{(0)}(\fR\fGamma)^j{}_i \nonumber \\
 && = (-4\partial A \bar{\partial}A dz\wedge d\bar{z})
 \wedge \Biggl[
 (-2e^{-2A})[(\bar{\bf K}\cdot\fr\cdot\gamma\cdot{\bf K})-({\bf K}\cdot\fr\cdot\gamma\cdot\bar{\bf K})]
 \nonumber \\
 &&\qquad\qquad\qquad\qquad \qquad\qquad 
 +(-2e^{-2A})^2
 (\bar{\bf K}\cdot{\bf K})\wedge[(\bar{\bf K}\cdot\Dhat{\bf K})+({\bf K}\cdot\Dhat\bar{\bf K})+
 (\bar{\bf K}\cdot \gamma\cdot{\bf K})+({\bf K}\cdot\gamma\cdot\bar{\bf K})]
 \Biggr] \, , \nonumber \\
  &&{}^{(d)}(\fGamma^2\fR)^z{}_z {}^{(d)}(\fR\fGamma)^z{}_z
  \nonumber \\
   && = (-4\partial A \bar{\partial}A dz\wedge d\bar{z})\wedge
   (-2e^{-2A})^2\Bigl[\fGamma_N\wedge(\bar{\bf K}\cdot{\bf K})^2
   -(\bar{\bf K}\cdot{\bf K})\wedge[(\bar{\bf K}\cdot\Dhat{\bf K}) -(\bar{\bf K}\cdot\gamma\cdot{\bf K})]
   \Bigr]\, , \nonumber \\
  &&{}^{(d)}(\fGamma^2\fR)^{\bar{z}}{}_{\bar{z}} {}^{(d)}(\fR\fGamma)^{\bar{z}}{}_{\bar{z}}
  \nonumber \\
   && = (-4\partial A \bar{\partial}A dz\wedge d\bar{z})\wedge
   (-2e^{-2A})^2\Bigl[\fGamma_N\wedge(\bar{\bf K}\cdot{\bf K})^2
   -(\bar{\bf K}\cdot{\bf K})\wedge[({\bf K}\cdot\Dhat\bar{\bf K}) -({\bf K}\cdot\gamma\cdot\bar{\bf K})]
   \Bigr]\, , \nonumber \\
   &&{}^{(d)}(\fGamma^2\fR)^z{}_j {}^{(d)}(\fR\fGamma)^j{}_z \nonumber \\
   && = (-4\partial A \bar{\partial}A dz\wedge d\bar{z})\wedge
   (-2e^{-2A})^2[-\fGamma_N\wedge (\bar{\bf K}\cdot{\bf K})^2+(\bar{\bf K}\cdot{\bf K})\wedge (\bar{\bf K}\cdot\Dhat{\bf K})
   ]\, , \nonumber \\
   &&{}^{(d)}(\fGamma^2\fR)^{\bar{z}}{}_j {}^{(d)}(\fR\fGamma)^j{}_{\bar{z}} \nonumber \\
   && = (-4\partial A \bar{\partial}A dz\wedge d\bar{z})\wedge
   (-2e^{-2A})^2[-\fGamma_N\wedge (\bar{\bf K}\cdot{\bf K})^2+(\bar{\bf K}\cdot{\bf K})\wedge ({\bf K}\cdot\Dhat\bar{\bf K})
   ]\, , \nonumber \\
   &&{}^{(d)}(\fGamma^2\fR)^i{}_z {}^{(d)}(\fR\fGamma)^z{}_i \nonumber \\
   && = (-4\partial A \bar{\partial}A dz\wedge d\bar{z})\wedge
   \Biggl[
   (-2e^{-2A})[\fR_N\wedge(\bar{\bf K}\cdot\gamma\cdot{\bf K})+(\bar{\bf K}\cdot\gamma^3\cdot{\bf K})]
   +(-2e^{-2A})^2[2(\bar{\bf K}\cdot{\bf K})\wedge(\bar{\bf K}\cdot\gamma\cdot{\bf K})]
   \Biggr]\, , \nonumber \\
   &&{}^{(d)}(\fGamma^2\fR)^i{}_{\bar{z}} {}^{(d)}(\fR\fGamma)^{\bar{z}}{}_i \nonumber \\
   && = (-4\partial A \bar{\partial}A dz\wedge d\bar{z})\wedge
   \Biggl[
   (-2e^{-2A})[\fR_N\wedge({\bf K}\cdot\gamma\cdot\bar{\bf K})-({\bf K}\cdot\gamma^3\cdot\bar{\bf K})]
   +(-2e^{-2A})^2[2(\bar{\bf K}\cdot{\bf K})\wedge({\bf K}\cdot\gamma\cdot\bar{\bf K})]
   \Biggr]\, , \nonumber \\
    &&{}^{(d)}(\fGamma^2\fR)^i{}_j {}^{(d)}(\fR\fGamma)^j{}_i \nonumber \\
     && = (-4\partial A \bar{\partial}A dz\wedge d\bar{z})\wedge
     (-2e^{-2A})^2\Bigl[
     -2\fGamma_N\wedge(\bar{\bf K}\cdot{\bf K})^2 -(\bar{\bf K}\cdot{\bf K})\wedge[
     (\bar{\bf K}\cdot\gamma\cdot{\bf K})+({\bf K}\cdot\gamma\cdot \bar{\bf K})]
     \Bigr]  \, . \nonumber \\
     &&
 \end{eqnarray}
 By summing all of them, after integration, we finally obtain the total of this type of contribution as 
    \begin{eqnarray}\label{eq:G3R2pApA}
  &&\int_{{\rm Cone}_{\Sigma, n}} \tr(\fGamma^3\wedge \fR^2)|_{\partial A \bar{\partial}A} \nonumber \\
&&\qquad 
= (4\pi\epsilon)\int_\Sigma
\Biggl[
-\fR_N\wedge[(\bar{\bf K}\cdot\gamma\cdot{\bf K})+({\bf K}\cdot\gamma\cdot\bar{\bf K})]
-2\fGamma_N\wedge(\bar{\bf K}\cdot{\bf K})^2
\nonumber \\
  &&\qquad \qquad\qquad\qquad 
 +(\bar{\bf K}\cdot{\bf K})\wedge[(\bar{\bf K}\cdot\Dhat{\bf K})+({\bf K}\cdot\Dhat\bar{\bf K})]
  +3(\bar{\bf K}\cdot{\bf K})\wedge[(\bar{\bf K}\cdot\gamma\cdot{\bf K})+({\bf K}\cdot\gamma\cdot \bar{\bf K})]
  \nonumber \\
  &&\qquad \qquad\qquad\qquad 
  -[(\bar{\bf K}\cdot\fr\cdot\gamma\cdot{\bf K})-({\bf K}\cdot\fr\cdot\gamma\cdot\bar{\bf K}) ]
  -[(\bar{\bf K}\cdot\gamma^3\cdot{\bf K})-({\bf K}\cdot\gamma^3\cdot\bar{\bf K})]
\Biggr]\, . \nonumber \\
\end{eqnarray}
 
 \noindent
 \underline{(3) $\partial A$ or $\bar{\partial}A$ Term} \\
 In a similar way to the previous cases, we can see that all terms of this type 
 contain ${\bf K}$ and $\bar{\bf K}$. Therefore, we conclude that there is no contribution 
 of this type: 
     \begin{eqnarray}\label{eq:G3R2pA}
  &&\int_{{\rm Cone}_{\Sigma, n}} \tr(\fGamma^3\wedge \fR^2)|_{\partial A\, ,  \bar{\partial}A}=0 \, . 
  \end{eqnarray}
 
  \noindent
  \underline{Summary for $\tr(\fGamma^3\wedge \fR^2)$} \\ 
    By summing up Eqs.~\eqref{eq:G3R2ppA}, \eqref{eq:G3R2pApA} and \eqref{eq:G3R2pA}, 
  we conclude that the order-$\epsilon$ contribution from $\tr(\fGamma^3\wedge \fR^2)$
  is given  by 
     \begin{eqnarray}\label{eq:G3R2sum}
  &&\left[\int_{{\rm Cone}_{\Sigma, n}} \tr(\fGamma^3\wedge \fR^2)\right]_{\epsilon} \nonumber \\
 &&\qquad 
= (4\pi\epsilon)\int_\Sigma
\Biggl[
-8\fGamma_N\wedge\fR_N\wedge(\bar{\bf K}\cdot{\bf K}) 
-3\fR_N\wedge[(\bar{\bf K}\cdot\gamma\cdot {\bf K})+({\bf K}\cdot\gamma\cdot\bar{\bf K})]
\nonumber \\
  &&\qquad \qquad\qquad\qquad 
  +\fGamma_N\wedge
  \Bigl[
  14(\bar{\bf K}\cdot{\bf K})^2 -[
  (\Dhat\bar{\bf K}\cdot\gamma\cdot{\bf K}) +(\Dhat{\bf K}\cdot\gamma\cdot\bar{\bf K})
  +(\bar{\bf K}\cdot\gamma\cdot\Dhat{\bf K})+({\bf K}\cdot\gamma\cdot\Dhat\bar{\bf K})
  ]   \Bigr]\nonumber \\
    &&\qquad \qquad\qquad\qquad 
    +
    5(\bar{\bf K}\cdot{\bf K})\wedge[(\Dhat\bar{\bf K}\cdot{\bf K})+(\Dhat{\bf K}\cdot\bar{\bf K})]
    +7(\bar{\bf K}\cdot{\bf K})\wedge[(\bar{\bf K}\cdot\gamma\cdot{\bf K})+({\bf K}\cdot\gamma\cdot\bar{\bf K})]
    \nonumber \\
    &&\qquad \qquad\qquad\qquad 
    -[(\bar{\bf K}\cdot\fr \cdot\gamma\cdot{\bf K})-({\bf K}\cdot\fr\cdot\gamma\cdot\bar{\bf K})]   
    -[(\bar{\bf K}\cdot\gamma^3\cdot{\bf K})-({\bf K}\cdot\gamma^3\cdot\bar{\bf K})]
    \nonumber \\
&&\qquad \qquad\qquad\qquad  
-[
(\Dhat\bar{\bf K}\cdot\gamma^2\cdot{\bf K})-(\Dhat{\bf K}\cdot\gamma^2\cdot\bar{\bf K})
+(\bar{\bf K}\cdot\gamma^2\cdot\Dhat{\bf K})-({\bf K}\cdot\gamma^2\cdot\Dhat\bar{\bf K})
]
\Biggr] \, . \nonumber \\
&&
  \end{eqnarray}
 
 \subsubsection{\fixform{$\tr(\fGamma\wedge \fR^3)$}}
 In the end, here we evaluate $\tr(\fGamma\wedge \fR^3)=(\fGamma\fR)^\mu{}_\nu \wedge(\fR^2)^\nu{}_\mu$ 
 on the regularized cone background. 
 
   \noindent
 \underline{(1) $\partial\bar{\partial}A$ Term} \\
The followings are potential nontrivial terms of this type: 
\begin{eqnarray}
&&{}^{(d^2)}(\fGamma\fR)^{{z}}{}_{{z}}{}^{(0)}(\fR^2)^{{z}}{}_{{z}}\, , \qquad 
{}^{(d^2)}(\fGamma\fR)^{\bar{z}}{}_{\bar{z}}{}^{(0)}(\fR^2)^{\bar{z}}{}_{\bar{z}}\,, \nonumber \\
&&{}^{(d^2)}(\fGamma\fR)^{{i}}{}_{{z}}{}^{(0)}(\fR^2)^{{z}}{}_{{i}}\, , \qquad 
{}^{(d^2)}(\fGamma\fR)^{i}{}_{\bar{z}}{}^{(0)}(\fR^2)^{\bar{z}}{}_{i}\,, \nonumber \\
&&{}^{(0)}(\fGamma\fR)^{{z}}{}_{{z}}{}^{(d^2)}(\fR^2)^{{z}}{}_{{z}}\, , \qquad 
{}^{(0)}(\fGamma\fR)^{\bar{z}}{}_{\bar{z}}{}^{(d^2)}(\fR^2)^{\bar{z}}{}_{\bar{z}}\,, \nonumber \\
&&{}^{(0)}(\fGamma\fR)^{{j}}{}_{{z}}{}^{(d^2)}(\fR^2)^{{z}}{}_{{j}}\, , \qquad 
{}^{(0)}(\fGamma\fR)^{j}{}_{\bar{z}}{}^{(d^2)}(\fR^2)^{\bar{z}}{}_{j}\,, \nonumber \\
&&{}^{(0)}(\fGamma\fR)^{{z}}{}_{i}{}^{(d^2)}(\fR^2)^{i}{}_{{z}}\, , \qquad 
{}^{(0)}(\fGamma\fR)^{\bar{z}}{}_{i}{}^{(d^2)}(\fR^2)^{i}{}_{\bar{z}}\,, \nonumber 
\end{eqnarray}
each of which is computed as 
\begin{eqnarray}
&&{}^{(d^2)}(\fGamma\fR)^{{z}}{}_{{z}}{}^{(0)}(\fR^2)^{{z}}{}_{{z}} \nonumber \\
&&=(-2\partial\bar{\partial}Adz\wedge d\bar{z})\wedge
\Bigl[\fGamma_N\wedge \fR_N^2 +(-2e^{-2A})[2\fGamma_N\wedge \fR_N \wedge(\bar{\bf K}\cdot{\bf K})
+\fGamma_N\wedge(\Dhat\bar{\bf K}\cdot\Dhat{\bf K})]
\nonumber \\
&&\qquad \qquad\qquad\qquad\qquad\qquad   
+(-2e^{-2A})^2\fGamma_N\wedge(\bar{\bf K}\cdot{\bf K})^2
\Bigr]\, , \nonumber \\ 
&&{}^{(d^2)}(\fGamma\fR)^{\bar{z}}{}_{\bar{z}}{}^{(0)}(\fR^2)^{\bar{z}}{}_{\bar{z}} \nonumber \\
&&=(-2\partial\bar{\partial}A dz\wedge d\bar{z})\wedge
\Bigl[\fGamma_N\wedge \fR_N^2 +(-2e^{-2A})[2\fGamma_N\wedge \fR_N \wedge(\bar{\bf K}\cdot{\bf K})
+\fGamma_N\wedge(\Dhat\bar{\bf K}\cdot\Dhat{\bf K})]
\nonumber \\
&&\qquad \qquad\qquad\qquad\qquad\qquad   
+(-2e^{-2A})^2\fGamma_N\wedge(\bar{\bf K}\cdot{\bf K})^2
\Bigr]\, , \nonumber \\ 
&&{}^{(d^2)}(\fGamma\fR)^{{i}}{}_{{z}}{}^{(0)}(\fR^2)^{{z}}{}_{{i}} \nonumber \\
&&=(-2\partial\bar{\partial}A dz\wedge d\bar{z})\wedge
\Biggl[
(-2e^{-2A})[\fR_N\wedge(\Dhat\bar{\bf K}\cdot{\bf K})+(\Dhat\bar{\bf K}\cdot\fr\cdot{\bf K})] 
+(-2e^{-2A})^2[2(\bar{\bf K}\cdot{\bf K})\wedge(\Dhat\bar{\bf K}\cdot{\bf K})]
\Biggr]\, , \nonumber \\
&&{}^{(d^2)}(\fGamma\fR)^{{i}}{}_{\bar{z}}{}^{(0)}(\fR^2)^{\bar{z}}{}_{{i}} \nonumber \\
&&=(-2\partial\bar{\partial}A dz\wedge d\bar{z})\wedge
\Biggl[
(-2e^{-2A})[\fR_N\wedge(\Dhat{\bf K}\cdot\bar{\bf K})-(\Dhat{\bf K}\cdot\fr\cdot\bar{\bf K})] 
+(-2e^{-2A})^2[2(\bar{\bf K}\cdot{\bf K})\wedge(\Dhat{\bf K}\cdot\bar{\bf K})]
\Biggr]\, , \nonumber \\
\end{eqnarray}
\begin{eqnarray}
&&{}^{(0)}(\fGamma\fR)^{{z}}{}_{{z}}{}^{(d^2)}(\fR^2)^{{z}}{}_{{z}} \nonumber \\
&&=(-2\partial\bar{\partial}Adz\wedge d\bar{z})\wedge
\Biggl[
2\fGamma_N\wedge\fR_N^2 + (-2e^{-2A})[4\fGamma_N\wedge\fR_N\wedge(\bar{\bf K}\cdot{\bf K})
+2\fR_N\wedge(\bar{\bf K}\cdot\Dhat{\bf K})] 
\nonumber \\
&&\qquad\qquad\qquad\qquad \qquad 
+(-2e^{-2A})^2[2\fGamma_N\wedge(\bar{\bf K}\cdot{\bf K})^2+2(\bar{\bf K}\cdot{\bf K})\wedge(\bar{\bf K}\cdot\Dhat{\bf K})]
\Biggr]\, , \nonumber \\
&&{}^{(0)}(\fGamma\fR)^{\bar{z}}{}_{\bar{z}}{}^{(d^2)}(\fR^2)^{\bar{z}}{}_{\bar{z}} \nonumber \\
&&=(-2\partial\bar{\partial}Adz\wedge d\bar{z})\wedge
\Biggl[
2\fGamma_N\wedge\fR_N^2 + (-2e^{-2A})[4\fGamma_N\wedge\fR_N\wedge(\bar{\bf K}\cdot{\bf K})
+2\fR_N\wedge({\bf K}\cdot\Dhat\bar{\bf K})] 
\nonumber \\
&&\qquad\qquad\qquad\qquad \qquad 
+(-2e^{-2A})^2[2\fGamma_N\wedge(\bar{\bf K}\cdot{\bf K})^2+2(\bar{\bf K}\cdot{\bf K})\wedge({\bf K}\cdot\Dhat\bar{\bf K})]
\Biggr]\, , \nonumber \\
&&{}^{(0)}(\fGamma\fR)^{j}{}_{{z}}{}^{(d^2)}(\fR^2)^{{z}}{}_{j} \nonumber \\
&&=(-2\partial\bar{\partial}A dz\wedge d\bar{z})\wedge
\Biggl
[(-2e^{-2A})[\fR_N\wedge({\bf K}\cdot\Dhat\bar{\bf K})+(\Dhat\bar{\bf K}\cdot\gamma\cdot \Dhat{\bf K})]
+(-2e^{-2A})^2(\bar{\bf K}\cdot{\bf K})\wedge(\Dhat\bar{\bf K}\cdot{\bf K})
\Biggr]\, , \nonumber \\
&&{}^{(0)}(\fGamma\fR)^{j}{}_{\bar{z}}{}^{(d^2)}(\fR^2)^{\bar{z}}{}_{j} \nonumber \\
&&=(-2\partial\bar{\partial}Adz\wedge d\bar{z})\wedge
\Biggl
[(-2e^{-2A})[\fR_N\wedge(\bar{\bf K}\cdot\Dhat{\bf K})-(\Dhat{\bf K}\cdot\gamma\cdot \Dhat\bar{\bf K})]
+(-2e^{-2A})^2(\bar{\bf K}\cdot{\bf K})\wedge(\Dhat{\bf K}\cdot\bar{\bf K})
\Biggr]\, , \nonumber \\
&&{}^{(0)}(\fGamma\fR)^{{z}}{}_{i}{}^{(d^2)}(\fR^2)^{i}{}_{{z}} \nonumber \\
&&=(-2\partial\bar{\partial}Adz\wedge d\bar{z})\wedge
\Biggl[
(-2e^{-2A})[\fGamma_N\wedge(\Dhat\bar{\bf K}\cdot\Dhat{\bf K})+(\bar{\bf K}\cdot\fr\cdot\Dhat{\bf K})]
+ (-2e^{-2A})^2(\bar{\bf K}\cdot{\bf K})\wedge(\bar{\bf K}\cdot\Dhat{\bf K})
\Biggr]\, , \nonumber \\
&&{}^{(0)}(\fGamma\fR)^{\bar{z}}{}_{i}{}^{(d^2)}(\fR^2)^{i}{}_{\bar{z}} \nonumber \\
&&=(-2\partial\bar{\partial}Adz\wedge d\bar{z})\wedge
\Biggl[
(-2e^{-2A})[\fGamma_N\wedge(\Dhat\bar{\bf K}\cdot\Dhat{\bf K})-({\bf K}\cdot\fr\cdot\Dhat\bar{\bf K})]
+ (-2e^{-2A})^2(\bar{\bf K}\cdot{\bf K})\wedge({\bf K}\cdot\Dhat\bar{\bf K})
\Biggr]\, . \nonumber \\
\end{eqnarray}
After summing these and doing the integration, we finally obtain the total of this type of contribution as   
    \begin{eqnarray}\label{eq:GR3ppA}
  &&\int_{{\rm Cone}_{\Sigma, n}} \tr(\fGamma \wedge \fR^3)|_{\partial\bar{\partial}A} \nonumber \\
&&\qquad 
= (4\pi\epsilon)\int_\Sigma
\Biggl[
3\fGamma_N\wedge\fR_N^2
-12\fGamma_N\wedge\fR_N\wedge(\bar{\bf K}\cdot{\bf K}) \nonumber \\
&&\qquad\qquad\qquad\qquad 
-4\fR_N\wedge[(\Dhat\bar{\bf K}\cdot{\bf K})+(\Dhat{\bf K}\cdot\bar{\bf K})] 
-4\fGamma_N\wedge(\Dhat\bar{\bf K}\cdot\Dhat{\bf K})+12\fGamma_N\wedge(\bar{\bf K}\cdot{\bf K})^2 \nonumber \\
&&\qquad\qquad\qquad\qquad 
+ 12(\bar{\bf K}\cdot{\bf K})\wedge[(\bar{\bf K}\cdot\Dhat{\bf K})+({\bf K}\cdot\Dhat\bar{\bf K})]
\nonumber \\
&&\qquad\qquad\qquad\qquad 
-[(\Dhat\bar{\bf K}\cdot\fr\cdot {\bf K})-(\Dhat{\bf K}\cdot\fr\cdot\bar{\bf K})+(\bar{\bf K}\cdot\fr\cdot\Dhat{\bf K})
-({\bf K}\cdot\fr\cdot\Dhat\bar{\bf K})] \nonumber \\
&&\qquad\qquad\qquad\qquad 
-[(\Dhat\bar{\bf K}\cdot\gamma\cdot\Dhat{\bf K})-(\Dhat{\bf K}\cdot\gamma\cdot\Dhat\bar{\bf K})]
\Biggr]\, . \nonumber \\
&&
\end{eqnarray}
 
  \noindent
 \underline{(2) $\partial A \bar{\partial}A$ Term} \\
In this case, there are following potential nontrivial possibilities: 
\begin{eqnarray}
&&{}^{(dd)}(\fGamma\fR)^{{z}}{}_{{j}}{}^{(0)}(\fR^2)^{{j}}{}_{{z}}\, , \qquad 
{}^{(dd)}(\fGamma\fR)^{\bar{z}}{}_{{j}}{}^{(0)}(\fR^2)^{{j}}{}_{\bar{z}}\, , \nonumber \\ 
&&{}^{(0)}(\fGamma\fR)^{{z}}{}_{{z}}{}^{(dd)}(\fR^2)^{{z}}{}_{{z}}\, , \qquad 
{}^{(0)}(\fGamma\fR)^{\bar{z}}{}_{\bar{z}}{}^{(dd)}(\fR^2)^{\bar{z}}{}_{\bar{z}}\, , \nonumber \\
&&{}^{(0)}(\fGamma\fR)^{{j}}{}_{{i}}{}^{(dd)}(\fR^2)^{{i}}{}_{{j}}\, , \nonumber \\
&&{}^{(d)}(\fGamma\fR)^{{z}}{}_{{z}}{}^{(d)}(\fR^2)^{{z}}{}_{{z}}\, , \qquad 
{}^{(d)}(\fGamma\fR)^{\bar{z}}{}_{\bar{z}}{}^{(d)}(\fR^2)^{\bar{z}}{}_{\bar{z}}\, , \nonumber \\ 
&&{}^{(d)}(\fGamma\fR)^{{z}}{}_{{j}}{}^{(d)}(\fR^2)^{{j}}{}_{{z}}\, , \qquad 
{}^{(d)}(\fGamma\fR)^{\bar{z}}{}_{j}{}^{(d)}(\fR^2)^{j}{}_{\bar{z}}\, , \nonumber \\ 
&&{}^{(d)}(\fGamma\fR)^{{i}}{}_{{z}}{}^{(d)}(\fR^2)^{{z}}{}_{{i}}\, , \qquad 
{}^{(d)}(\fGamma\fR)^{i}{}_{\bar{z}}{}^{(d)}(\fR^2)^{\bar{z}}{}_{i}\, , \nonumber \\ 
&&{}^{(d)}(\fGamma\fR)^{{j}}{}_{{i}}{}^{(d)}(\fR^2)^{{i}}{}_{{j}}\, . \nonumber 
\end{eqnarray}
We can compute them as 
\begin{eqnarray}
&&{}^{(dd)}(\fGamma\fR)^{{z}}{}_{{j}}{}^{(0)}(\fR^2)^{{j}}{}_{{z}}\nonumber \\
&&= (-4\partial A \bar{\partial}A dz\wedge d\bar{z})\wedge
\Biggl[
(-2e^{-2A})[\fR_N\wedge(\bar{\bf K}\cdot\Dhat{\bf K})+(\bar{\bf K}\cdot\fr\cdot\Dhat{\bf K})]
+(-2e^{-2A})^2[2(\bar{\bf K}\cdot{\bf K})\wedge(\bar{\bf K}\cdot\Dhat{\bf K})]
\Biggr]\, , \nonumber \\
&&{}^{(dd)}(\fGamma\fR)^{\bar{z}}{}_{{j}}{}^{(0)}(\fR^2)^{{j}}{}_{\bar{z}}\nonumber \\
&&= (-4\partial A \bar{\partial}A dz\wedge d\bar{z})\wedge
\Biggl[
(-2e^{-2A})[\fR_N\wedge({\bf K}\cdot\Dhat\bar{\bf K})-({\bf K}\cdot\fr\cdot\Dhat\bar{\bf K})]
+(-2e^{-2A})^2[2(\bar{\bf K}\cdot{\bf K})\wedge({\bf K}\cdot\Dhat\bar{\bf K})]
\Biggr]\, , \nonumber \\
&&{}^{(0)}(\fGamma\fR)^{{z}}{}_{{z}}{}^{(dd)}(\fR^2)^{{z}}{}_{{z}}\nonumber \\
&&= (-4\partial A \bar{\partial}A dz\wedge d\bar{z})\wedge
\Biggl[
(-2e^{-2A})[-\fGamma_N\wedge\fR_N\wedge(\bar{\bf K}\cdot{\bf K})]
\nonumber \\
&&\qquad\qquad\qquad\qquad\qquad \qquad 
+(-2e^{-2A})^2[-\fGamma_N\wedge(\bar{\bf K}\cdot{\bf K})^2-(\bar{\bf K}\cdot{\bf K})\wedge(\bar{\bf K}\cdot\Dhat{\bf K})]
\Biggr]\, , \nonumber \\
&&{}^{(0)}(\fGamma\fR)^{\bar{z}}{}_{\bar{z}}{}^{(dd)}(\fR^2)^{\bar{z}}{}_{\bar{z}}\nonumber \\
&&= (-4\partial A \bar{\partial}A dz\wedge d\bar{z})\wedge
\Biggl[
(-2e^{-2A})[-\fGamma_N\wedge\fR_N\wedge(\bar{\bf K}\cdot{\bf K})]
\nonumber \\
&&\qquad\qquad\qquad\qquad\qquad \qquad 
+(-2e^{-2A})^2[-\fGamma_N\wedge(\bar{\bf K}\cdot{\bf K})^2-(\bar{\bf K}\cdot{\bf K})\wedge({\bf K}\cdot\Dhat\bar{\bf K})]
\Biggr]\, , \nonumber \\
&&{}^{(0)}(\fGamma\fR)^{j}{}_{i}{}^{(dd)}(\fR^2)^{i}{}_{j}\nonumber \\
&&= (-4\partial A \bar{\partial}A dz\wedge d\bar{z}) \nonumber \\
&&\qquad\qquad
\wedge\Biggl[
(-2e^{-2A})[(\bar{\bf K}\cdot\gamma\cdot\fr\cdot{\bf K})-({\bf K}\cdot\gamma\cdot\fr\cdot\bar{\bf K})]
\nonumber \\
&&\qquad\qquad\qquad
+(-2e^{-2A})^2\Bigl[
(\bar{\bf K}\cdot{\bf K})\wedge[(\bar{\bf K}\cdot\gamma\cdot{\bf K})+({\bf K}\cdot\gamma\cdot\bar{\bf K})]
+(\bar{\bf K}\cdot{\bf K})\wedge [(\Dhat\bar{\bf K}\cdot{\bf K})+(\Dhat{\bf K}\cdot\bar{\bf K})]
\Bigr]
\Biggr]\, , \nonumber
\end{eqnarray} 
\begin{eqnarray}
&&{}^{(d)}(\fGamma\fR)^{{z}}{}_{{z}}{}^{(d)}(\fR^2)^{{z}}{}_{{z}} \nonumber \\
&&= (-4\partial A \bar{\partial}A dz\wedge d\bar{z})
\wedge\Bigl[(-2e^{-2A})\fR_N\wedge (\bar{\bf K}\cdot\Dhat{\bf K})
+(-2e^{-2A})^2[2(\bar{\bf K}\cdot{\bf K})\wedge(\bar{\bf K}\cdot\Dhat{\bf K})]
\Bigr]\, , \nonumber \\
&&{}^{(d)}(\fGamma\fR)^{\bar{z}}{}_{\bar{z}}{}^{(d)}(\fR^2)^{\bar{z}}{}_{\bar{z}} \nonumber \\
&&= (-4\partial A \bar{\partial}A dz\wedge d\bar{z})
\wedge\Bigl[(-2e^{-2A})\fR_N\wedge ({\bf K}\cdot\Dhat\bar{\bf K})
+(-2e^{-2A})^2[2(\bar{\bf K}\cdot{\bf K})\wedge({\bf K}\cdot\Dhat\bar{\bf K})]
\Bigr]\, , \nonumber 
\end{eqnarray}
\begin{eqnarray}
&&{}^{(d)}(\fGamma\fR)^{{z}}{}_{j}{}^{(d)}(\fR^2)^{j}{}_{{z}} \nonumber \\
&&= (-4\partial A \bar{\partial}A dz\wedge d\bar{z})
\wedge\Biggl[
(-2e^{-2A})[-\fGamma_N\wedge\fR_N\wedge(\bar{\bf K}\cdot{\bf K})-\fGamma_N\wedge(\bar{\bf K}\cdot\fr\cdot{\bf K})]
\nonumber \\
&&\qquad\qquad\qquad\qquad\qquad \qquad 
+(-2e^{-2A})^2[-2\fGamma_N\wedge(\bar{\bf K}\cdot{\bf K})^2]
\Biggr]\, , \nonumber \\
&&{}^{(d)}(\fGamma\fR)^{\bar{z}}{}_{j}{}^{(d)}(\fR^2)^{j}{}_{\bar{z}} \nonumber \\
&&= (-4\partial A \bar{\partial}A dz\wedge d\bar{z})
\wedge\Biggl[
(-2e^{-2A})[-\fGamma_N\wedge\fR_N\wedge(\bar{\bf K}\cdot{\bf K})-\fGamma_N\wedge({\bf K}\cdot\fr\cdot\bar{\bf K})]
\nonumber \\
&&\qquad\qquad\qquad\qquad\qquad \qquad 
+(-2e^{-2A})^2[-2\fGamma_N\wedge(\bar{\bf K}\cdot{\bf K})^2]
\Biggr]\, , \nonumber \\
&&{}^{(d)}(\fGamma\fR)^{{i}}{}_{{z}}{}^{(d)}(\fR^2)^{{z}}{}_{{i}} \nonumber \\
&&= (-4\partial A \bar{\partial}A dz\wedge d\bar{z})
\wedge\Biggl[
(-2e^{-2A})[\fR_N\wedge(\bar{\bf K}\cdot\gamma\cdot{\bf K})+(\bar{\bf K}\cdot\fr\cdot\gamma\cdot{\bf K})]
+(-2e^{-2A})^2[2(\bar{\bf K}\cdot{\bf K})\wedge(\bar{\bf K}\cdot\gamma\cdot{\bf K})]
\Biggr]\, , \nonumber \\
&&{}^{(d)}(\fGamma\fR)^{{i}}{}_{\bar{z}}{}^{(d)}(\fR^2)^{\bar{z}}{}_{{i}} \nonumber \\
&&= (-4\partial A \bar{\partial}A dz\wedge d\bar{z})
\wedge\Biggl[
(-2e^{-2A})[\fR_N\wedge({\bf K}\cdot\gamma\cdot\bar{\bf K})-({\bf K}\cdot\fr\cdot\gamma\cdot\bar{\bf K})]
+(-2e^{-2A})^2[2(\bar{\bf K}\cdot{\bf K})\wedge({\bf K}\cdot\gamma\cdot\bar{\bf K})]
\Biggr]\, , \nonumber \\
&&{}^{(d)}(\fGamma\fR)^{{i}}{}_{j}{}^{(d)}(\fR^2)^{j}{}_{{i}} 
= (-4\partial A \bar{\partial}A dz\wedge d\bar{z})\wedge
(-2e^{-2A})^2\Bigl[
-(\bar{\bf K}\cdot{\bf K})\wedge[(\Dhat{\bf K}\cdot\bar{\bf K})+(\Dhat\bar{\bf K}\cdot {\bf K })]
\Bigr]\, . 
\end{eqnarray}
Therefore, by summing all of them and doing integration, we finally obtain 
     \begin{eqnarray}\label{eq:GR3pApA}
  &&\int_{{\rm Cone}_{\Sigma, n}} \tr(\fGamma \wedge \fR^3)|_{\partial A \bar{\partial}A} \nonumber \\
&&\qquad 
= (4\pi\epsilon)\int_\Sigma
\Biggl[
4\fGamma_N\wedge\fR_N\wedge(\bar{\bf K}\cdot{\bf K}) 
-\fR_N\wedge[(\bar{\bf K}\cdot\gamma\cdot{\bf K})+({\bf K}\cdot\gamma\cdot\bar{\bf K})]  \nonumber \\
&&\qquad\qquad\qquad\qquad
-2\fR_N\wedge[(\bar{\bf K}\cdot\Dhat{\bf K})+({\bf K}\cdot\Dhat\bar{\bf K})]
+\fGamma_N\wedge[(\bar{\bf K}\cdot\fr\cdot{\bf K})+({\bf K}\cdot\fr\cdot\bar{\bf K})]
-6\fGamma_N\wedge(\bar{\bf K}\cdot{\bf K})^2
\nonumber \\ 
&&\qquad\qquad\qquad\qquad
+3(\bar{\bf K}\cdot{\bf K})\wedge[(\bar{\bf K}\cdot\gamma\cdot{\bf K})+({\bf K}\cdot\gamma\cdot\bar{\bf K})]
+3(\bar{\bf K}\cdot{\bf K})\wedge[(\Dhat{\bf K}\cdot\bar{\bf K})+(\Dhat\bar{\bf K}\cdot{\bf K})]
\nonumber \\
&&\qquad\qquad\qquad\qquad
-[(\bar{\bf K}\cdot\gamma\cdot\fr\cdot{\bf K})-({\bf K}\cdot\gamma\cdot \fr \cdot \bar{\bf K})
+(\bar{\bf K}\cdot\fr\cdot\gamma\cdot{\bf K}) -({\bf K}\cdot\fr\cdot\gamma\cdot\bar{\bf K})
]\nonumber \\
&&\qquad\qquad\qquad\qquad
-[(\bar{\bf K}\cdot\fr \cdot\Dhat{\bf K}) -({\bf K}\cdot\fr\cdot\Dhat\bar{\bf K})
\Biggr]\, . \nonumber \\
&&
\end{eqnarray}
 
 \noindent
 \underline{(3) $\partial A$ or $\bar{\partial}A$ Term} \\
In this case, there are nontrivial terms of the form 
\begin{eqnarray}
&&{}^{(d)}(\fGamma\fR)^z{}_z {}^{}(\fR)^z{}_z d\fGamma^z{}_z 
= (2\partial A dz)\wedge \fR_N^2\wedge \, d\fGamma^z{}_z +\ldots 
=  (-2\partial\bar{\partial} A dz\wedge d\bar{z})\wedge \fR_N^2\wedge \fGamma_N +\ldots \, , \nonumber \\
&&{}^{(d)}(\fGamma\fR)^{\bar{z}}{}_{\bar{z}} {}^{}(\fR)^{\bar{z}}{}_{\bar{z}} d\fGamma^{\bar{z}}{}_{\bar{z}} 
= (-2\bar{\partial} A d\bar{z})\wedge \fR_N^2\wedge \, d\fGamma^{\bar{z}}{}_{\bar{z}} +\ldots 
=  (-2\partial\bar{\partial} A dz\wedge d\bar{z})\wedge \fR_N^2\wedge \fGamma_N +\ldots \, . \nonumber \\
\end{eqnarray}
 Here the second equality corresponds to the integration by part explained around Eq.~\eqref{eq:integrationbypart}. 
 Therefore, we finally obtain the total of this type of contribution as 
   \begin{eqnarray}\label{eq:GR3pA}
  &&\int_{{\rm Cone}_{\Sigma, n}} \tr(\fGamma \wedge \fR^3)|_{\partial A,\,   \bar{\partial}A} 
  = (4\pi\epsilon)\int_\Sigma \fGamma_N\wedge \fR_N^2\, . 
  \end{eqnarray} 

  \noindent
  \underline{Summary for $\tr(\fGamma\wedge \fR^3)$} \\
      By summing up Eqs.~\eqref{eq:GR3ppA}, \eqref{eq:GR3pApA} and \eqref{eq:GR3pA}, 
  we obtain the order-$\epsilon$ contribution from ${\tr(\fGamma\wedge \fR^3)}$ as 
  \begin{eqnarray}\label{eq:GR3sum}
    &&\left[\int_{{\rm Cone}_{\Sigma, n}} \tr(\fGamma \wedge \fR^3)\right]_{\epsilon} \nonumber \\
&&\qquad 
= (4\pi\epsilon)\int_\Sigma
\Biggl[
4\fGamma_N\wedge \fR_N^2 -8\fGamma_N\wedge \fR_N\wedge(\bar{\bf K}\cdot{\bf K})
-\fR_N\wedge[(\bar{\bf K}\cdot\gamma\cdot {\bf K})+({\bf K}\cdot\gamma\cdot\bar{\bf K})] 
\nonumber \\
&&\qquad\qquad\qquad\qquad 
-6\fR_N\wedge[(\bar{\bf K}\cdot\Dhat{\bf K})+({\bf K}\cdot\Dhat\bar{\bf K})]
-4 \fGamma_N\wedge(\Dhat\bar{\bf K}\cdot\Dhat{\bf K}) 
\nonumber \\
&&\qquad\qquad\qquad\qquad 
+\fGamma_N\wedge[(\bar{\bf K}\cdot \fr\cdot {\bf K})+({\bf K}\cdot \fr\cdot \bar{\bf K})]
+6\fGamma_N\wedge (\bar{\bf K}\cdot{\bf K})^2 \nonumber \\
&&\qquad\qquad\qquad\qquad 
+3(\bar{\bf K}\cdot{\bf K})\wedge[(\bar{\bf K}\cdot\gamma\cdot{\bf K})+({\bf K}\cdot\gamma\cdot\bar{\bf K})]
+15(\bar{\bf K}\cdot{\bf K})\wedge[(\Dhat{\bf K}\cdot\bar{\bf K})+(\Dhat\bar{\bf K}\cdot{\bf K})]
\nonumber \\
&&\qquad\qquad\qquad\qquad 
-[
(\bar{\bf K}\cdot \gamma\cdot\fr\cdot {\bf K})- ({\bf K}\cdot\gamma\cdot \fr\cdot\bar{\bf K})
+(\bar{\bf K}\cdot\fr\cdot \gamma\cdot {\bf K}) - ({\bf K}\cdot \fr\cdot \gamma\cdot \bar{\bf K})
] \nonumber \\
&&\qquad\qquad\qquad\qquad 
-[(\Dhat\bar{\bf K}\cdot \fr \cdot {\bf K})-(\Dhat{\bf K}\cdot \fr\cdot\bar{\bf K})
+2(\bar{\bf K}\cdot \fr\cdot \Dhat{\bf K}) -2 ({\bf K}\cdot \fr \cdot \Dhat\bar{\bf K})
] \nonumber \\
&&\qquad\qquad\qquad\qquad
-[(\Dhat\bar{\bf K}\cdot \gamma\cdot \Dhat{\bf K})-(\Dhat{\bf K}\cdot\gamma\cdot\Dhat\bar{\bf K})]
\Biggr]\, . \nonumber \\
&&
  \end{eqnarray} 
 
 \subsection{Final Result} 
 By substituting Eqs.~\eqref{eq:G7sum}, \eqref{eq:G5Rsum}, \eqref{eq:G2RGRsum}, \eqref{eq:G3R2sum}, \eqref{eq:GR3sum} 
 into Eq.~\eqref{eq:cssingletrace7dtransgression}, we obtain the order-$\epsilon$ contribution to 
 $\ICS^{7d,\, single}$ on the regularized cone background as 
 \begin{eqnarray}
&&\left[\int_{{\rm Cone}_{\Sigma, n}} \ICS^{7d\, single}\right]_\epsilon \nonumber \\
&&\qquad = (16\pi\epsilon) \int_\Sigma
\Biggl[
\fGamma_N\wedge \fR_N^2 + 
[-2\fR_N + 3(\bar{\bf K}\cdot{\bf K})]\wedge [({\bf K}\cdot \Dhat\bar{\bf K})+(\bar{\bf K}\cdot\Dhat{\bf K})]
\nonumber \\
&&\qquad\qquad\qquad\qquad
+ (\Dhat{\bf K}\cdot \fr \cdot \bar{\bf K}) - (\Dhat\bar{\bf K}\cdot \fr \cdot {\bf K})   
\nonumber \\
&&\qquad\qquad \qquad \qquad
+ d\Bigl[
\frac{1}{2}\fGamma_N\wedge[(\bar{\bf K}\cdot \Dhat{\bf K})+({\bf K}\cdot\Dhat\bar{\bf K})]
-\frac{3}{20}\fGamma_N\wedge[(\bar{\bf K}\cdot\gamma\cdot{\bf K})+({\bf K}\cdot \gamma\cdot\bar{\bf K})]
\nonumber \\ 
&&\qquad\qquad \qquad\qquad \qquad  
+ \frac{1}{4}[(\Dhat\bar{\bf K}\cdot\gamma\cdot {\bf K})-(\Dhat{\bf K}\cdot\gamma\cdot\bar{\bf K})]
-\frac{3}{20}[(\bar{\bf K}\cdot\gamma^2\cdot{\bf K})+({\bf K}\cdot\gamma^2\cdot\bar{\bf K})]
\Bigr]
\Biggr] \, . \nonumber \\
&&
 \end{eqnarray}
 Here we have used the fact that $D\gamma^i{}_j = \fr^i{}_j + (\gamma^2)^{i}{}_j$. 
 Therefore, by neglecting the total derivative term, we finally obtain the holographic 
 entanglement entropy formula for the 7d single-trace Chern-Simons term as follows:
 \begin{eqnarray}
&&\SeeCS^{7d,\, single} = (16\pi) \int_\Sigma
\Biggl[
\fGamma_N\wedge \fR_N^2 + 
[-2\fR_N + 3(\bar{\bf K}\cdot{\bf K})]\wedge [({\bf K}\cdot \Dhat\bar{\bf K})+(\bar{\bf K}\cdot\Dhat{\bf K})]
\nonumber \\
&&\qquad\qquad\qquad\qquad\qquad\qquad
+ (\Dhat{\bf K}\cdot \fr \cdot \bar{\bf K}) - (\Dhat\bar{\bf K}\cdot \fr \cdot {\bf K})   \Biggr]\, . 
 \end{eqnarray}
 This result is consistent with the result obtained from the anomaly polynomial in 
 \S\ref{sec:anomalymethod1}.


\bibliographystyle{utphys}
\bibliography{CSEEDraft}

\end{document}